\documentclass[10pt]{article}

\usepackage{latexsym}
\usepackage[dvips]{epsfig} 
\usepackage{amssymb,amsmath,amsthm,mathrsfs} 
\usepackage{graphicx}
\usepackage{lineno}

\usepackage[authoryear,sort &compress]{natbib}
\usepackage{url}
\usepackage{enumerate}
\usepackage{colordvi,color,pspicture}
\usepackage{psfrag}
\usepackage{rotating}
\usepackage{algorithmic}

\newcommand{\Var}{\mathbf{Var}}
\newcommand{\Cov}{\mathbf{Cov}}

\DeclareGraphicsExtensions{.jpg, .eps}
\theoremstyle{plain}

\theoremstyle{definition}

\theoremstyle{remark}

\begin{document}

\title{Technical Report \# KU-EC-08-3:\\
Analysis of Metric Distances and Volumes of Hippocampi Indicates Different
Morphometric Changes over Time in Dementia of Alzheimer Type and
Nondemented Subjects}
\author{
E. Ceyhan$^{1,2^\ast}$, Can Cerito\~{g}lu$^{2}$, M. Faisal Beg$^{3}$, Lei
Wang$^{4}$, John C. Morris$^{5,6}$, John G.\\
Csernansky$^{4,7}$, Michael I. Miller$^{2,8}$, John Tilak Ratnanather$^{2,8}$
}
\date{\today}

\maketitle
\begin{center}
$^{1}$\textit{Dept. of Mathematics, Ko\c{c} University, 34450, Sar{\i}yer, Istanbul, Turkey.}\\
$^{2}$\textit{Center for Imaging Science, The Johns Hopkins University, Baltimore, MD 21218.}\\
$^{3}${\it School of Engineering Science, Simon Fraser University, Burnaby, V5A 1S6, Canada.}\\
$^{4}${\it Dept. of Psychiatry, Washington University School of Medicine, St. Louis, MO 63110.}\\
$^{5}${\it Dept. of Neurology, Washington University School of Medicine, St. Louis, MO 63110.}\\
$^{6}${\it Alzheimer's Disease Research Center, Washington University School of Medicine, St.Louis, MO 63110.}\\
$^{7}${\it Dept. of Anatomy {\&} Neurobiology, Washington University School of Medicine, St.Louis, MO 63110.}\\
$^{8}${\it Institute for Computational Medicine, The Johns Hopkins University, Baltimore, MD 21218.}
\end{center}

\noindent
*\textbf{corresponding author:}\\
Elvan Ceyhan,\\
Dept. of Mathematics, Ko\c{c} University,\\
Rumelifeneri Yolu, 34450 Sar{\i}yer,\\
Istanbul, Turkey\\
{\bf e-mail:} elceyhan@ku.edu.tr\\
{\bf phone:} +90 (212) 338-1845\\
{\bf fax:} +90 (212) 338-1559\\

\noindent
\textbf{short title:} Metric distances between hippocampi predict shape changes

\noindent
\textbf{keywords:} morphometry, computational anatomy, Large
Deformation Diffeomorphic Metric Mapping (LDDMM), hippocampus,
dementia of Alzheimer's type

\newpage

\begin{abstract}
\noindent
In this article, we analyze the morphometry (shape and size) of
hippocampus in subjects with very mild dementia of Alzheimer's type
(DAT) and nondemented controls and how the morphometry changes over a two-year period.
Morphometric differences with respect to a
template hippocampus were measured by the metric distance obtained
from the Large Deformation Diffeomorphic Metric Mapping (LDDMM)
algorithm which was previously used to calculate dense one-to-one
correspondence vector fields between the shapes.
LDDMM assigns metric distances on the space of anatomical images thereby allowing
for the direct comparison and quantization of morphometric changes.
We characterize what additional information the metric distances provide in terms of
size and shape given the volume measurements of the hippocampi.
Moreover, we demonstrate how metric distances can be used in cross-sectional,
longitudinal, and left-right asymmetry comparisons.
We perform a principal component analysis on metric distances and hippocampus,
brain, and intracranial volumes.
We use repeated measures ANOVA models to test the main effects of
and interaction between the diagnosis, duration, and hemisphere
factors to see which factors significantly explain the differences in metric distances.
When a factor is found to be significant,
we use classical parametric and non-parametric tests to compare the
metric distances for that factor.
The analysis of metric distances is then used to compare the effects of aging in the hippocampus.
At baseline, the metric distances for demented subjects are found not
to be significantly different from those for nondemented subjects.
At follow-up, the metric distances for demented subjects were
significantly larger compared to nondemented subjects.
The metric distances for demented subjects increased significantly from
baseline to follow-up but not for nondemented subjects.
We also demonstrate that metric distances can be used in a logistic
regression model for diagnostic discrimination of subjects.
We compare metric distances with the volumes and obtain similar results
in cross-sectional and longitudinal comparisons.
In classification,
the model that uses volume, metric distance, and volume loss over time together
performs better in detecting DAT.
Thus, metric distances with
respect to a template computed via LDDMM can be a powerful tool in
detecting differences in shape in cross-sectional as well as
longitudinal studies.
\end{abstract}

\section{Introduction}
\label{sec:intro}
Numerous post-mortem studies have shown that neurofibrillary tangles
and amyloid plaques characteristic of Alzheimer's Disease (AD) are
prominent within the hippocampus of individuals with mild dementia
of the Alzheimer's type (DAT) and that the distribution of these
neuropathological markers becomes more widespread to include several
regions of the neocortex as the disease process
progresses [1-7].
The accumulation of neurofibrillary tangles and amyloid plaques
characteristic of AD are associated with neuronal damage and
death [8]. Furthermore, macroscopic gray matter
losses from the accumulation of microscopic scale neuronal
destruction are detectable in living subjects using currently
available magnetic resonance (MR) imaging. Specifically, volume
losses within the
hippocampus [9-14]
have recently been reported in subjects with mild-to-moderate AD. In
an unusual study where the antemortem MR scans and post-mortem
material was available for the same subjects, hippocampal volume
losses were shown to be powerful antemortem predictors of AD
neuropathology [15]. Progressive atrophy of the
entire brain has been observed in AD
cases [16]. However, due to the complexity of
the human brain and the non-uniform distribution of AD
neuropathology early in the course of disease, detailed examination
of specific brain regions known to be affected early in the AD
disease process (e.g., hippocampus) may be preferred for
distinguishing preclinical and very mild forms of AD from normal aging [17-19].

Methods developed in the field of Computational Anatomy (CA) that
enable quantification of brain structure volumes and shapes between
and within groups of individuals with and without various
neurological diseases~have emerged from several groups in recent
years [20-25]. Based on the mathematical principles of general
pattern theory [18, 19, 23, 26, 27], these methods combine image-based diffeomorphic maps
between MR scans with representations of brain structures as smooth
manifolds. Because of their high repeatability and sensitivity to
changes in neuroanatomical shapes, they can be especially sensitive
to abnormalities of brain structures associated with early forms of
AD. Using such methods, we previously demonstrated that the combined
assessment of hippocampal volume loss and shape deformity optimally
distinguished subjects with very mild DAT from both elder
nondemented subjects and younger healthy subjects
 [10]. These methods also allowed us to
demonstrate that hippocampal shape deformities associated with very
mild DAT and nondemented aging were distinct [28].
These methods were also extended to quantify changes in
neuroanatomical volumes and shapes within the same individuals over time [29].
Other longitudinal neuroimaging analysis of hippocampal structures in individuals with
AD have also emerged [30-41].

An important task in CA~is the study of neuroanatomical variability.
The anatomic model is a quadruple $(\Omega ,{\cal G},{\cal I},{\cal P}$)
consisting of $\Omega$ the template coordinate space (in ${\rm
R}^3$), defined as the union of 0, 1, 2, and 3-dimensional manifolds,
${\cal G}:\Omega \leftrightarrow \Omega$ a set of diffeomorphic
transformations on $\Omega$, ${\cal I}$ the space of anatomies is
the orbit of a template anatomy $I_0$ under ${\cal G}$, and ${\cal P}$ the
family of probability measures on ${\cal G}$.
In this framework, a geodesic $\phi:[0,1] \to {\cal G}$ is computed where each point
$\phi_t \in {\cal G},t\in [0,1]$ is a diffeomorphism of the domain $\Omega$.
The evolution of the template image $I_0$ along path is
given by $\phi_t I_0 = I_0 \circ \phi_t^{-1}$ such that the end
point of the geodesic connects the template $I_0$ to the target
$I_1$ via $I_1 = \phi_1 I_0 = I_0 \circ \phi_1^{-1}$.   Thus,
anatomical variability in the target is encoded by these geodesic
transformations when a template is fixed.

\begin{figure}[htbp]
\centering
\rotatebox{90}{ \resizebox{5. in}{!}{\includegraphics{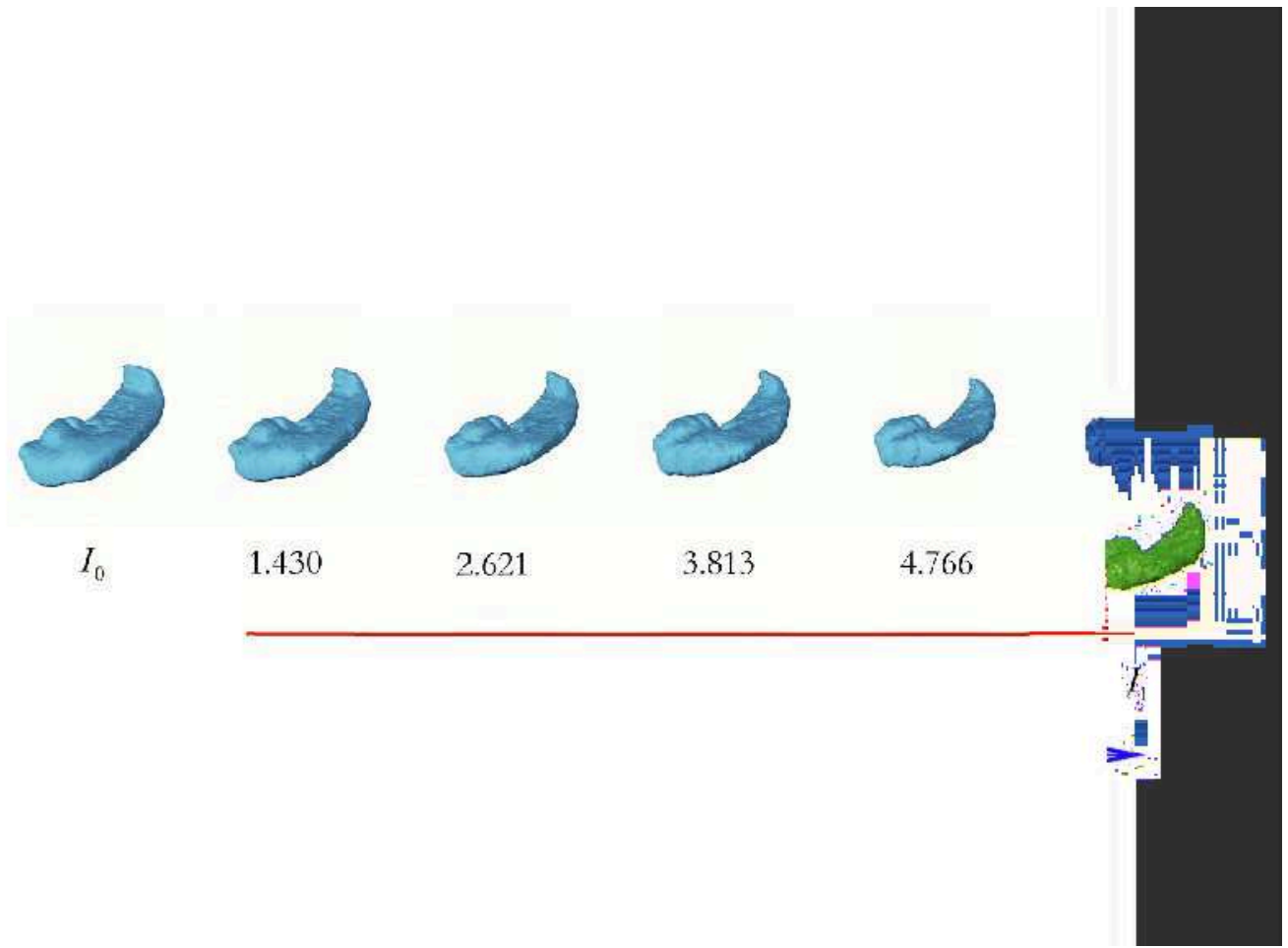}}}
\caption{
\label{fig:change-in-D-diffeo-flow}
Change in metric distance during diffeomorphic flow from
template ($I_0$) to target ($I_1 = \phi_1 I_0 = I_0 \circ \phi
_1^{-1}$).
}
\end{figure}

Furthermore, geodesic curves induce metric distances between the
template and the target shapes in the orbit as follows. The
diffeomorphisms are constructed as a flow of ordinary differential
equations $\dot {\phi}_t = v_t (\phi_t$), $t\in [0,1]$ with $\phi
_0 = id$ the identity map, and associated vector fields, $v_t ,\;t\in
[0,1]$. The optimal velocity vector field parameterizing the
geodesic path is found by solving
\begin{equation}
\label{eqn:optimal-velocity}
\widehat {v} = \mathop {\arg \inf}\limits_{\begin{array}{c}
 v:\phi = \int_0^1 {v_t (\phi_t)dt,} \\
 \phi_0 = id \\
 \end{array}} \int_0^1 {\left\| {v_t} \right\|_V^2 dt\mbox{ such that}I_0 \circ \phi_1^{-1}} = I_1 ,
\end{equation}
where $v_t \in V$, the Hilbert space of smooth vector fields with
norm $\left\| \cdot \right\|_V $ defined through a differential
operator enforcing smoothness. The length of the minimal length path
through the space of transformations connecting the given anatomical
configurations in $I_0$ and $I_1$ defines a metric distance
between anatomical shapes in $I_0$ and $I_1$ via
\begin{equation}
\label{eqn:dist-between-I0andI1}
d(I_0 ,I_1) = \int_0^1 {\left\| {\widehat {v}_t} \right\|_V dt},
\end{equation}
where $\widehat {v}_t $ is the optimizer calculated from the Large
Deformation Diffeomorphic Metric Mapping (LDDMM) algorithm
[42]. Here, the metric distance does not have any units. The
construction of such a metric space allows one to quantify
similarities and differences between anatomical shapes in the orbit.
This is the vision laid out by D'Arcy W. Thompson almost one hundred
years ago.
Figure \ref{fig:change-in-D-diffeo-flow} exemplifies the change in the
metric distance during the evolution of the diffeomorphic map from
the template shape to the target shape.

The notion of mathematical biomarker in the form of metric distance
can be used in different ways. One is to generate metric distances
of shapes relative to a template [42, 43]. Another
is to generate metric distances between each shape within a
collection [44]. The latter approach allows for
sophisticated pattern classification analysis; it is however
computationally expensive. We present an analysis based on the
former approach which could provide a powerful tool in analyzing
subtle shape changes over time with considerably less computational
load. This approach may allow detecting the subtle morphometric
changes observed in the hippocampus in DAT subjects in particular
for those previously analyzed [29, 45].
These studies compared rates of change in hippocampal volume and
shape in subjects with very mild DAT and matched (for age and
gender) nondemented subjects. The change in hippocampal shape over
time was defined as a residual vector field resulting from
rigid-body motion registration, and changes in patterns along
hippocampal surfaces were visualized and analyzed via a statistical
measure of individual and group change in hippocampal shape over
time and used to distinguish between the subject groups. Hence the
motivation to analyze LDDMM generated metric distances between
binary hippocampus images at baseline and at follow-up with respect
to the same template hippocampus image. That is, the template was
compared again, and not propagated between time points. One might
wonder why we do not track changes within a subject directly, rather
than via a reference template, as it could give a more sensitive
measure of shape change since the small difference in shape would
make finding correspondence more accurate. Although we have
considered doing this, the difficulty is that since the template (or
origin) is different for each longitudinal computation, how to
correctly perform statistical comparison of group change is not
completely settled. This is actively being developed by using the
concept of ``parallel transport'' [46, 47].
In this study, we compute and analyze metric distances based on the
data used in [29].

We briefly describe the data set in Section 2.1, computation of
metric distances via LDDMM in Section 2.2, statistical methods we
employ in Section 2.3, and results and findings in Section 3, which
include descriptive summary statistics of the metric distances,
comparison of metric distances of hippocampi of non-demented
subjects and subjects with very mild dementia, correlation between
metric distances, comparison of distributions of metric distances,
and discriminative power of metric distances. We perform similar
analysis on hippocampal volumes in Section 4, compare volumes and
LDDMM distances in Section 5, and analyze annual percentage rate of
change in volumes and distances in Section 6. In the final section,
we discuss the use of metric distances for baseline-followup
studies, group comparisons, and discrimination analysis.

\section{Methods}
\label{sec:methods}
\subsection{Subjects and Data Acquisition}
\label{sec:subjects-data-acquisition}
Detailed description of subjects can be found in
[29] where 18 very mild DAT subjects
(Clinical Dementia Rating Scale, CDR0.5) and 26 age-matched
nondemented controls (CDR0) were each scanned approximately two
years apart. Clinical Dementia Rating (CDR) Scale assessments which
detect the severity of dementia symptoms were performed annually in
all subjects by experienced clinicians without reference to
neuropsychological tests or in-vivo neuroimaging data. The
experienced clinician conducted semi-structured interviews with an
informant and the subject to assess the subject's cognitive and
functional performance; a neurological examination was also
obtained. The clinician determined the presence or absence of
dementia and, when present, its severity with the CDR. Overall CDR
scores of 0 indicate no dementia, while CDR scores of 0.5, 1, 2, and
3 indicate very mild, mild, moderate and severe dementia,
respectively [48]. CDR assessments have been shown to
have an inter-rater reliability of $\kappa = 0.74$
(weighted kappa coefficient [49] $\kappa$ of 0.87) [50], and
this high degree of inter-rater reliability has been confirmed in
multi-center dementia studies [51]. Elderly
subjects with no clinical evidence of dementia (i.e., CDR0 subject) have
been confirmed with normal brains at autopsy with 80\% accuracy;
i.e., approximately 20\% of such individuals show evidence of
AD~[52]. CDR0.5 subjects have subtle
cognitive impairment, and 93\% of them progress to more severe
stages of illness (i.e., CDR $>$ 0.5) and show neuropathological
signs of AD at autopsy ([53],
[54], and [52]).
Although elsewhere the CDR0.5 individuals in our sample may be
considered to have MCI [55], they
fulfill our diagnostic criteria for very mild DAT and at autopsy
overwhelmingly have neuropathologic AD [56].
A summary of subject information is listed in Table \ref{tab:summary-stat-dist}.

The scans were obtained using a Magnetom SP-4000 1.5 Tesla imaging
system, a standard head coil, and a magnetization prepared rapid
gradient echo (MPRAGE) sequence. The MPRAGE sequence
(TR/TE - 10/4, ACQ - 1, Matrix - $256\times 256$, Scanning time - 11.0 min)
produced 3D data with a $1\,mm\times 1\,mm$ in-plane resolution and
1 {\it mm} slice thickness across the entire cranium.

\begin{figure}[htbp]
\centering
\rotatebox{90}{ \resizebox{5. in}{!}{\includegraphics{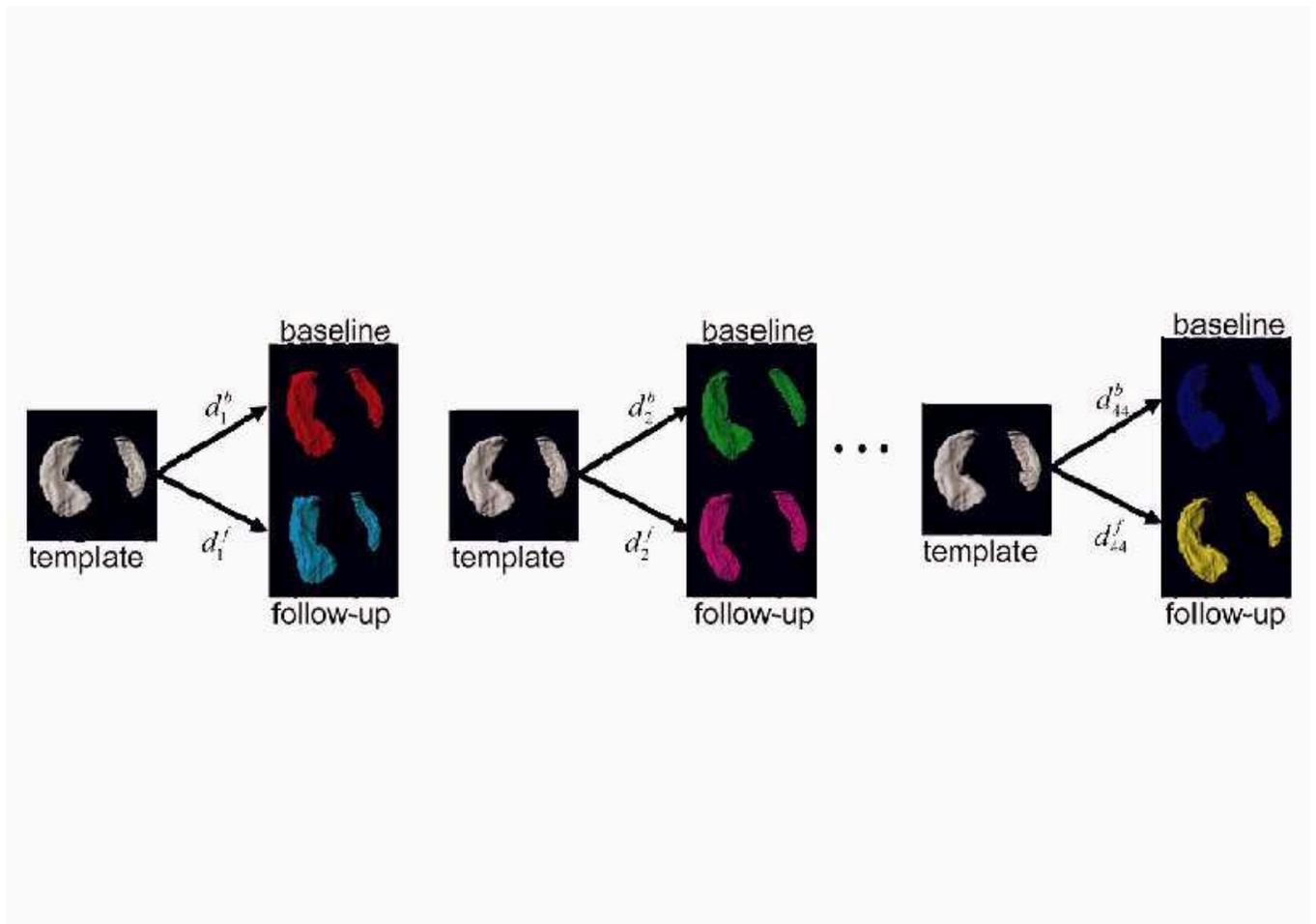}}}
\caption{
\label{fig:generation-D-baseline-followup}
Generation of metric distances $d_k^{\{b,f\}}$ for
subjects $k = 1,\ldots ,44$ at baseline ($b$) and at follow-up ($f$).
}
\end{figure}

A neuroanatomical template was produced using an MR image from an
additional elder control (i.e., CDR0 or non-demented) subject (male,
age = 69). The choice and a detailed description of the template is
provided in [57]. The subject selected to produce
this template was obtained from the same source as the other
subjects in the study, but was not otherwise included in the data
analysis. Data used are the left and right hippocampal surfaces in
the template scan created from expert-produced manual outlines using
methods previously described [28,58], and the left and right hippocampal surfaces of each subject
generated at baseline and follow-up. These surfaces were converted
to binary hippocampus volumetric images by flood filling the inside
of the surface and giving it label 1, and the outside of the surface
was labeled as 0, or background. Each individual hippocampal
surface was first scaled by a factor of 2 and aligned with the
template surface, which was also scaled by a factor of 2, via a
rigid-body rotation and translation before converting to volumetric
binary images. In [58] we showed that mapping accuracy could be
enhanced at higher resolution because of smaller voxels -- voxels at
the periphery of the structure (i.e., surface) account for much more
of the structural volume at $1\;mm^3$ voxel resolution versus $0.5\;mm^3$.
Since then we have adapted this as part of the standard
mapping procedure. These surfaces were then converted into binarized
image of dimension $64\times 112\times 64$ with voxel resolutions of
$0.5\times 0.5\times 0.5\;mm^3,$ followed by smoothing by a Gaussian
filter of $9\times 9\times 9$-voxel window and one voxel standard
deviation to smooth out the edges for LDDMM, which was then applied
to each template-subject pair to compute metric distances, $d_k^b $,
$d_k^f (k = 1,\ldots ,44$), in each hemisphere at baseline ($b$) and
at follow-up ($f$) as illustrated in Figure \ref{fig:generation-D-baseline-followup}.

Controlling for brain size is important because people with bigger
brains tend to have bigger hippocampus and we want our results to
not reflect that very uninteresting fact; we inherently correct for
brain size by first rigid-aligning the subject brain to the
prototype brain prior to LDDMM. Segmentation of hippocampal MRI
shapes across subjects, especially in diseased states, is a
challenging problem. However the accuracy of the segmentation is not
the point of this paper and has been demonstrated before [10, 57, 58].

In addition to the metric distances, our data set also consists of the following variables:
gender, age, education in years
(these variables are used for controlling the confounding affects of these factors on hippocampus morphometry,
so the subjects are taken to be similar or evenly distributed in these variables).
Furthermore, we have brain and intracranial volumes at baseline and followup,
and hippocampus volumes for left and right hippocampi at baseline and follow-up.

\subsection{Computing Metric Distance via Large Deformation Diffeomorphic Metric Mapping}
Metric distances between the binary images and the template image
are obtained by computing diffeomorphisms between the images.
Computation and analysis of these diffeomorphic mappings have been
previously described [57].
Diffeomorphisms are
estimated via the variational problem that, in the space of smooth
velocity vector fields $V$ on domain $\Omega$, takes the
form [42]:
\begin{equation}
\label{eqn:variational-problem}
\widehat {v} = \mathop {\mbox{arg}\,\mbox{min}}\limits_{v:\dot
{\phi}_t = v_t (\phi_t)} \,\left( {\int_0^1 {\left\| {v_t}
\right\|_V^2} dt+\frac{1}{\sigma^2}\left\| {I_0 \circ \phi_1^{-1}
-I_1} \right\|_{L^2}^2} \right).
\end{equation}
The optimizer of this cost generates the optimal change of
coordinates $\varphi = \phi_1^{\widehat {v}}$ upon integration
$d\mathop {\widehat {\phi}}\nolimits_t^v \big/dt = \mathop {\widehat
{v}}\nolimits_t \left( \mathop {\widehat {\phi}}\nolimits_t^v \right)$, $\phi_0 = id$,
where the subscript $v$ in $\phi ^v$ is used to explicitly
denote the dependence of $\phi $ on the associated velocity field
$v$. Enforcing a sufficient amount of smoothness on the elements of
the space $V$ of allowable velocity vector fields ensures that the
solution to the differential equation $\mathop {\dot {\phi
}}\nolimits_t = v_t (\phi_t),\;\,t\in [0,1],\;v_t \in V$ is in the
space of diffeomorphisms~[59, 60]. The required smoothness is
enforced by defining the norm on the space $V$ of smooth velocity
vector fields through a differential operator $L$ of the type
$L = (-\alpha \Delta +\gamma)^\alpha I_{n\times n}$ where $\alpha
>1.5$ in 3-dimensional space such that $\left\| f \right\|_V
 = \left\| {Lf} \right\|_{L_2}$ and $\left\| \cdot \right\|_{L_2}$
is the standard $L_2 $ norm for square integrable functions defined
on $\Omega$. The gradient of this cost is given by
\begin{equation}
\label{eqn:gradient-of-cost}
\nabla_v E_t = 2\mathop {\widehat {v}}\nolimits_t -K\left(
{\frac{2}{\sigma^2}\left\vert D\phi_{t,1}^{\widehat {v}} \right\vert \nabla J_t^0
(J_t^0 -J_t^1)} \right)
\end{equation}
where $J_t^0 = I_0 \circ \phi_t $ and $J_t^1 = I_1 \circ \phi_t^{-1}
$, $\vert Dg\vert $ is the determinant of the Jacobian matrix for $g$ and
$K$ is a compact self-adjoint operator $K:L_2 \left( \Omega ,{\rm R}^d \right)\to V$
uniquely defined by $<a,b>_{L_2} = <Ka,b>_V $ such that for
any smooth vector field $f\in V$, $K(L^\dag L)f = f$ holds. The metric
distance is then calculated via Equation (2).

\subsection{Statistical Methods}
\label{subsec:statistical}
First, we investigate what LDDMM metric distance measures and how it is related
to hippocampus, brain, and intracranial volumes.
That is, as a compound measure of morphometry,
how much of the metric distance is related to shape and size and how it is associated with the volume,
which is mostly a measure of size.
Along this line, we provide the correlation between volume and metric distance measures
by the pairs plots at baseline and follow-up of left and right hippocampi.
Furthermore, we perform a principal component analysis (PCA) on metric distance and volumes
to characterize the major traits these quantities measure.
Then we provide a statistical methodology for the analysis of LDDMM distances.
We compute and interpret simple
summary statistics, such as, mean, standard deviation (SD), minimum,
first quartile ($Q_1$), median, third quartile ($Q_3$), and
maximum for $d_k^{\{b,f\}}$. Then we apply repeated measures
analysis of metric distances with diagnosis group as main effect and
timepoint as the repeated factor, side (i.e., hemisphere) as main
effect and timepoint as the repeated factor, and diagnosis group as
main effect and side-by-timepoint as the repeated factor, since
there are within-subject dependence of metric distances for left and
right hemispheres and at baseline and follow-up. We apply four
possible competing models each assuming a different
variance-covariance structure to obtain the model that best fit to
our data set. The first model assumes compound symmetry, in which
the diagonals (i.e., the variances) are equal, and so are the off
diagonals (i.e., the covariances). The other three models assume
unstructured, autoregressive (AR), and autoregressive heterogeneous
variances, respectively. In the unstructured model, each variance
and covariance term is different, in the AR model, the variances are
assumed to be equal but the covariances change by time, and in the
ARH model, the variances are also different and the covariances
change by time.
The corresponding variance-covariance (Var-Cov) structures [61-63] for the models are shown in
Table \ref{tab:var-cov-for-RepeatedMeasuresANOVA}, where $\sigma^2$ is the common variance
term, $\sigma_i^2$ is the variance for repeated factor $i$,
$\sigma_{ij}$ is the covariance between repeated factors $i$ and
$j$, and $\rho $ is the correlation coefficient of first order in an
autoregressive model. We use various model selection criteria
(Akaike Information Criterion (AIC), Bayesian Information Criterion
(BIC), Log-likelihood) to compare competing models to see which
model best fits our data [64].

For post-hoc comparison of CDR0.5 vs CDR0, our null hypothesis for
the comparison of diagnosis groups, CDR0 and CDR0.5, is $H_o:\mu
_{CDR0} = \mu_{CDR0.5}$ for each baseline left, baseline right, and
follow-up left, follow-up right hippocampi. For the $t$-test, among
the underlying assumptions are the normality of the distributions
and homogeneity of the variances of the independent samples. We
employ Lilliefor's test of normality [65]; and Brown and
Forsythe's (B-F) test (i.e., Levene's test with absolute deviations
from the median) for homogeneity of the variances [66]. If
there is lack of significant deviation from normality of
distribution of a metric distances for a group, we will state it as
``the metric distances for the group can be assumed to come from a
normally distributed population'', henceforth.

We compare metric distances at baseline and follow-up. The LB-CDR0.5
metric distances and LF-CDR0.5 metric distances are dependent as
they come from the same person at baseline and follow-up. Likewise
for LB-CDR0 and LF-CDR0 pairs. Hence our null hypothesis for the
comparison of baseline and follow-up groups is $H_o:\delta (B,F) = 0$
where $\delta (B,F$) is the mean difference of metric distances
between hippocampi at baseline and follow-up for each of CDR0 left,
CDR0 right, CDR0.5 left, CDR0.5 right hippocampi. We also compare
the metric distances for the left and right hippocampi are dependent
(for each left-right hippocampi pair comes from the same subject).
Hence our null hypothesis for the comparison of left and right
metric distances is $H_o:\delta (L,R) = 0$ where $\delta (L,R$) is
the mean difference of metric distances between left and right
metric distances for each of CDR0 baseline, CDR0 follow-up, CDR0.5
baseline, CDR0.5 follow-up hippocampi.

We also calculate and interpret correlation coefficients between
metric distances. Since metric distances of all groups can be
assumed to have normal distribution based on Lilliefor's test of
normality, we use Pearson's correlation coefficient, denoted $r_P$,
between baseline and follow-up (overall and by diagnosis group) and
for the left and right hippocampi and the corresponding tests of
$H_o:r_P = 0\;\;\mbox{vs}\;\;H_a:r_P >0$ for inference [67,68].

We also estimate the empirical cumulative distribution functions
(cdf) of the metric distances and compare them by Kolmogorov-Smirnov
(K-S) test, Cram\'{e}r's test, and Cram\'{e}r-von Mises test. The
null hypothesis for the comparison of cdfs of the metric distances
per diagnosis groups, CDR0 and CDR0.5, is $H_o:F_{CDR0} = F_{CDR0.5}
$ for each baseline left, baseline right, follow-up left, follow-up
right hippocampi. For calculation of the critical value of
Cram\'{e}r's test the kernel $\phi_C(x) = \frac{\sqrt{x}}{2}$
(which is recommended for location alternatives) is used. The
estimated $p$-values are based on $\alpha = 0.05$ and 10000
ordinary bootstrap replicates.

We apply logistic discrimination with metric distances and other variables,
since the diagnosis have only two levels, namely CDR0 and CDR0.5.
We use logistic regression to estimate or predict the risk or probability of
having DAT using metric distances, together with side (i.e., hemisphere)
and timepoint (baseline vs follow-up) factors.
In other words, we model the probability that the subject is CDR0.5
given the metric distance of the subject for left or right hippocampus at baseline or follow-up.
In standard logistic regression the model-parameters are obtained via maximum likelihood estimators.
For more on logistic regression and logistic discrimination, see [69] and [70], respectively.
We consider the logistic model with the response where
(i.e., the probability that the subject is diagnosed with CDR0.5).
First we model with one predictor variable at a time from side, timepoint,
and metric distance, etc., if the variable is not significant at .05 level,
we omit that variable from further consideration.
We consider the full logistic model with the response
$\mbox{logit } p = \log \left[ {p \mathord{\left/ {\vphantom {p {(1-p)}}} \right. \kern-\nulldelimiterspace}
{(1-p)}} \right]$ where $p = P(Y = 1$) (i.e., the probability
that the subject is diagnosed with CDR0.5); the remaining variables
with all possible interactions as the predictor
variables. On this full model, we choose a reduced model by AIC in a
stepwise algorithm, and then use stepwise backward elimination
procedure on the resulting model [64]. We stop the elimination
procedure when all the remaining variables are significant at
$\alpha = 0.05$ level.

Based on the final model with significant predictors, we apply
logistic discrimination. In logistic discrimination the
log-odds-ratio of the conditional classification and therefore
indirectly the conditional probabilities of being CDR0.5 and CDR0
are modeled. In general, if this estimated probability is larger
than a prespecified probability $p_o$, the subject is classified as
CDR0.5, otherwise the subject is classified as CDR0 (i.e., normal).
This means our decision function reduces to
\begin{equation}
\label{eqn:classification-rule}
\widehat {p}_k = P\left( {Y = 1\vert d_{ijk}} \right)\left\{
{\begin{array}{l}
 >p_o \Rightarrow \mbox{ classify  }CDR0.5, \\
 \le p_o \Rightarrow \mbox{ classify  }CDR0, \\
 \end{array}} \right.
 \end{equation}
where $p_o$ is usually taken to be 0.5.
This threshold probability $p_o$ can also be optimized with respect to a cost function
which incorporates correct classification rates, sensitivity, and/or specificity [71].

We apply the same analysis procedure on hippocampal volumes to
compare the results with LDDMM metric distances. Furthermore, we
find the differential volume loss and metric distance change by
using the annual percentage rate of change (APC) in volume and
metric distance (see [71] for APC in volume for entorhinal cortex).
We also consider the logistic
discrimination models that incorporate volume and metric distance
together and APC in volumes and metric distances together.

\section{Analysis of LDDMM Distances of Hippocampi}
\label{sec:analysis}

\subsection{Preliminary Analysis of LDDMM Distances and Other Variables}
The summary measures for the variables are provided in Table \ref{tab:summary-stat-dist}.
Observe that the subjects are evenly
distributed in terms of gender, years of education, scan intervals, and age
for the diagnostic groups.
The brain and intracranial volumes are much larger
in scale, then come the hippocampal volumes, and then the metric distances.
Notice that brain and hippocampal volumes all decrease by time and are
smaller in CDR0.5 subjects compared to CDR0 subjects.
On the other hand, the
metric distances tend to increase by time and are larger for the CDR0.5 subjects.
Also presented in Table \ref{tab:summary-stat-dist} are the
$p$-values for Lilliefor's test of normality and Wilcoxon rank sum test for
differences between the diagnostic groups. Notice that most variables can be
assumed to follow a Gaussian distribution, but since a few fails to do so,
we apply the Wilcoxon rank sum test instead of Welch's $t$-test.
The diagnostic groups do not significantly differ in age, education, brain and
intracranial volumes.
Furthermore, among the metric distances, we see that only right follow-up metric
distances are significantly different between the diagnostic groups.

\begin{figure}[htbp]
\centering
\rotatebox{0}{ \resizebox{7 in}{!}{\includegraphics{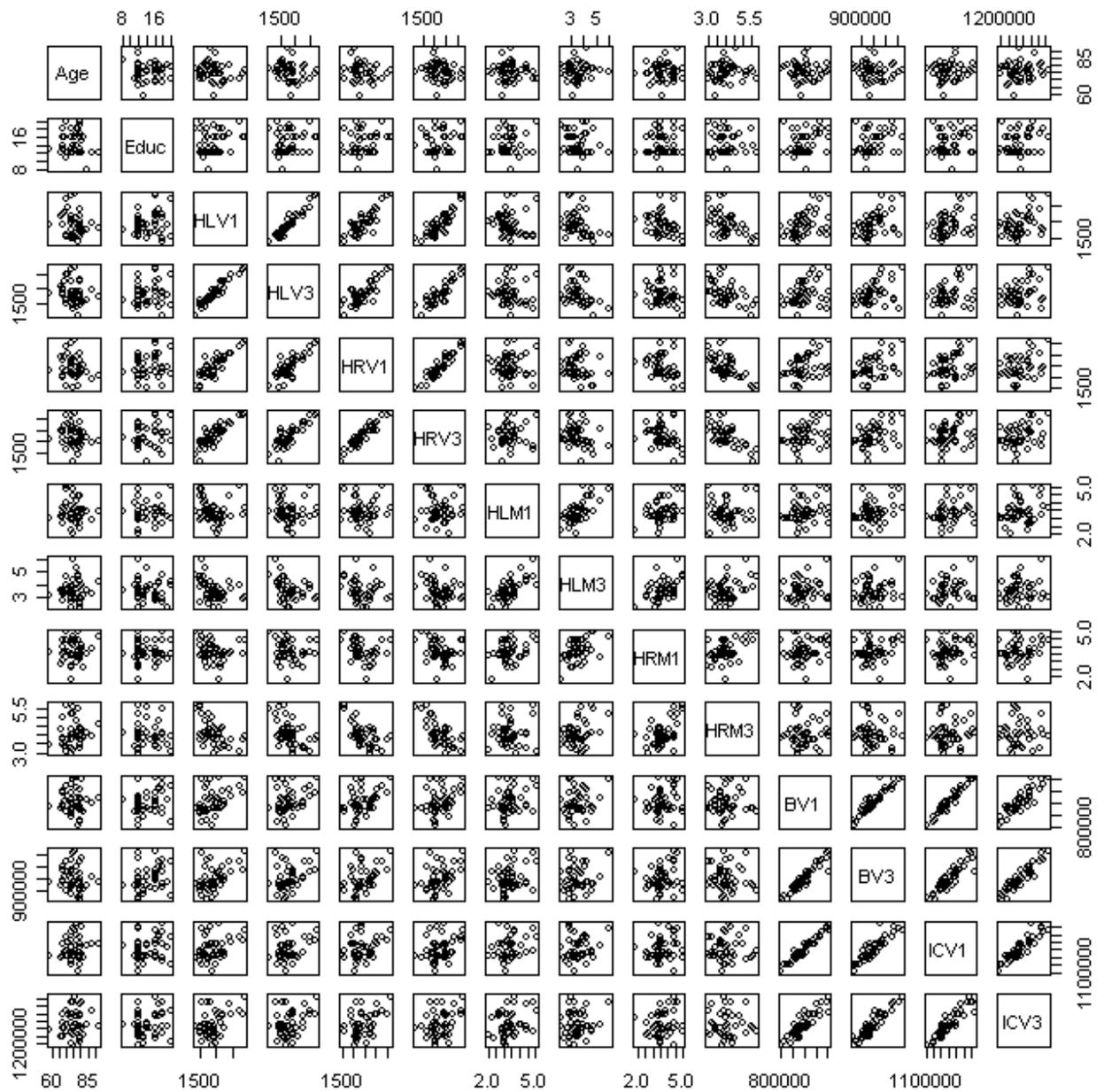}}}
\caption{
\label{fig:pairs-plot}
Pairs plots of the continuous variables for the hippocampi at baseline and follow-up.
HLV: volume of left hippocampus;
HRV: volume of right hippocampus;
HLM: metric distance for left hippocampus;
HRM: metric distance for right hippocampus;
BV: brain volume;
ICV: intracranial volume.
The numbers 1 and 3 stand for year 1 (i.e.,
baseline) and year 3 (i.e., follow-up), respectively.
}
\end{figure}

We present the pairs plot (scatter plot of each pair) of continuous variables in
Figure \ref{fig:pairs-plot} and also calculate the correlation
coefficients between each pair of the variables (not presented).
We observe that age and education are not significantly correlated with any of the
other variables.
Hence we discard them in our prospective analysis (except for logistic discrimination).
We observe significant correlation between each pair of
hippocampal volumes, and between each pair of brain and intracranial volumes.
The metric distances are only moderately correlated with each other.
Hippocampal volumes are mildly correlated
with brain and intracranial volumes.
The same holds for the metric distances but to a lesser extent.

Summary statistics of population mean, standard
deviation (SD), minimum, first quartile ($Q_1$), median, third
quartile ($Q_3$), and maximum for $d_k^{\{b,f\}}$ are presented in Table \ref{tab:summary-stat-dist}.

Baseline metric distances seem to be different in distribution
(location and spread) from follow-up metric distances follow-up
distances being larger than baseline distances for both left and
right hippocampi; likewise left metric distances seem to be
different from right metric distances with right distances being
larger than left for both baseline and follow-up. Let LDB be the
metric distances for left hippocampi at baseline, LDF be the metric
distances for left hippocampi at follow-up. Let RDB and RDF be
similarly defined for right hippocampi. One-tailed $t$-tests
revealed that the order of these measures is LDB $<$ LDF $<$ RDB $<$
RDF with all inequalities being significant at .05 level. This
implies that the morphometric differences of left hippocampi with
respect to the left hippocampus of the template subject at baseline
are significantly smaller than those at follow-up, i.e., at
baseline, left hippocampi are more similar to the left template
hippocampus, and by follow-up left hippocampi tend to become more
different in morphometry (shape and size) from the template
hippocampus. This is not surprising, as the template hippocampus is
one from the baseline hippocampi. That is the template was based on
a baseline scan. Although this should seem to be irrelevant in view
of the wide age variation, it is not the age that is the main point
here, when baseline and follow-up are compared, we use matched pair
(i.e., dependent) tests, which would reveal differences that would
otherwise be concealed by the independent two-sample tests. For
example, when all the subjects age about two years, their
morphometric alterations accumulate to render their relative
difference from the template more significant.

The right hippocampi reveal similar morphometric differences and
change over time. Furthermore, we observe that the morphometric
difference of right hippocampi from the right template hippocampus
is significantly larger compared to the morphometric difference of
left hippocampi from the left template at both baseline and follow-up.
The summary statistics (means and standard deviations (SD)) for
left and right metric distances by group are provided in
Table \ref{tab:summary-stat-dist}.

Observe that CDR0 distances are smaller than CDR0.5 distances at
baseline and at follow-up for both left and right hippocampi. This
suggests that the morphometric differences of CDR0 hippocampi with
respect to the template hippocampus are smaller than those of CDR0.5
hippocampi. This is not surprising, considering the template
hippocampus being one of the CDR0 hippocampi. Furthermore, the
standard deviations of the distances for CDR0 subjects tend to be
smaller than those of CDR0.5 subjects. That is, the morphometric
variability of CDR0 hippocampi with respect to the template
hippocampus is smaller than that of CDR0.5 hippocampi. The
statistical significance of these results will be provided in the
following sections.

\begin{figure}[htbp]
\centering
\rotatebox{-90}{ \resizebox{5.5 in}{!}{\includegraphics{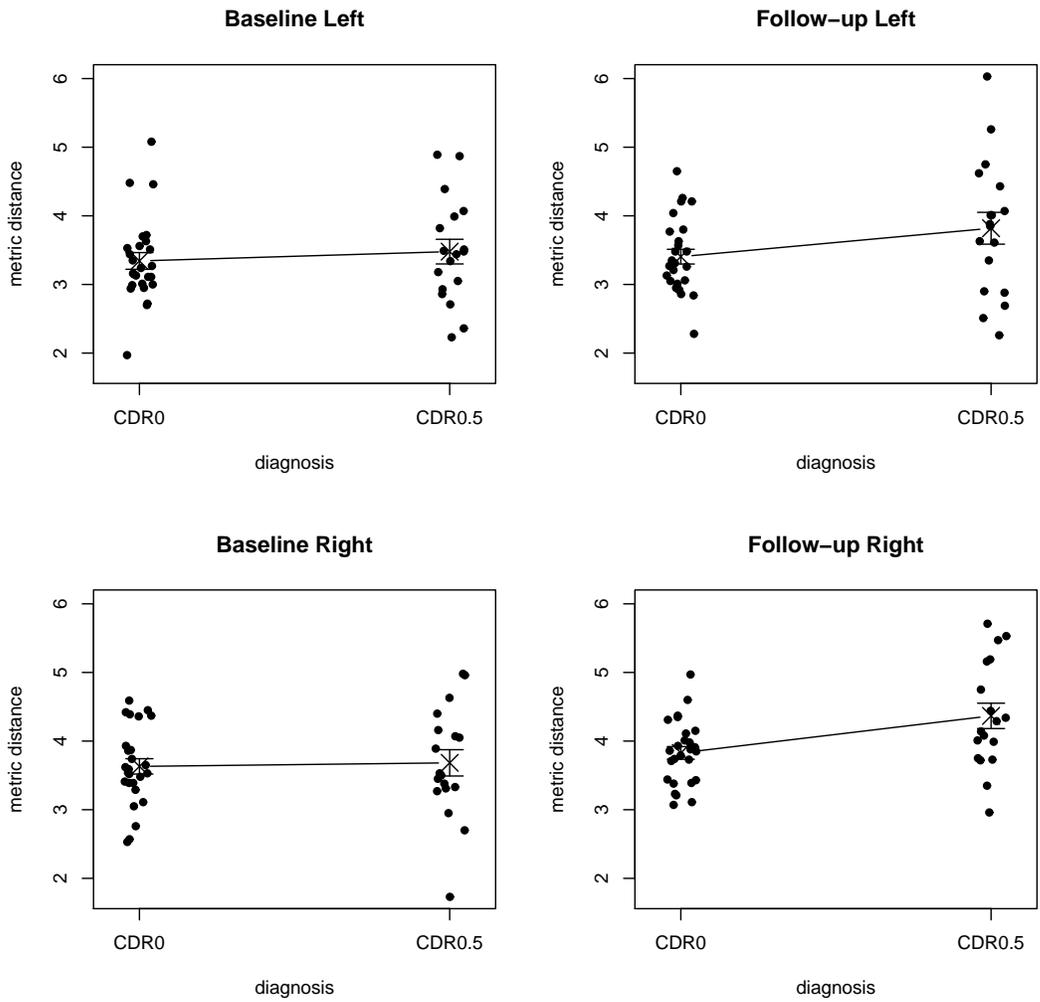}}}
\caption{
\label{fig:scatter-plot-Metric-Dist}
Scatter plots of the metric distances for the left and
right distances at baseline and follow-up. The metric distances are
jittered for better visualization and the crosses represent the mean
distance values.
}
\end{figure}

See also Figure \ref{fig:scatter-plot-Metric-Dist} for the (jittered) scatter plots of
the metric distances by group, where the crosses are centered at the
mean distances and the points are jittered (scattered) along the
horizontal axis in order to avoid frequent point concurrence and
tight clustering of points, thereby making the plot better for
visualization.

\subsection{Principal Component Analysis for the Volumes and Metric Distances}
The volumes and metric distances measure different but related aspects of
morphometry, so some of the variables are highly correlated with each other
(see Figure \ref{fig:pairs-plot}).
We perform principal component
analysis (PCA) to obtain a set of uncorrelated variables that hopefully
represent some identifiable aspect of the morphometry.
See [70,76] for more on PCA.
When we perform PCA
of metric distances and volumes of left hippocampi at baseline with
eigenvalues based on the covariance matrix, we observe that the first
principal component (PC) accounts for almost all the variation
(see Table \ref{tab:PCA-left-covariance}).
Considering the variable loadings in Table \ref{tab:PCA-left-covariance},
we see that PC1 is the head size component,
PC2 is the contrast between brain and intracranial volumes,
PC3 is the hippocampus size,
and PC4 is the metric distance component.
However,
the volumes are in $mm^3$ and metric distances are unitless, hence the data are not to scale.
In particular, the brain and intracranial volumes have the
largest variation in the data set, hence dominate the PCs.
To remove the influence of the scale (or unit), we apply PCA with eigenvalues based on the
correlation matrix (i.e., PCA on the standardized variables).
The importance scores of principal components and variable loadings from the PCA of metric
distances and volumes of left hippocampi at baseline and followup with
eigenvalues based on the correlation matrix are presented in
Table 3. Notice that with the correlation matrix,
the first three PCs account for almost all the variation in the variables.
Comparing the variable loadings, PC1 seems to be the head size component,
PC2 is the hippocampus shape, PC3 is the hippocampus size and the contrast
between hippocampus and head size, and PC4 is the contrast between brain and
intracranial volume.
The PCA on variables for right hippocampi yields
similar results (see Table 4).

The variable loadings in the PCA of variables at baseline and follow-up
suggest that brain and intracranial volumes are mostly measures of head
size, metric distance is mostly a measure of hippocampus shape and partly is a
measure of head and hippocampus size, and hippocampus volume is mostly a
measure of hippocampus size and partly related to hippocampus shape and head
size. That is, volumes and the metric distance convey information that is
related but not identical. Volumes mostly emphasize the size differences,
while metric distances emphasize the shape differences. Hence, one should
use both of them in morphometric analysis of brain tissues.

\subsection{Repeated Measures Analysis of LDDMM Distances}
Due to within-subject dependence of metric distances for left and
right hemispheres and for baseline and follow-up measures, we apply
repeated-measures analysis with group and side as main effects and
timepoint as the repeated factor, and group as main effect and
side-by-timepoint as the repeated factor (see below). For the left
data, metric distances at baseline for CDR0.5 subjects are labeled
as LB-CDR0.5, at follow-up are labeled as LF-CDR0.5. CDR0
individuals are labeled as LB-CDR0 and LF-CDR0 accordingly. Similar
labeling is done for the right metric distances. Hence, we have four
measurements for each subject, so repeated measures analysis can be
performed on our data set.

\subsubsection{Modeling LDDMM Distances with Group as Main Effect with Compound Symmetry in
Var-Cov Structure}
\begin{figure}[htbp]
\centering
\rotatebox{-90}{ \resizebox{2.5 in}{!}{\includegraphics{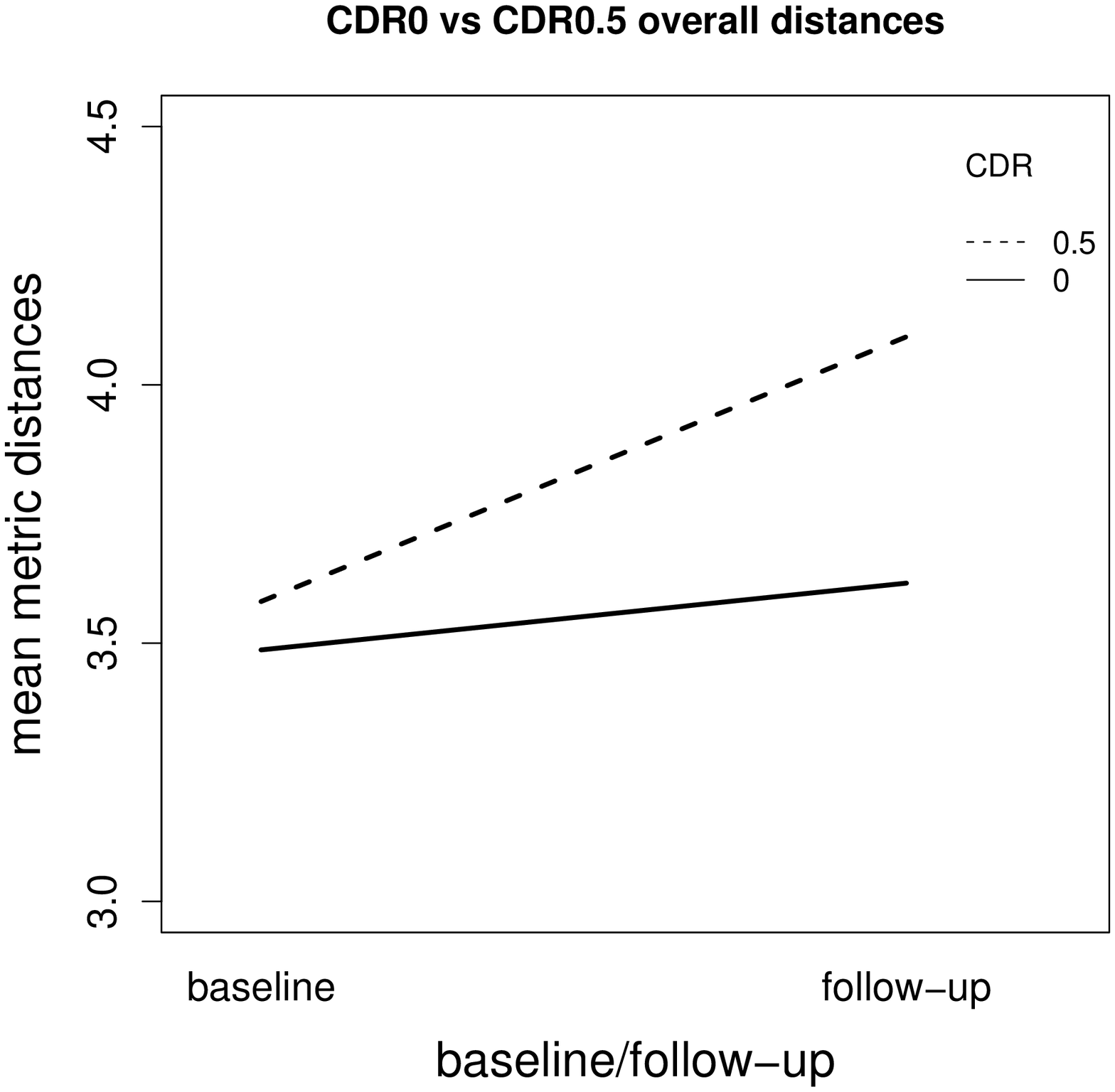}}}
\rotatebox{-90}{ \resizebox{2.5 in}{!}{\includegraphics{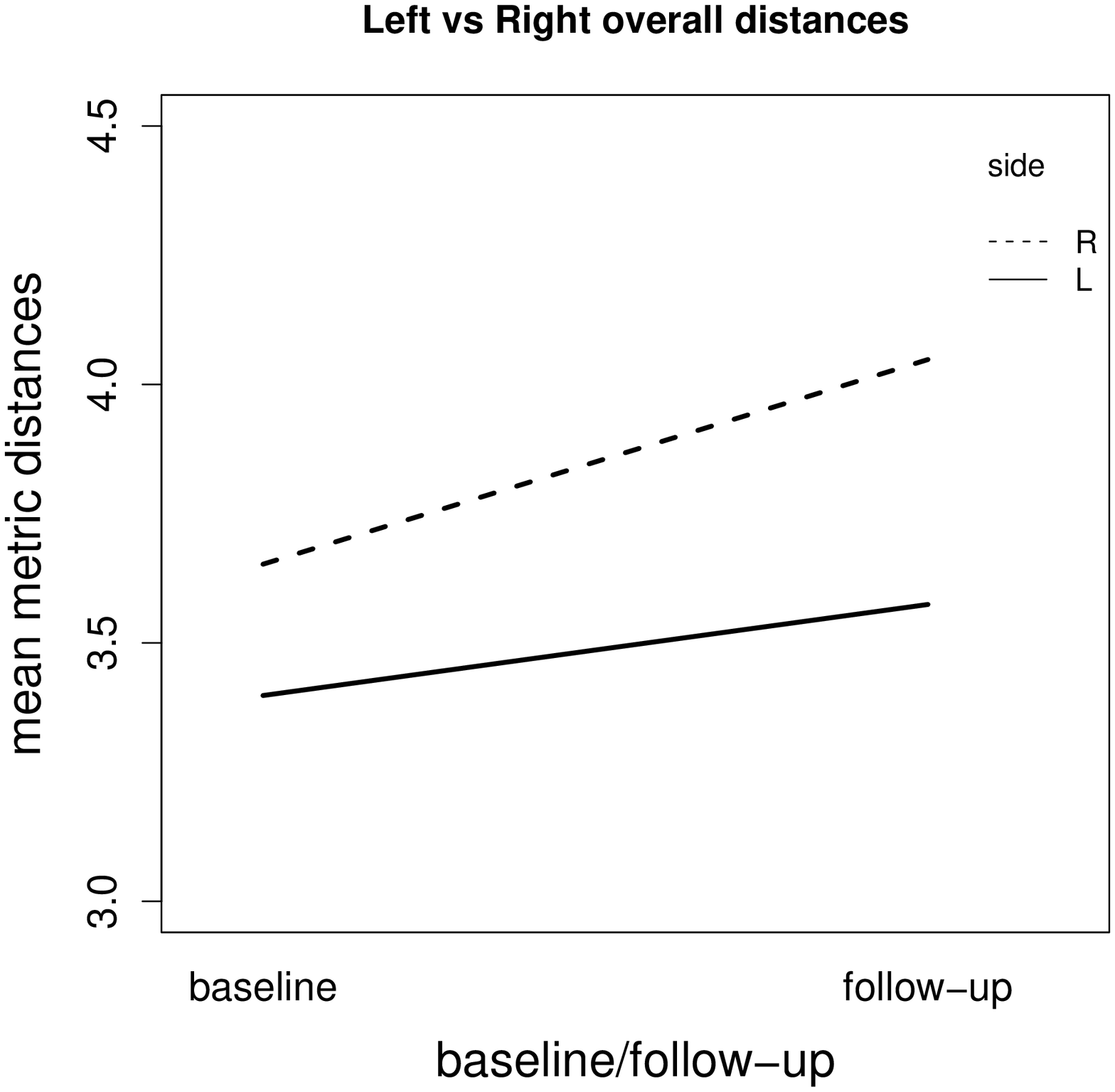}}}
\caption{
\label{fig:interaction-plot-overall}
The left interaction plot is for diagnosis levels over
the timepoint levels (the effect of sides is ignored). The slope
(i.e., the rate of change in morphometry for CDR0.5 subjects) is
significantly larger than that of CDR0 subjects. The right
interaction plot for side levels over the timepoint levels (the
effect of diagnosis is ignored). The slopes seem to not
significantly differ between Left and Right hippocampi.
}
\end{figure}

For the repeated measures ANOVA with group as
main effect and compound symmetry repeated over time, for each
subject, we will denote diagnosis, timepoint, and hemisphere factors
as numerical subscripts for convenience.
The corresponding model is
\begin{equation}
\label{eqn:model-dist-with-group}
d_{ijk} = \mu +\alpha_i^D +\alpha_j^T +\alpha_{ij}^{DT}+\varepsilon_{ijk}
\end{equation}
where $d_{ijk}$ is the distance for subject $k$ with diagnosis $i$ ($i = 1$ for CDR0; 2 for CDR0.5)
at timepoint $j$ ($j=1$ for baseline; 2 for follow-up), $\mu$ is the overall mean, $\alpha_i^D$ is the
effect of diagnosis level $i$,
$\alpha_j^T $ is the effect of timepoint level $j$,
$\alpha_{ij}^{DT}$ is the diagnosis-by-timepoint interaction, i.e., part
of the mean distance not attributable to the additive effect of
diagnosis and timepoint, and $\varepsilon_{ijk}$ is the error
term. The Var-Cov structure for the error term is
$$\Var(\varepsilon_{ijk}) = \sigma^2 \text{ and } \Cov(\varepsilon_{ijk} ,\varepsilon_{ij'k}) = \sigma_1^T.$$
Notice that the effect of side (left or right) is ignored in this model.
There is no significant group main effect ($F = 3.36,\; df = 1,42,\; p = 0.0739$).
However, the within group time-point main
effect ($F = 11.16,\; df = 1,130,\; p = 0.0011$) and the
group-by-timepoint interaction ($F = 4.84,\; df = 1,130,\;p = 0.0295$)
are both significant, which imply that the two groups should
be compared at the different time points.
In Figure \ref{fig:interaction-plot-overall},
we present the interaction plots for diagnosis over time, where the
end points of the line segments are located at the mean metric
distances at baseline and follow-up years. We see that both lines
increase over time, but are not parallel; the increase of the line
for CDR0.5 group is steeper.

\subsubsection{Modeling LDDMM Distances with Side as Main Effect with Compound Symmetry in
Var-Cov Structure}
\label{sec:modeling}
For the repeated
measures ANOVA with side as main effect and compound symmetry
repeated over time, the corresponding model is
\begin{equation}
\label{eqn:model-dist-with-side}
d_{ijk} = \mu +\alpha_i^S +\alpha_j^T +\alpha_{ij}^{ST}+\varepsilon_{ijk}
\end{equation}
where $d_{ijk}$ is the distance for subject $k$ for side $i$ ($i=1$ for left; 2 for right)
at timepoint $j$ ($j=1$ for baseline; 2 for follow-up), $\mu$ is the overall
mean, $\alpha_i^S $ is the effect of side level $i$, $\alpha_j^T $
is the effect of timepoint level $j$, $\alpha_{ij}^{ST}$ is
the side-by-timepoint interaction, and $\varepsilon_{ijk}$ is the error term.
The Var-Cov structure for the error term is
$$\Var(\varepsilon_{ijk}) = \sigma^2\text{ and } \Cov(\varepsilon_{ijk} ,\varepsilon_{ij'k}) = \sigma_1^T.$$
Notice that the effect of diagnosis (CDR0 or CDR0.5) is ignored in
this model. The side and timepoint main effects are both significant
($F = 20.25, df = 1,129, $p $< 0.0001 and $F = 12.51, {\it
df = }1,129, $p$ = 0.0006, respectively), but side-by-timepoint
interaction is not significant (F = 1.85, df = 1,129, $p$ = 0.1766).
Consequently, we see that the lines are parallel but far apart, the
main effect of side comparison is meaningful and about the same at
each timepoint. Moreover, the sides do change in morphometry over time.
In Figure \ref{fig:interaction-plot-overall}, we see that both lines increase
over time and are parallel, but the slope for right side seems to be
steeper, which will eventually make the slope estimates
significantly different.

\subsubsection{Modeling LDDMM Distances with Group, Side, and Group-by-Side Interaction}
\label{subsubsec:mylabel2}
Looking at models including only the main
effects of side or group separately does not answer all our
questions. We would also like to know, for example, if the metric
distances of left hippocampi of CDR0.5 subjects are different from
those of left CDR0 subjects. In order to address these types of
questions we need to look at a model that includes the interaction
of diagnosis and side. First, we need to model the Var-Cov structure
for the repeated measures for each subject. We have four correlated
measures per subject, namely LDB, LDF, RDB, and RDF. Below is the
estimated Var-Cov matrix for these variables:
\[
\left[ {\begin{array}{l}
 \mbox{0.46 0.35 0.18 0.02} \\
 \mbox{0.35 0.60 0.24 0.18} \\
 \mbox{0.18 0.24 0.45 0.23} \\
 \mbox{0.02 0.18 0.23 0.44} \\
 \end{array}} \right]
\]
We start with compound symmetry for our model, and then try
unstructured, autoregressive (AR), and autoregressive heterogeneous
(ARH) Var-Cov structures. The variances (in the diagonal) suggest
heterogeneity between them, and also, covariances seem to differ.
This suggests that either an unstructured or ARH model might fit
this data best. See Table \ref{tab:model-selection-criteria} for the comparison of
model selection criteria such as AIC, BIC, and Log-likelihood and
likelihood ratio test $p$-value.

The most promising model is the unstructured model based on
likelihood ratio test, since -2 Log Likelihood scores are
significantly smaller than the -2 Log Likelihood scores of other
models. However, BIC and AIC favor the model with the AR
variance-covariance structure. Besides, the log-likelihood approach
gives the second smallest -2 Log Likelihood score for this model.
Hence, we choose the model with AR Var-Cov structure. The
corresponding model is
\begin{equation}
\label{eqn:model-dist-with-group-side-interaction}
d_{ijkl} = \mu +\alpha_i^S +\alpha_j^D +\alpha_k^T +\alpha
_{ij}^{SD} +\alpha_{ik}^{ST} +\alpha_{jk}^{DT} +\alpha
_{ijk}^{SDT} +\varepsilon_{ijkl},
\end{equation}
where $d_{ijkl}$ is the distance for subject $l$ for side $i$ (1 for left; 2 for right)
with diagnosis $j$ ($j=1$ for CDR0; 2 for CDR0.5) at timepoint $k$ ($k=1$ for baseline; 2 for follow-up),
$\mu$ is the overall mean,
$\alpha_i^S $ is the effect of side level $i$,
$\alpha_j^D$ is the effect of diagnosis level $j$, $\alpha_k^T $
is the effect of timepoint level $k$,
$\alpha_{ij}^{SD}$ is the side-by-diagnosis interaction, $\alpha_{ik}^{ST}$ is the
side-by-timepoint interaction, $\alpha_{jk}^{DT}$ is the
diagnosis-by-timepoint interaction, $\alpha_{ijk}^{SDT}$ is the
side-by-diagnosis-by-timepoint interaction, and $\varepsilon_{ijkl}$
is the error term. The Var-Cov structure for the error term is
\[
\Cov(\varepsilon_{ijkl} ,\varepsilon_{i'jk'l}) = \left[
{{\begin{array}{*{20}c}
 {\sigma^2} \hfill & \hfill & \hfill & \hfill \\
 {\sigma \rho} \hfill & {\sigma^2} \hfill & \hfill & \hfill \\
 {\sigma \rho^2} \hfill & {\sigma \rho} \hfill & {\sigma^2} \hfill &
\hfill \\
 {\sigma \rho ^3} \hfill & {\sigma \rho^2} \hfill & {\sigma \rho} \hfill &
{\sigma^2} \hfill \\
\end{array}}} \right].
\]
The three way interaction of side-by-group-by-timepoint is not
significant ($F = 0.50,\; df = 1,168,\; p = 0.4823$), and
neither are the two way side-by-group ($F = 0.76,\; df = 1,168,\;p = 0.3860$),
and side-by-timepoint interactions ($F = 2.25,\; df = 1,168,\; p =  0.1359$).
On the other hand, the group-by-timepoint interaction is significant
($F = 8.47,\; df =1,168,\; p = 0.0041$).
The main effects of side, group, and timepoint are all significant
($F = 6.12,\; df = 1,168,\; p = 0.0143$; $F = 4.05,\; df = 1,168,\;p = 0.0457$; and
$F = 19.52,\; df = 1,168,\; p < 0.0001$, respectively), but due to
interaction, the main effect for diagnosis (i.e., group) is close to
clinically meaningless; i.e., the groups should be compared at each
time point instead of an overall comparison of group means. But, the
main effects of timepoint and side being significant is
interpretable between baseline and follow-up.

Below we perform various post-hoc tests to see which groups are
significantly different or significantly change over time. To
accomplish this, we test for differences at each timepoint, between
baseline and follow-up, and between left and right distances.

\subsection{Post-Hoc Comparison of LDDMM Distances of CDR0.5 vs CDR0 Hippocampi}
\label{subsec:mylabel1}
For the $p$-values regarding the comparison
of independent groups, see Table \ref{tab:comparison-lddmm-distance}.
The significant values at $\alpha = 0.05$ are marked with *. None of the distance
groups deviate significantly from normality (all $p$-values greater
than 0.10). That is, distance distribution of each group can be
assumed to come from a Gaussian distribution. Moreover, LB-CDR0.5
and LB-CDR0 distances can be assumed to have equal variances ($p = 0.2948$),
and so can RB-CDR0.5 and RB-CDR0 ($p = 0.2273$). But, the
variance of LF-CDR0 distances is significantly smaller than that of
LF-CDR0.5 distances ($p = 0.0294$), and similarly for RF-CDR0 versus
RF-CDR0.5 ($p =  0.0262$). Therefore, for comparisons at baseline, we
can use the $p$-values from the $t$-tests [68], while for
follow-up comparisons, it is more appropriate to use the $p$-values
from Wilcoxon rank sum tests [67].

Observe that RF-CDR0.5 mean distances are significantly larger than
RF-CDR0 mean distances at .05 level ($p = 0.0106$), and LF-CDR0.5
distances are significantly larger than LF-CDR0 distances at 0.10
level ($p = 0.0813$). On the other hand, LB-CDR0.5 and LB-CDR0
distances are not significantly different ($p = 0.5362$), and
likewise for RB-CDR0.5 and RB-CDR0 distances ($p = 0.8176$). This
implies that at baseline, the morphometric differences of CDR0.5 and
CDR0 hippocampi with respect to the template hippocampus are about
same, which might indicate no significant shape differences in the
left and right hippocampi due to dementia. However, since the metric
distances do not necessarily provide direction in either shape or
size, this is not a decisive implication. At follow-up, left and
right hippocampi of CDR0.5 subjects tend to significantly differ in
morphometry from the template compared to those of CDR0 subjects.
Moreover, this significance emanates over time; that is, right
hippocampi of CDR0.5 subjects tend to undergo more alteration in
morphometry compared to those of CDR0 subjects over time.

\subsection{Comparison of Baseline and Follow-up Metric Distances}
For the comparison of dependent groups by paired difference method,
see Table \ref{tab:comparison-lddmm-distance}.
The paired differences in
Table \ref{tab:comparison-lddmm-distance} can all be assumed to be normal based on
Lilliefor's test of normality. Hence, we use the more powerful
$t-$test for paired differences [68].

Observe that LB-CDR0.5 metric distances are significantly smaller
than LF-CDR0.5 distances at $\alpha = .05$ ($p = 0.0259$). Likewise
for RB-CDR0 vs RF-CDR0.5 distances ($p = 0.0002$). That is, CDR0.5
hippocampi tend to become more different in morphometry from the
template, which implies that for both left and right distances there
is significant change in morphometry (perhaps reduction in size) of
CDR0.5 hippocampi over time. In fact, significant volume reduction
over time is detected [29]. The
morphometric changes in CDR0.5 right hippocampi from baseline to
follow-up is barely significantly larger than those of CDR0.5 left
hippocampi ($p = 0.0445$). The associated $p$-value here is obtained
by testing the difference sets (LB-CDR0.5)-(LF-CDR0.5) versus
(RB-CDR0)-(RF-CDR0.5) using the usual paired $t$-test. On the other
hand, only RB-CDR0 is almost significantly less than RF-CDR0 at .05
level ($p = 0.0621$), which implies there is not strong evidence for
shape change in control subjects over time, but some weak evidence
for mild change in right hippocampi can be detected as a result of
aging. Furthermore, the morphometric changes in CDR0 right
hippocampi from baseline to follow-up are not significantly
different from those of CDR0 left hippocampi ($p = 0.3817$).

The morphometric changes in CDR0.5 left hippocampi from baseline to
follow-up are not significantly different from those of CDR0 left
hippocampi ($p = 0.1337$), while the morphometric changes in CDR0.5
right hippocampi from baseline to follow-up are significantly larger
from those of CDR0 right hippocampi ($p = 0.0074$).
Therefore, over time, DAT influences the morphometry of right hippocampi more
compared to left hippocampi.

\subsection{Comparison of LDDMM Distances of Left and Right Hippocampi}
As for left vs right comparisons, LB-CDR0.5 and RB-CDR0.5 distances
are not significantly different from each other ($p = 0.3046$),
LF-CDR0.5 distances are significantly smaller than RF-CDR0.5
distances at .05 level ($p = 0.0179$), the same holds for LB-CDR0 vs
RB-CDR0.5 ($p = 0.0215$) and LF-CDR0 vs RF-CDR0 ($p = 0.0021$)
comparisons. This implies that at baseline morphometric differences
of CDR0.5 left hippocampi from the left template are about the same
as those of CDR0.5 right hippocampi from the right template. On the
other hand at follow-up, morphometric differences of CDR0.5 left
hippocampi are smaller than those of CDR0.5 right hippocampi. At
baseline and follow-up, morphometric differences of CDR0 left
hippocampi from the left template are smaller than those of CDR0
right hippocampi. That is, CDR0 left hippocampi are more similar in
morphometry to the left template when compared to CDR0 right
hippocampi to the right template. These distance comparisons for
left versus right hippocampi would imply left-right morphometric
asymmetry, only if the left and right hippocampi of the template
subject were very similar (up to a reflection). Otherwise, these
comparisons are only suggestive of morphometric differences from the
respective hemisphere (side) of the hippocampi.

\subsection{Analysis of the Correlation between Metric Distances of Dependent Hippocampi}
Correlation coefficients between metric distances for baseline and
follow-up (overall and by group) and for the left and right
hippocampi are provided in Table \ref{tab:correl-baseline-followup} and
Table \ref{tab:correl-left-right}, respectively, where Pearson's product-moment
correlation coefficient is denoted as $r_P$, Spearman's rank
correlation coefficient is denoted as $\rho_S$, and Kendall's rank
correlation coefficient is denoted as $\tau_K$ [67, 68]. The
corresponding null and alternative hypotheses are
$H_o :\mbox{correlation} = 0\quad \mbox{vs}\quad H_a :\mbox{correlation}>0$.

The values in the parentheses right of the correlation coefficients
are the corresponding $p$-values. The significant $p$-values at
level $\alpha = 0.05$ are marked with an asterisk (*). Since all
groups can be assumed to be normal, the more powerful Pearson's
correlation test will be used for inference.

Notice that from the correlation analysis of baseline vs follow-up,
we see that the overall distances, L-CDR0, and R-CDR0 are
significantly correlated at 0.05 level. But LB-CDR0 and LF-CDR0 are
significantly correlated at .05 level based on Pearson's test, and
Spearman's test, and at 0.10 by Kendall's tests. However, RB-CDR0
and RF-CDR0 are significantly correlated at 0.10 level by only
Pearson's test. This implies that except for the CDR0 right
hippocampi, the distances tend to increase at baseline together with
distances at follow-up. That is, as the morphometric differences
from the template hippocampus increase at baseline, so do the
differences from the template at follow-up (except for CDR0 right
hippocampi).

Notice also that from the correlation analysis of left and right
distances, we observe that overall left and right at baseline (LDB
and RDB) distances are significantly correlated at .05 level, but
LDF and RDF are correlated at .05 level by Pearson's test only, and
0.10 by Spearman's test. And LB-CDR0 and RB-CDR0 have significant
correlation structure at .05 level. However, the correlation
coefficients are not that large, which suggests mild correlation
between left and right metric distances. That is, as the
morphometric differences of left hippocampi from the left template
increase, differences of right hippocampi from the right template
tend to increase slightly.

\subsection{Comparison of Distributions of the Metric Distances}
The samples (groups) should be independent for these tests to be
valid, so we only compare LB-CDR0.5 vs LB-CDR0, RB-CDR0.5 vs
RB-CDR0, LF-CDR0.5 vs LF-CDR0, and RF-CDR0.5 vs RF-CDR0. The
corresponding $p$-values for the two-sided and one-sided cdf
comparison tests are provided in the Table \ref{tab:cdf-compare-distance},
where $p_{KS}$ is the $p$-value for the two-sided K-S test, with (l) and
(g) are abbreviations of first cdf less than the second and first
cdf greater than the second, respectively, $p_C$ is the $p$-value
for Cram\'{e}r's test, and $p_{CvM}$ is the $p$-value for
Cram\'{e}r-von Mises test [65, 72].

\begin{figure}[htbp]
\centering
\rotatebox{-90}{ \resizebox{2.5 in}{!}{\includegraphics{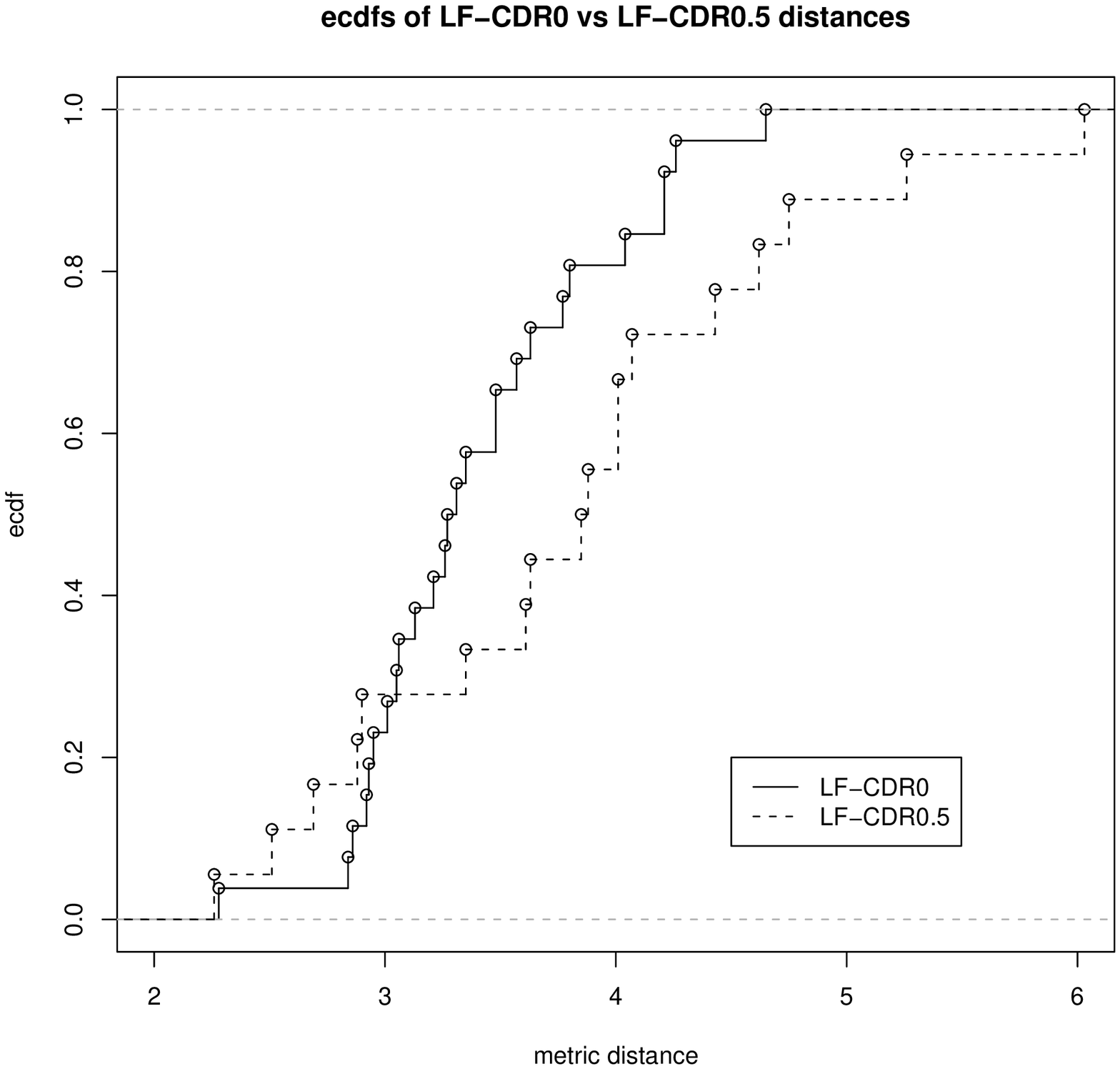}}}
\rotatebox{-90}{ \resizebox{2.5 in}{!}{\includegraphics{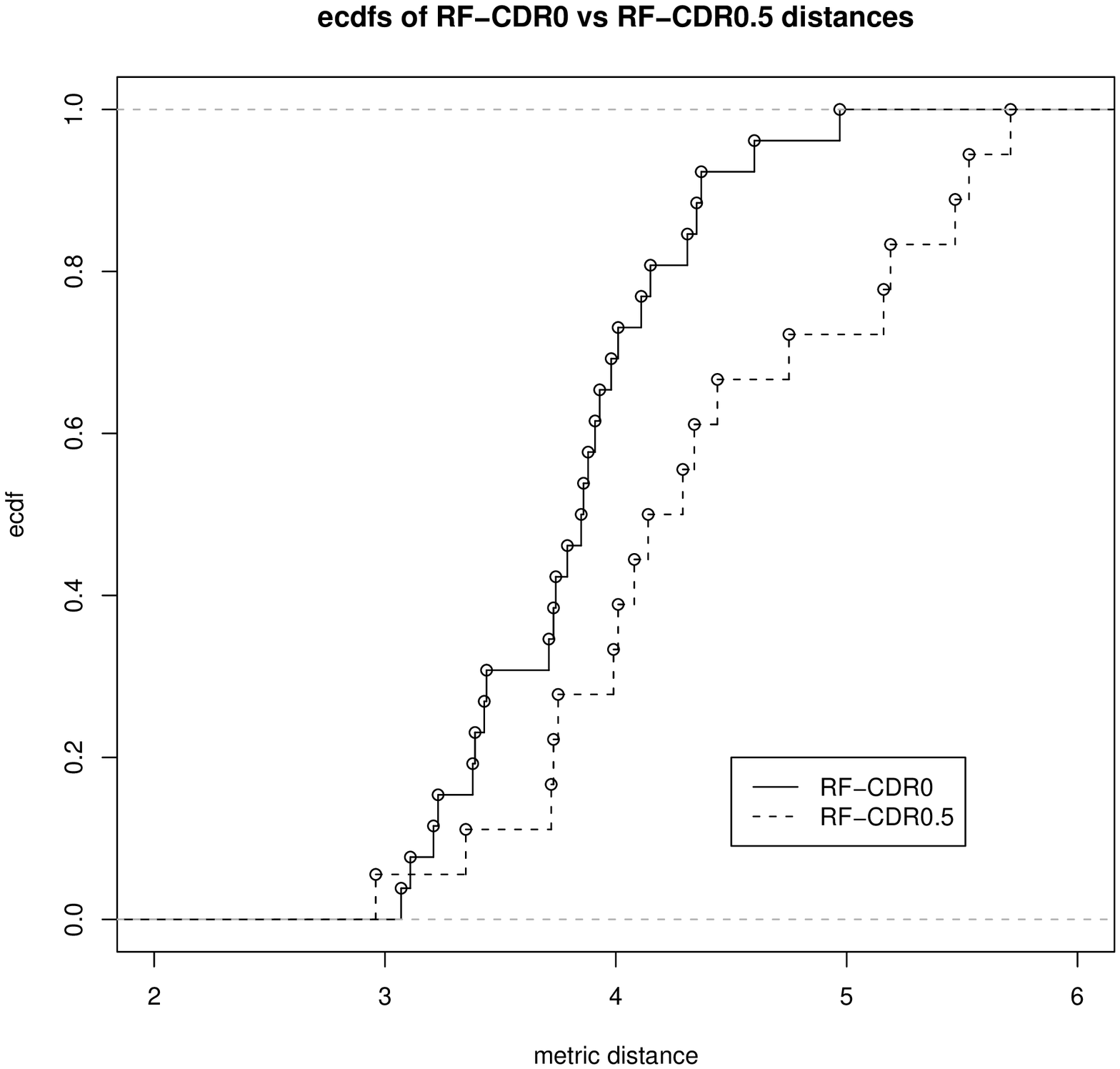}}}
\caption{
\label{fig:cdf-plot-metric-distance}
Empirical cdfs of the metric distances for the CDR0.5 vs
CDR0 Left and Right hippocampus at follow-up.
}
\end{figure}

Notice that at $\alpha = 0.05$ level, the cdf of RF-CDR0.5 distances
is significantly smaller than the cdf of RF-CDR0 distances ($p = 0.0259$ for K-S test).
That is, RF-CDR0.5 metric distances are
stochastically larger than RF-CDR0 right metric distances. In other
words, RF-CDR0.5 hippocampus shapes are more likely to be different
than the template hippocampus compared to RF-CDR0 hippocampus
shapes. Furthermore, the cdf of LF-CDR0.5 distances is significantly
smaller than the cdf of LF-CDR0 distances ($p = 0.0604$ by K-S test
and $p = 0.0495$ by Cram\'{e}r's test); that is, LF-CDR0.5 metric
distances are stochastically larger than LF-CDR0 metric distances.
See Figure \ref{fig:cdf-plot-metric-distance} for the corresponding cdf plots.
Observe that these results are in agreement with the ones in
Table \ref{tab:comparison-lddmm-distance}.

\subsection{Logistic Discrimination with Metric Distances}
\begin{figure}[htbp]
\centering
\rotatebox{-90}{ \resizebox{2.5 in}{!}{\includegraphics{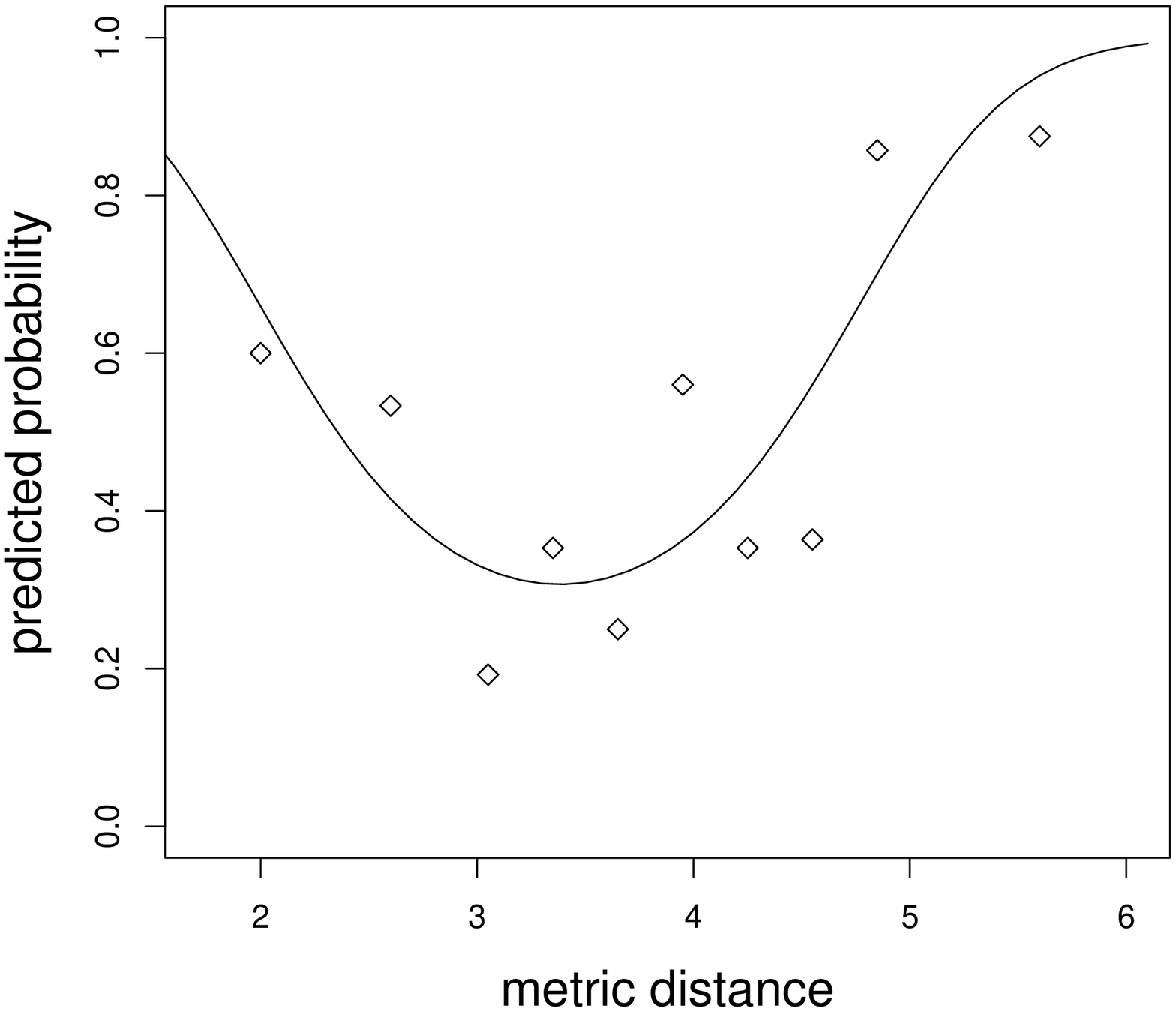}}}
\rotatebox{-90}{ \resizebox{2.5 in}{!}{\includegraphics{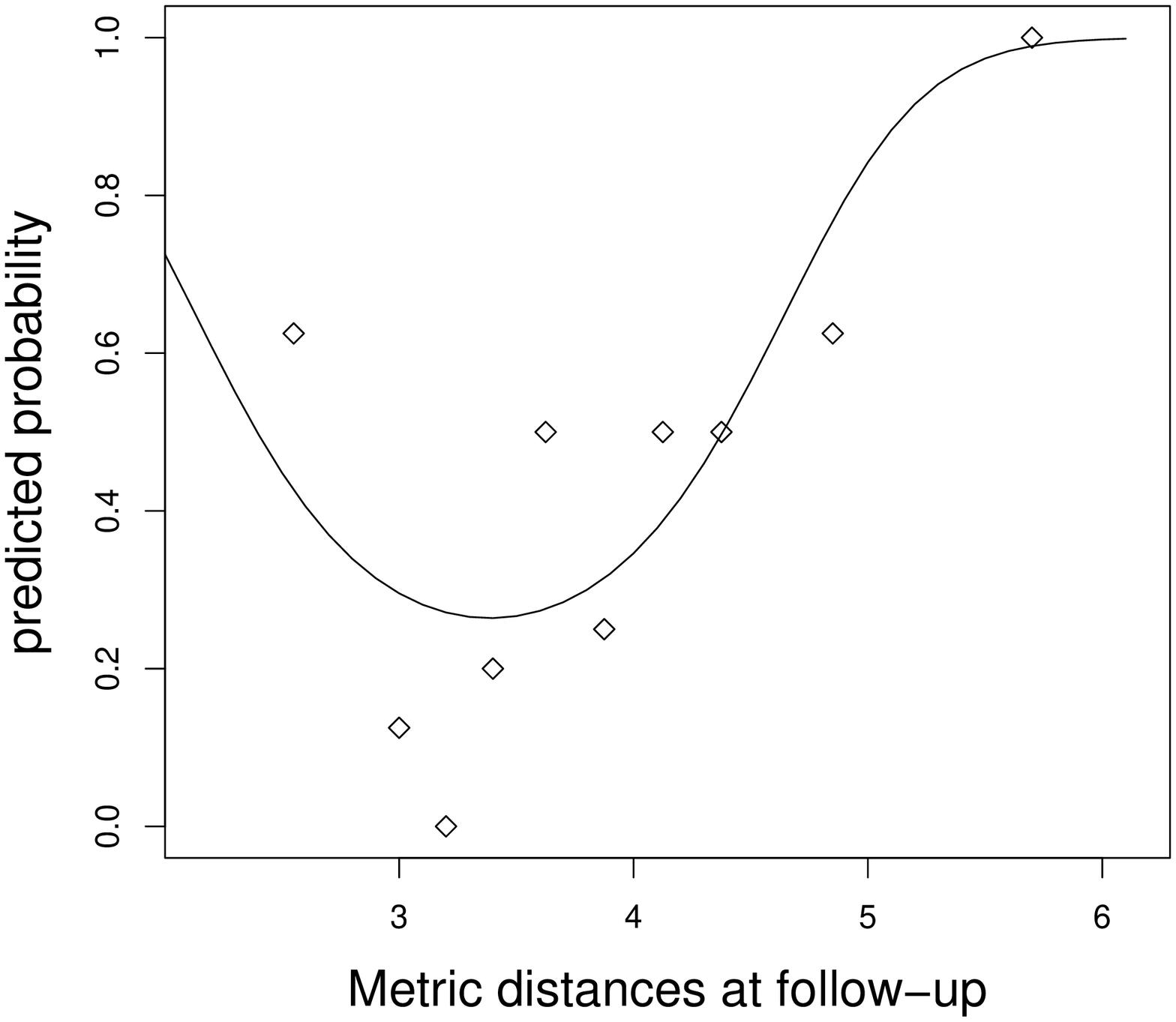}}}
\rotatebox{-90}{ \resizebox{2.5 in}{!}{\includegraphics{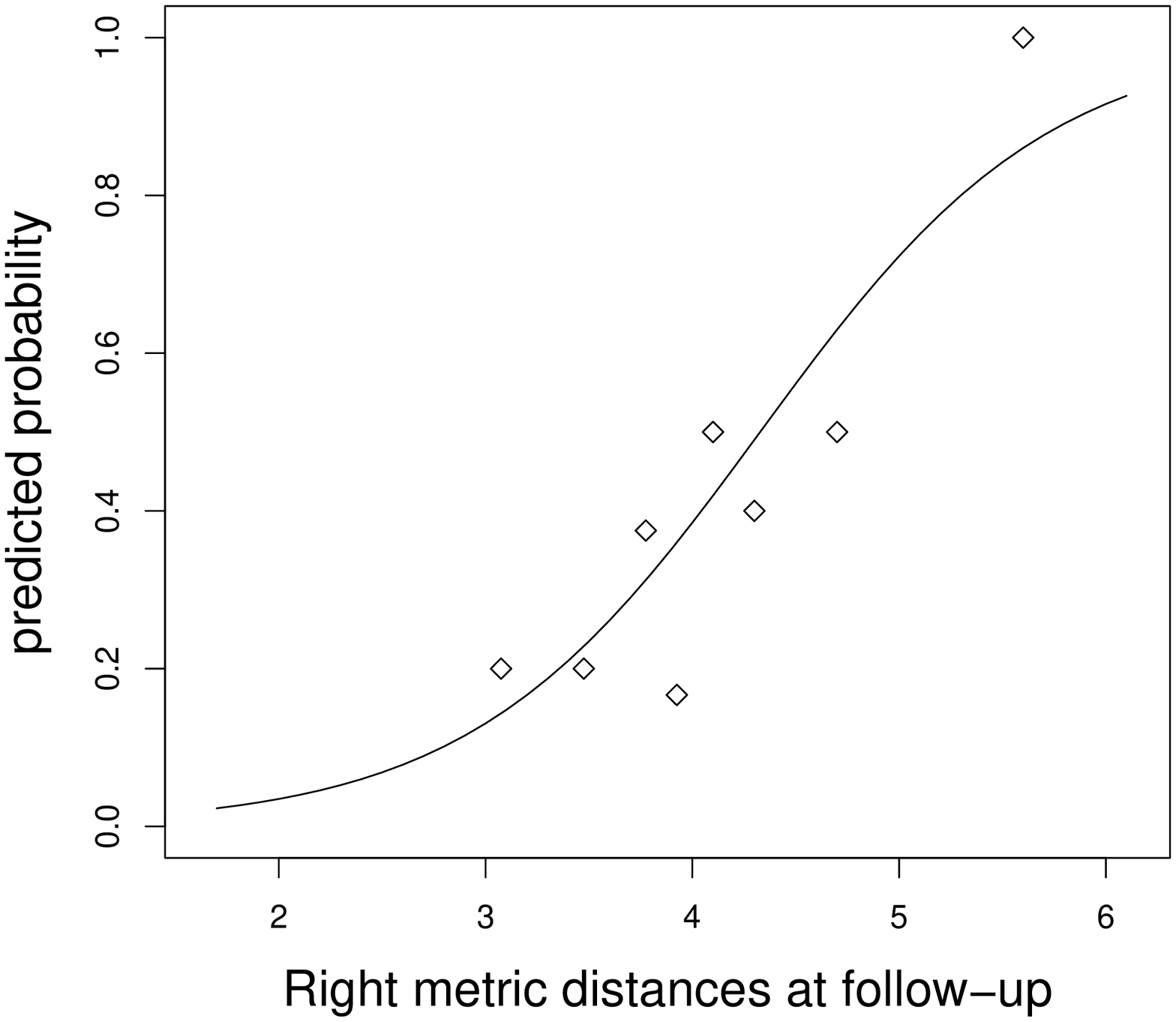}}}
\caption{
\label{fig:fitted-probability}
Fitted probability for having mild dementia (CDR0.5) and
observed proportion in metric distances with model (9) (top-left);
model (10) (top-right); and model (11) (bottom),
}
\end{figure}

We model the probability that the subject has CDR0.5 given the
hippocampal LDDMM distances of the subject for left and right
hippocampi at baseline and follow-up. First we consider the full
logistic model (designated as $M_I(D)$) with the response
$\mbox{logit }p = \log \left[ {p \mathord{\left/ {\vphantom {p
{(1-p)}}} \right. \kern-\nulldelimiterspace} {(1-p)}} \right]$ where
$p = P(Y = 1$) (i.e., the probability that condition of the subject is
CDR0.5); side, timepoint, and distance with all possible
interactions are the predictor variables. When the stepwise model
selection procedure is applied, the resulting model is
$\mbox{logit }p_k = \beta_0 +\beta_1 d_{ijk}$ where $p_k$ is the probability of
subject $k$ having DAT and $d_{ijk}$ the distance for subject $k$
with diagnosis $i$ ($i$ = 1 for CDR0 and 2 for CDR0.5) at timepoint
$j$ ($j$ = 1 for baseline and 2 for follow-up), $\beta_0$ is the
intercept and $\beta_1$ is the slope of the fitted line.
However, the graph of the proportions of CDR0.5 subjects for grouped metric
distances in Figure \ref{fig:fitted-probability} suggests that the relationship
is a quadratic one (in fact, we found that the higher order distance
terms are not significant). That is, the analysis of deviance table
indicates that only the linear and quadratic terms are significant
($p = 0.001$ and $p = 0.010$). So the resulting model is
\begin{equation}
\label{eqn:MIID}
M_{II}(D)\mbox{: logit }p_k = \beta_0 +\beta_1 d_{ijk}
+\beta_2 d_{ijk}^2
\end{equation}
where $\beta_2 $ is the coefficient of the quadratic term.

Using this logistic classifier with $p_o = 0.5$, we obtain
classification summary matrix A in Table \ref{tab:classification-matrices} for the 176
hippocampi MRI images in this data set. The labels on the left
margin show the groups to which the hippocampi MRIs are classified
into, while the top margin shows the groups from which these MRI images come.
Observe that 95 out of 104 (91\%) of the
hippocampus MRIs from CDR0 subjects would be classified correctly
and 20 out of 72 (28\%) of the hippocampus MRIs from CDR0.5
subjects are classified correctly. However, in this logistic
discrimination procedure, we treat each hippocampus from left,
right, baseline or follow-up hippocampi as a distinct subject. From
a clinical point of view, each subject has four hippocampus MRIs in
this study, and one MRI classified as CDR0.5 would suffice to
classify the subject as CDR0.5, while all four MRIs should be
classified as CDR0 for the subject to be classified as CDR0. With
this classification rule, we obtain the classification matrix B in
Table \ref{tab:classification-matrices}. Notice that 18 out of 26 (69.2\%) of the
CDR0 subjects would be classified correctly and that 10 out of 18
(55.6\%) of the CDR0.5 subjects are classified correctly.

However, as we have seen in Section 3.1.3, due to group-by-timepoint
interaction, we need to consider diagnosis groups at each time point.
When we use $d_{ik}^B$ and $d_{ik}^F$ one at a time in a
logistic model, we see that only the model
\begin{equation}
\label{eqn:MIIID}
M_{III}(D)\mbox{: logit }p_k = \beta_0 +\beta_1 d_{ik}^F +\beta_2 \left( {d_{ik}^F} \right)^2
\end{equation}
has significant coefficients for the distance terms. Using this
logistic model in the logistic classifier, we get the classification
matrix C in Table \ref{tab:classification-matrices}.
Notice that 22 out of 26 (85\%) of the CDR0 subjects would be classified correctly and
that 10 out of 18 (56\%) of the CDR0.5 subjects are classified correctly.

Moreover, when we use $d_k^{LB} ,\;d_k^{LF} ,\;d_k^{RB} ,$ and
$d_k^{RF}$ one at a time in a logistic model, we see that only the
model
\begin{equation}
\label{eqn:MIVD}
M_{IV}(D)\mbox{: logit }p_k = \beta_0 +\beta_1 d_k^{RF}
\end{equation}
has a significant coefficient for the distance term. Using $d_k^{RF}
$ and this logistic model in the logistic classifier, we get the
classification matrix D in Table \ref{tab:classification-matrices}.
Notice that 22 out of 26 (85\%) of the CDR0 subjects would be classified correctly
and that 8 out of 18 (44\%) of the CDR0.5 subjects are
classified correctly. The above classification matrices are almost
the same with leave-one-out cross-validation with logistic
discrimination (not presented).

We also calculate the sensitivity and specificity of the
classification procedures summarized in Table \ref{tab:class-rates-for-matrices}.
Sensitivity is the proportion of subjects that are classified to be
CDR0.5 (i.e., positive) of all CDR0.5 subjects. That is, sensitivity
is defined as $\displaystyle P_{sens} = \frac{T_{CDR0.5}}{N_{CDR0.5}}\times
100\,\% $ where $T_{CDR0.5}$ is the number of correctly classified
CDR0.5 subjects and $N_{CDR0.5}$ is the total number of CDR0.5
subjects in the data set (i.e., $N_{CDR0.5} = 18$ in our data).
Notice that the higher the sensitivity, the fewer real cases of DAT
go undetected. Specificity is the proportion of subjects that are
classified CDR0 (i.e., negative, control, or healthy) of all CDR0
subjects; that is $\displaystyle P_{spec} = \frac{T_{CDR0}}{N_{CDR0}}\times 100\,\% $
where $T_{CDR0}$ is the number of correctly classified
CDR0 subjects and $N_{CDR0}$ is the total number of CDR0 subjects
in the data set (i.e., $N_{CDR0} = 26$ in our data). Notice that the
higher the specificity, the fewer healthy people are labeled as
sick. The correct classification rates, sensitivity and specificity
percentages for the classification matrices A-D are presented in
Table \ref{tab:class-rates-for-matrices}. Observe that best classification performance
is with the logistic model $M_{III}(D)$ in Equation (10) with one
CDR0.5-labeled hippocampus enough to label the subject to have
CDR0.5 (see matrix C in Table \ref{tab:classification-matrices}). Furthermore, in
these classification procedures, specificity rates are
(significantly) larger than the sensitivity rates.

We could also change the threshold probability $p_o$ in Equation (5).
The correct classification rates, sensitivity, and specificity
percentages with models $M_I(D)-M_{IV}(D)$ and $p_o \in \left\{
{1/2,18/44} \right\}$ are presented in Table \ref{tab:class-rates-dist}.
Observe that with $p_o = 1/2$ the best classifier is based on
$M_{III}(D)$ and with $p_o = 18/44$ the best classifier is based on
$M_{IV}(D)$. Setting $p_o = 18/44$ (the proportion of CDR0.5
subjects in the data set) we get higher sensitivity rates than those with $p_o = 1/2$.
However, as $p_o$ decreases, the correct
classification rate and specificity tend to decrease.

One can optimize the threshold value of $p_o$ in Equation (5) to
maximize the correct classification rates and minimize the
misclassification rates using an appropriately chosen cost function.
For example one can consider the cost function
\begin{equation}
\label{eqn:cost1}
C_1(p_o ,w_1 ,w_2) = -\left( {T_{CDR0} -F_{CDR0}} \right)^{w_1}\left( {T_{CDR0.5} -F_{CDR0.5}} \right)^{w_2},
\end{equation}
where $w_1 \le w_2 $ are positive odd numbers, $F_{CDR0}$ is the
number of CDR0.5 subjects classified (falsely) as CDR0 and
$F_{CDR0.5}$ is the number of CDR0 subjects classified (falsely) as
CDR0.5. Notice that minimizing this cost function will maximize the
correct classification rates and minimize the misclassification rates.
The correct classification rates, sensitivity and specificity
rates are provided in Table \ref{tab:class-rates-dist}.
Using $w_1 = w_2 = 1$,
optimal threshold values are $p_o = 0.5$ for model $M_{II}(D)$ in
Equation (9), $p_o = 0.45$ for model $M_{III}(D)$ in Equation (10),
and optimal $p_o = 0.38$ for model $M_{IV}(D)$ in Equation (11). The
specificity rates are 69\%, 73\%, and 69\%, respectively.
The sensitivity rates are 56\%, 67\%, and 72\%,
respectively. Obviously, from a clinical point of view,
misclassifying a CDR0.5 subject as CDR0 (i.e., classifying a
diseased subject as healthy) might be less desirable, since a
subject labeled as CDR0.5 will undergo further screening but a
subject labeled as CDR0 will be released. So the parameters $w_1$
and $w_2 $ could be modified to reflect such practical concerns and
then a different set of threshold $p_o$ values could be found. For
example, we set $w_1 = 1$ and $w_2 = 3$ which favors correct
classification of CDR0.5 subjects more than that of CDR0 subjects
(i.e., favors higher sensitivity).
Observe that with $w_1 = w_2 = 1$ the best classifier is based on model $M_{IV}
(D)$ and with $w_1 = 1$ and $w_2 = 3$ the best classifier is based on
model $M_{III}(D)$.

Alternatively we can maximize the sensitivity and specificity rates
by minimizing the following cost function
\begin{equation}
\label{eqn:cost2}
C_2(p_o ,\eta_1 ,\eta_2) = -\left[ {\eta_1 \frac{\left(
{T_{CDR0} -F_{CDR0}} \right)}{N_{CDR0}}+\eta_2 \frac{\left(
{T_{CDR0.5} -F_{CDR0.5}} \right)}{N_{CDR0.5}}} \right],
\end{equation}
where $\eta_1 ,\eta_2 \ge 0$ and $\eta_1 +\eta_2 = 1$. Notice
that as either of sensitivity or specificity increases, the cost
function $C_2(p_o ,\eta_1 ,\eta_2$) decreases. With $\eta_1
 = \eta_2 = 0.5$ the best classifier is based on model $M_{IV}(D)$ and
with $\eta_1 = .3,\;\eta_2 = 0.7$ the best classifier is based on
model $M_{III}(D)$. Observe that from $\eta_1 = \eta_2 = 0.5$ to
$\eta_1 = .3,\;\eta_2 = 0.7$, sensitivity increases, correct
classification rate and specificity tend to decrease.

\section{Analysis of Hippocampal Volumes}
\label{sec:mylabel2}
The LDDMM distance gives one number reflecting
the global size and shape. Volume measurements were presented in
detail in [29]. The LB-CDR0 subjects
had an average hippocampal volume of 2081 ( $\pm$ 354) {\it
mm}$^{3}$ while RB-CDR0 subjects had 2600 ( $\pm$ 481) {\it
mm}$^{3}$. The LB-CDR0.5 subjects had an average hippocampal volume
of 1717 ( $\pm$ 224) {\it mm}$^{3}$ and RB-CDR0.5 had 2186 ( $\pm$
370) {\it mm}$^{3}$. On the other hand, LF-CDR0 subjects showed a
volume reduction of 82 {\it mm}$^{3}$ (4.0\%, NS) and RF-CDR0
subjects showed a reduction of142 {\it mm}$^{3}$ (5.5\%, NS) where
NS stands for ``not significant''. LF-CDR0.5 subjects had
hippocampal volume reduction of 164 {\it mm}$^{3}$ (8.3\%, $p = 0.03$)
and RF-CDR0.5 subjects had reduction of 236 {\it mm}$^{3}$
(10.2\%, $p = 0.05$) on the right side.
Repeated-measures ANOVA showed both significant change over time (within group,
$F =  98.97,\; df = 1,42,\; p<.0001$) and significant time-group interaction
($F =  7.81,\; df = 1,42,\; p = 0.0078$) in the hippocampal volumes.
The time-group interaction persisted when covaried with
baseline total cerebral brain volume ($p = 0.0066$) or with baseline
total intracranial volume ($p = 0.0077$).
In order to take into account variations in between-visit intervals (mean 2.11 $\pm$ 0.47
years), scan interval was also used as a covariate in the volumes comparison.
Again, the significant time $\times $group interaction
after covarying for scan intervals ($p = 0.015$) persisted.

\subsection{Repeated Measures Analysis of Hippocampal Volumes}
We repeat the same modeling procedure of Section 3.1 on the
hippocampal volumes. For modeling hippocampal volumes using the
repeated measures ANOVA with group as main effect and compound
symmetry in Var-Cov structure and volume measurements repeated over
time for each subject, the model is
\begin{equation}
\label{eqn:model-vol-with-group}
V_{ijk} = \mu +\alpha_i^D +\alpha_j^T +\alpha_{ij}^{DT} +\varepsilon_{ijk},
\end{equation}
where $V_{ijk}$ is the volume for subject $k$ with diagnosis $i$ at
timepoint $j$, $\mu$ is the overall mean, $\alpha_i^D$ is the
effect of diagnosis level $i$,
$\alpha_j^T $ is the effect of timepoint level $j$,
$\alpha_{ij}^{DT}$ is the diagnosis-by-timepoint interaction, i.e.,
part of the mean volume not attributable to the additive effect of
diagnosis and timepoint, and $\varepsilon_{ijk}$ is the error term.
Notice that the effect of side (left or right) is ignored in
this model. There is significant group main effect
($F = 17.54,\; df = 1,42,\;p = 0.0001$) and within group time-point main effect
($F = 9.87,\; df = 1,130,\; p = 0.0021$) but the
group-by-timepoint interaction is not significant
($F = 0.84,\;df = 1,130,\; p = 0.3624$).
This implies that the main effect of
group comparison is meaningful and about the same at each timepoint.
Moreover, the groups do change in morphometry over time.

For modeling volumes using the repeated measures ANOVA with side as
main effect and compound symmetry in Var-Cov structure and volume
measurements repeated over time, the corresponding model is
\begin{equation}
\label{eqn:model-vol-with-side}
V_{ijk} = \mu +\alpha_i^S +\alpha_j^T +\alpha_{ij}^{ST}+\varepsilon_{ijk},
\end{equation}
where $V_{ijk}$ is the volume for subject $k$ for side $i$ ($i = 1$
for left; 2 for right) at timepoint $j$, $\mu$ is the overall mean,
$\alpha_i^S $ is the effect of side level $i$, $\alpha_j^T $ is
the effect of timepoint level $j = 1,2$, $\alpha_{ij}^{ST}$ is the
side-by-timepoint interaction, and $\varepsilon_{ijk}$ is the
error term. Notice that the effect of diagnosis (CDR0 or CDR0.5) is
ignored in this model. The side and timepoint main effects are both significant
($F = 377.21,\; df = 1,129,\; p < .0001$ and $F = 38.31,\; df = 1,129,\; p < .0001$, respectively),
but side-by-timepoint interaction is not significant ($F = 1.84,\;df = 1,129,\;p = 0.1769$).
Consequently, we conclude that the lines
that join mean volumes in the interaction plot are parallel and far
apart, the main effect of side comparison is meaningful, and about
the same at each timepoint. Moreover, the left and right hippocampi
do change in morphometry over time.

For the model that includes the diagnosis, side, and
diagnosis-by-side interaction, we use the same model selection
criteria in Section 3.1.3. We find that the most promising model
based on likelihood ratio test, BIC, and AIC is the one with
unstructured Var-Cov matrix. The corresponding model with
significant terms at $\alpha = .05$ level is
\begin{equation}
\label{eqn:model-vol-with-group-side-interaction}
V_{ijkl} = \mu +\alpha_i^S +\alpha_j^D +\alpha_k^T
+\alpha_{ik}^{ST} +\alpha_{jk}^{DT} +\varepsilon_{ijkl},
\end{equation}
where $V_{ijkl}$ is the volume for subject $l$ for side $i$
with diagnosis $j$ at timepoint $k$,
$\mu$ is the overall mean,
$\alpha_i^S $ is the effect of side level $i$,
$\alpha_j^D$ is the effect of diagnosis level $j$,
$\alpha_k^T $ is the effect of timepoint level $k$ = 1,3,
$\alpha_{ik}^{ST}$ is the side-by-timepoint interaction,
$\alpha_{jk}^{DT}$ is the diagnosis-by-timepoint interaction,
and $\varepsilon_{ijkl}$ is the error term.
The (unstructured) Var-Cov structure for the error term is
\[
\Cov(\varepsilon_{ijkl} ,\varepsilon_{i'jk'l}) = \left[
{{\begin{array}{*{20}c}
 {\sigma^2} \hfill & \hfill & \hfill & \hfill \\
 {\sigma_{21}} \hfill & {\sigma^2} \hfill & \hfill & \hfill \\
 {\sigma_{31}} \hfill & {\sigma_{32}} \hfill & {\sigma^2} \hfill &
\hfill \\
 {\sigma_{41}} \hfill & {\sigma_{42}} \hfill & {\sigma_{43}} \hfill &
{\sigma^2} \hfill \\
\end{array}}} \right].
\]
The main effects of side, group, and timepoint are all significant
($F = 120.10,\; df = 1,170,\; p < .0001$; $F = 25.25, \; df = 1,170,\; p < .0001$;
and $F = 89.53,\; df = 1,170,\;p < 0.0001$, respectively).
But due to interaction, the main effect for
diagnosis (i.e., group) is close to clinically meaningless, i.e.,
the group means should be compared at each time point or hemisphere
instead of comparing the overall means of the groups. But, the main
effects of group and side being significant are interpretable
between baseline and follow-up.

\subsection{Post-Hoc Comparison of Hippocampal Volumes for Differences in Group, Time, and Hemisphere}
\label{subsec:mylabel2}
We repeat the analysis
procedure of Section 3 on hippocampal volumes also. We find that
left hippocampus volumes are significantly smaller than the right
hippocampus volumes at both baseline and follow-up years (i.e.,
there is significant volumetric left-right asymmetry in hippocampi);
baseline volumes are larger than follow-up volumes for both left and
right hippocampi (i.e., there is significant reduction in volume by
time) ($p<.0001$ for each comparison). The means and standard
deviations of the volumes for left and right hippocampi of each
group are provided in Table \ref{tab:summary-stat-dist}.
We observe the same trend in the overall comparison for each group also.
However, left-right volumetric asymmetry significantly reduces by time in
CDR0.5 group ($p = .0407$); but the same holds only barely in CDR0
group ($p = .0524$). The level of left-right volumetric asymmetry is
about the same in both CDR0 and CDR0.5 groups at baseline
($p = .3495$) and follow-up ($p = .4853$). The volumes decrease
significantly by time in CDR0 group for both left and right
hippocampi ($p<.0001$ for both); the same holds for CDR0.5 group
also ($p = .0001$ for both). The volumetric reduction is significantly
larger in CDR0.5 right hippocampi compared to CDR0.5 left hippocampi
($p = .0407$); but the same holds only barely in CDR0 group
($p = .0524$). On the other hand, the volumetric reduction is
significantly larger in CDR0.5 left hippocampi compared to CDR0 left
hippocampi ($p = .0108$); the same holds for right hippocampi also
($p = .0418$). The variances of volumes are not significantly
different for (LB-CDR0.5, LB-CDR0), (RB-CDR0.5, RB-CDR0), and
(RF-CDR0.5, RF-CDR0) groups, but volumes of LF-CDR0 hippocampi are
significantly larger than LF-CDR0.5 hippocampi ($p = .0268$). The
CDR0.5 volumes are significantly smaller than CDR0 volumes in left
hippocampi at baseline ($p = .0001$) and follow-up ($p<.0001$), and
for right hippocampi at baseline ($p = .0071$) and follow-up
($p = .0001$). The CDR0.5 volumes are stochastically smaller than CDR0
volumes for left hippocampi at baseline ($p = .0007$) and follow-up
($p = .0003$), and for right hippocampi at baseline ($p = .0064$) and
follow-up ($p = .0028$).

\subsection{Logistic Discrimination with Hippocampal Volumes}
We apply the logistic discrimination methods of Section 3.7 on
hippocampal volumes. First we consider the full logistic model
(designated as $M_I(V)$) with side, timepoint, and volume with all
possible interactions being the predictor variables. We apply the
same stepwise elimination procedure as in Section 3.7 and get the
following reduced model:
\begin{equation}
\label{eqn:MIIV}
M_{II}(V)\mbox{: logit }p_l = \beta_0 +\alpha_k^S +\beta_1 V_{ijkl},
\end{equation}
where $p_l $ is the probability of subject $l$ having DAT and
$V_{ijkl}$ the volume for subject $l$ with diagnosis $i$ ( $i = 1$ for CDR0 and 2 for CDR0.5)
at timepoint $j$ ($j = 1$ for baseline and 2 for follow-up) with side $k$ ($k = 1$ for left and 2 for right),
$\beta_0$ is the overall intercept, $\alpha_k^S $ is the effect of side level $k$,
and $\beta_1$ is the slope of the fitted line.

However, as we have seen in Section 3.1.3, due to group-by-timepoint
interaction, we need to consider diagnosis groups at each time
point. When we use $V_{ikl}^B$ and $V_{ikl}^F$ one at a time in a
logistic model, we see that the model
\begin{equation}
\label{eqn:MIIIV}
M_{III}(V)\mbox{: logit }p_l = \beta_0 +\alpha_k^S +\beta_1 V_{ikl}^F
\end{equation}
has the most significant coefficients for the volume terms.

Moreover, when we use $V_{kl}^{LB} ,\;V_{kl}^{LF} ,\;V_{kl}^{RB}$, and $V_{kl}^{RF}$
one at a time in a logistic model, we see that the following model has the best fit.
\begin{equation}
\label{eqn:MIVV}
M_{IV}(V)\mbox{: logit }p_l = \beta_0 +\beta_1 V_{kl}^{LF}
\end{equation}
The classification rates are presented in Table \ref{tab:class-rates-vol}.
Observe that with $p_o = 1/2$
the best classifier is based on model $M_{III}(V)$ and with $p_o
 = 18/44$ the best classifier is based on model $M_{IV}(V)$.
Furthermore, as $p_o$ decreases from 1/2, sensitivity increases but
the correct classification rate and specificity decreases. We use
the cost function $C_1(p_o ,w_1 ,w_2$) with $w_1 = w_2 = 1$ and with
$w_1 = 1$ and $w_2 = 3$ to calculate the optimal $p_o$ values for
each of the models $M_I(V)-M_{IV}(V)$.
Observe that with $w_1 = w_2 = 1$ the best
classifier is based on model $M_{IV}(V)$ and with $w_1 = 1$ and $w_2
 = 3$ the best classifier is based on model $M_{III}(V)$.
We find the optimal $p_o$ values based on the cost function $C_2(p_o ,\eta_1 ,\eta_2$)
with $\eta_1 = \eta_2 = 0.5$ and with $\eta_1 = .3,\;\eta_2 = 0.7$ for each of models $M_I(V)-M_{IV}(V)$.
With $\eta_1 = \eta_2 = 0.5$ the best
classifier is based on model $M_{IV}(V)$ and with $\eta_1 = .3,\;\eta_2 = 0.7$
the best classifier is based on model $M_I(V)$.
Observe that from $\eta_1 = \eta_2 = 0.5$ to $\eta_1 = .3,\;\eta_2 = 0.7$,
sensitivity increases, correct classification rate and specificity tend to decrease.

\section{Comparison of Hippocampal Volumes and Metric Distances}
\label{sec:comparison}
Although volume is a measure of size and
metric distance is a measure of overall morphometric difference from
a template, the repeated measure analysis and post-hoc analysis of
volumes and metric distances provide similar results. The main
difference is that volumes tend to decrease, while LDDMM distances
tend to increase by time.

The logistic discrimination models are similar, except model $M_{IV}(D)$
for metric distances contains right follow-up distances, while
model $M_{IV}(V)$ for volumes contains left follow-up volumes.
The classification performances with $p_o = 1/2$ and $p_o = 18/44$
suggest that volume models have better performance
than the metric distance models (see Tables \ref{tab:class-rates-dist} and \ref{tab:class-rates-vol}).
Using the optimal $p_o$ values with the cost functions
$C_1(p_o,w_1 ,w_2$) and $C_2(p_o ,\eta_1 ,\eta_2$), the classification
performances are significantly different for models $M_I(V)-M_{IV}(V)$
of volumes and $M_I(D)-M_{IV}(D)$ metric distances.
Comparing Tables \ref{tab:class-rates-dist} and \ref{tab:class-rates-vol}, we see that
logistic discrimination with volumes has better performance.

We apply the logistic discrimination using both volume and metric
distance as predictors. The models we consider are the full logistic
model (designated as model $M_I(V,D)$) with side, timepoint,
volume, and metric distances with all possible interactions being
predictor variables. We apply the same stepwise elimination
procedure as in Section 3.7 and get
$$M_{II} (V,D)\mbox{: logit }p_l = \beta_0 +\alpha_k^S +\beta_1 V_{ijkl}
+\beta_2 d_{ijkl}^9 +\beta_3 V_{ijkl} d_{ijkl}$$
where $p_l $ is the probability of subject $l$ having DAT and $V_{ijkl}$ the volume
and $d_{ijkl}$ the distance for subject $l$ with diagnosis $i$
($i = 1$ for CDR0 and 2 for CDR0.5) at timepoint $j$ ($j = 1$ for baseline and 2 for follow-up)
with side $k$ ($k = 1$ for left and 2 for right),
$\beta_0$ is the overall intercept,
$\alpha_k^S $ is the effect of side level $k$,
$\beta_1$ is the coefficient for volume,
$\beta_2 $ is the coefficient for ninth power of
the distance, $\beta_3 $ is the coefficient for the interaction
between volume and distance. When we use baseline or
follow-up measures one at a time in a logistic model, we see that the model
$$M_{III}(V,D)\mbox{: logit }p_l = \beta_0 +\alpha_k^S +\beta_1 V_{ikl}^F +\beta_2 \left(d_{ikl}^F\right)^5$$
has the most significant coefficients.
When we use side-by-timepoint
combinations one at a time in a logistic model, we see that the
following model has the best fit:
$$M_{IV}(V,D)\mbox{: logit }p_l = \beta_0 +\beta_1 \left(d_{ikl}^F\right)^3+\beta_2 V_{kl}^{LF}.$$
The corresponding classification rates are presented in
Table \ref{tab:class-rates-Vol-Dist}.
Observe that considering metric distance and
volume together in the logistic discrimination procedure with the
cost functions $C_1(p_o ,w_1 ,w_2$) and $C_2(p_o ,\eta_1 ,\eta_2$),
we get better classification rates compared to logistic
models with only one of metric distance or volume being the predictors.

\section{Annual Percentage Rates of Change in Hippocampal Volumes and Metric Distances}
Our volume and LDDMM metric comparisons are
cross-sectional or longitudinal by construction. However these
measures might need to be adjusted for anatomic variability, since
intersubject variability might add substantial amount of noise to
volume or distance measurements at baseline or follow-up. There is
no simple way to correct for this noise in practice. Differential
volume loss or distance change over time might be self-correcting
for such variability. For example, entorhinal cortex volume loss
over time was shown to be a better indicator for DAT than
cross-sectional measurements [73].

\subsection{Annual Percentage Rate of Change in Hippocampal Volume}
The hippocampal volume change over time can be written as the
following annual percentage rate of change (APC) [73]:
\begin{equation}
\label{eqn:APCinVol}
V^{APC} = \frac{V_k^b -V_k^f}{V_k^b \times T}\times 100\,\%,
\end{equation}
where $T$ is the interscan interval in years ($T \approx 2$ in our data).

For modeling annual percentage rate of change in volume $V^{APC}$
using the repeated measures ANOVA with group as main effect and
compound symmetry in Var-Cov structure and $V^{APC}$ measures
repeated over side for each subject, the model is
\begin{equation}
\label{eqn:model-VAPC-with-group}
V_{ijk}^{APC} = \mu +\alpha_i^D +\alpha_j^S +\alpha_{ij}^{DS} +\varepsilon_{ijk},
\end{equation}
where $V_{ijk}^{APC}$ is the APC in volume for side $j$
of subject $k$ with diagnosis $i$, $\mu$ is the overall mean, $\alpha_i^D$
is the effect of diagnosis level $i$ ($i = 1$ for CDR0; 2 for CDR0.5),
$\alpha_j^S $ is the effect of side level $j$ ($j=1$ for left and 2 for right),
$\alpha_{ij}^{DS}$ is the diagnosis-by-side interaction,
and $\varepsilon_{ijk}$ is the error term.
The diagnosis main effect is significant
($F = 18.62,\; df = 1,84,\; p < .0001$) but neither side main
effect ($F = 0.72,\; df = 1,84,\; p = .3754$) nor
diagnosis-by-side interaction is significant
($F = 0.11,\; df = 1,84,\; p = 0.7384$).
Consequently, we conclude that the lines that
join the mean $V^{APC}$ values in the interaction plot are parallel
and far apart, the main effect of diagnosis comparison is
meaningful, and about the same at each hemisphere. The post hoc
comparison of $V^{APC}$ values indicate that the APC in CDR0.5
volumes are significantly larger than APC in CDR0 volumes ($p = .0001$) .

We apply the logistic discrimination methods of Section 3.7 on APC
in hippocampal volumes. First we consider the full logistic model
(designated as $M_I \left( {V^{APC}} \right)$) with side
and APC in volume with all possible interactions being the predictor
variables. We apply the same stepwise elimination procedure as in
Section 3.7 and get
\begin{equation}
\label{eqn:MIIVAPC}
M_{II} \left( {V^{APC}} \right)\mbox{: logit }p_k = \beta_0
+\beta_1 V_{ijk}^{APC} +\beta_2 \left( {V_{ijk}^{APC}}\right)^3
\end{equation}
where $p_k$ is the probability of subject $k$ having DAT.

Furthermore, when we use $V_{ijk}^{APC,L}$ and $V_{ijk}^{APC,R}$
as predictors in a logistic model, we see that the following model
has the best fit.
\begin{equation}
\label{eqn:MIIIVAPC}
M_{III} \left( {V^{APC}} \right)\mbox{: logit }p_k =
\beta_0+\beta_1 V_{ik}^{APC,L}+\beta_2 V_{ik}^{APC,R}+\beta_3 \left( {V_{ik}^{APC,L}} \right)^2.
\end{equation}
The classification rates with $p_o = 1/2$ and $p_o = 18/44$ and optimal
$p_o$ values with respect to the cost functions are presented in
Table \ref{tab:class-rates-APCinVolumes}.
Observe that the classifier using the cost
function $C_2(p_o ,\eta_1 ,\eta_2$) with $\eta_1 = .3,\;\eta_2  = 0.7$
in model $M_I \left( {V^{APC}} \right)$ has the best performance.
Comparing Tables \ref{tab:class-rates-vol} and \ref{tab:class-rates-Vol-Dist},
we observe that correct classification
rates, sensitivity, and specificity percentages with
the classifiers based on APC in volume
are about the same as those with volume only.
Unlike the findings of [73] hippocampal volume loss
over time is not a better indicator for DAT than
cross-sectional measurements.
On the other hand, the classifier based on volume and distance
together performs better compared to models based on only
one of volume, distance, or APC in volume values.

\subsection{Annual Percentage Rate of Change in LDDMM Metric Distance}
The hippocampal LDDMM metric distance change over time can be
written as the following annual percentage rate of change:
\begin{equation}
\label{eqn:APCinDist}
D^{APC} = \frac{d_k^f -d_k^b}{d_k^b \times T}\times 100\,\%.
\end{equation}
Notice that to make APC in metric distance positive,
we take the difference $d_k^f -d_k^b$
as opposed to the order in the APC in volume definition
For modeling $D^{APC}$ using the repeated measures ANOVA with group
as main effect and compound symmetry in the Var-Cov structure and
$D^{APC}$ measures repeated over side for each subject, the model is
\begin{equation}
\label{eqn:model-DAPC-with-group}
D_{ijk}^{APC} = \mu +\alpha_i^D +\alpha_j^S +\alpha_{ij}^{DS} +\varepsilon_{ijk},
\end{equation}
where $D_{ijk}^{APC}$ is the APC in LDDMM metric distance for
side $j$ of subject $k$ with diagnosis $i$.
The other terms are as in Equation (21).
The diagnosis main effect is significant ($F = 4.75,\; df = 1,84,\;p = .0320$)
but neither side main effect ($F = 2.29,\; df = 1,84,\; p = .1338$) nor diagnosis-by-side
interaction is significant ($F = 0.87,\; df = 1,84,\; p = 0.3532$).
Consequently, we conclude that the lines that join the
means in the interaction plot are parallel and far apart, the main
effect of diagnosis comparison is meaningful, and about the same at
each hemisphere. The APC in CDR0.5 LDDMM distances is significantly
larger (in absolute value) than APC in CDR0 distances ($p = .0036$) .

We apply the logistic discrimination methods of Section 3.7 on APC
in hippocampal LDDMM distances. First we consider the full logistic
model (designated as $M_I \left( {D^{APC}} \right)$) with side
and APC in distance with all possible interactions being
the predictor variables. We apply the same stepwise elimination
procedure as in Section 3.7 and get
\begin{equation}
\label{eqn:MIIDAPC}
M_{II} \left( {D^{APC}} \right)\mbox{: logit }p_k = \beta_0+\beta_1 D_{ijk}^{APC}.
\end{equation}

Furthermore, when we use $D_{ijk}^{APC,L}$ and $D_{ijk}^{APC,R}$
as predictors in a logistic model, we see that the following model
has the best fit.
\begin{equation}
\label{eqn:MIIIDAPC}
M_{III} \left( {D^{APC}} \right)\mbox{: logit }p_k = \beta_0+\beta_1 D_{ik}^{APC,R}.
\end{equation}

The classification rates with $p_o = 1/2$ and $p_o = 18/44$ and
optimal $p_o$ values with respect to the cost functions are
presented in Table \ref{tab:class-rates-APCinDistances}.
Observe that these classifiers have relatively poor performance.
Comparing each of Tables \ref{tab:class-rates-dist} and \ref{tab:class-rates-APCinDistances},
we observe that the classifiers with the optimal $p_o$ values have much larger
sensitivity rates but this increase comes at the expense of
substantial decrease in correct classification rate and specificity.
Hence hippocampal LDDMM change over time is not a better indicator for DAT than
cross-sectional distance comparisons.

\subsection{Annual Percentage Rate of Change in Hippocampal Volume and Metric Distances}
We apply the logistic discrimination based on distance and APC in volumes.
First we consider the full logistic model (designated as
$M_I \left( V^{APC},D \right))$ with side and APC in volume,
and distances with all possible interactions being the predictor variables.
We apply the same stepwise elimination procedure as in Section 3.7 and get
\begin{equation}
\label{eqn:MIIVAPC-dist}
M_{II} \left( V^{APC},D \right)\mbox{: logit }p_k =\beta _0 +\beta _1 V_{ijk}^{APC}
+\beta _2 \left( {V_{ijk}^{APC} } \right)^3+\beta _3 \left( {d_{ik}^F } \right)^2.
\end{equation}
Furthermore, when we use and left and right measures as predictors in a
logistic model, we see that the following model has the best fit.
\begin{equation}
\label{eqn:MIIIVAPC-dist}
M_{III} \left( V^{APC},D \right)\mbox{: logit }p_k =
\beta _0 +\beta _1 V_{ik}^{APC,L}+ \beta _2 V_{ik}^{APC,R} + \beta _3 d_{ik}^{RF}.
\end{equation}

The classification rates with $p_o =1/2 $ and $p_o = 18/44$ and optimal $p_o$
values with respect to the cost functions are presented in
Table \ref{tab:class-rates-APCinVolume-Distance}.
With the cost function $C_1 (p_o ,w_1 = 1,w_2 = 1)$, the best classifier is based on $M_{III} \left( V^{APC},D \right)$
for which the optimal threshold value is $p_o \approx .37$, the correct
classification rate is 80\%, sensitivity is 78\%, and specificity is 81\%.
Likewise, with the cost function $C_1 (p_o ,w_1 =1,w_2 =3)$, the best
classifier is based on $M_I \left( V^{APC},D \right)$ for which
the optimal threshold value is $p_o =.56$, the correct classification rate
is 80\%, sensitivity is 78\%, and specificity is 81\%.
On the other hand, with cost function $C_2 (p_o ,\eta _1 =.5,\eta _2 =.5)$
the best classifier is based on
$M_I^{APC}\left( {V,D} \right)$ for which the optimal threshold value is $p_o =.64$,
the correct classification rate is 84\%, sensitivity is 72\%, and specificity is 92\%.
With cost function $C_2 (p_o,\eta _1 =.3,\eta _2 =.7)$, the best classifier is again
based on $M_I^{APC}\left( {V,D} \right)$ for which the optimal threshold value is $p_o =.56$,
the correct classification rate is 80\%, sensitivity is 78\%, and
specificity is 81\%.
Comparing Tables \ref{tab:class-rates-Vol-Dist} and \ref{tab:class-rates-APCinVolume-Distance},
we observe that the classifiers based on
metric distance and volume usually perform better compared to the
classifiers based on metric distance and APC in volume.
Comparing Table \ref{tab:class-rates-APCinVolumes} and \ref{tab:class-rates-APCinVolume-Distance},
we observe that adding the metric distance to the logistic model with APC in
volume improves the classification performance.
Hence the model with hippocampal volume loss over time and metric distance is a better indicator
for DAT compared to either variable used separately in logistic
discrimination.

We also apply the logistic discrimination based on volume, distance, and APC in volumes.
First we consider the full logistic model (designated as
$M_I \left( V, V^{APC},D \right))$ with side, volume, and APC in volume,
and distances with all possible interactions being the predictor variables.
We apply the same stepwise elimination procedure as in Section 3.7 and get
\begin{equation}
\label{eqn:MIIVVAPC-dist}
M_{II} \left( V, V^{APC},D \right)\mbox{: logit }p_k =\beta _0 +\beta _1 V_{ijk}^{B} +\beta _2 V_{ijk}^{APC}
+\beta _3 \left( {V_{ijk}^{APC} } \right)^3+\beta _4 \left( {d_{ijk}^F } \right)^3.
\end{equation}
Furthermore, when we use and left and right measures as predictors in a
logistic model, we see that the following model has the best fit.
\begin{equation}
\label{eqn:MIIIVVAPC-dist}
M_{III} \left( V,V^{APC},D \right)\mbox{: logit }p_k =
\beta _0 + \beta _1 V_{ik}^{RF}+ \beta _2 d_{ik}^{LF}+\beta _3 V_{ik}^{APC,R}+
\beta _4 \left(V_{ik}^{APC,L} \right)^3 + \beta _5 \left(d_{ik}^{RF}\right)^3.
\end{equation}

The classification rates with $p_o =1/2 $ and $p_o = 18/44$ and optimal $p_o$
values with respect to the cost functions are presented in
Table \ref{tab:class-rates-Volume-Distance-APCinVolume}.
With the cost function $C_1 (p_o ,w_1 = 1,w_2 = 1)$, the best classifier is based on $M_{III} \left(V, V^{APC},D \right)$.
Comparing Table \ref{tab:class-rates-Volume-Distance-APCinVolume} with
Tables \ref{tab:class-rates-dist}, \ref{tab:class-rates-vol}, \ref{tab:class-rates-APCinVolumes},
\ref{tab:class-rates-APCinDistances}, \ref{tab:class-rates-Vol-Dist}, and \ref{tab:class-rates-APCinVolume-Distance},
we observe that the classifiers based on
metric distance, volume, and APC in volumes usually perform better compared to the
classifiers based on other models.
Hence the model with volume, hippocampal volume loss over time, and metric distance is a better indicator
for DAT compared to other models based on subsets of these variables.

\section{Discussion and Conclusions}
\label{sec:disc}
In this study, we used the Large Deformation Diffeomorphic Metric
Mapping (LDDMM) algorithm to generate metric distances between
hippocampi in groups of subjects with and without Dementia of
Alzheimer's type (DAT) in its mild form (labeled as CDR0.5 and CDR0
patients, respectively) at baseline and at follow-up. The subjects
in this paper have been previously analyzed using related but
different tools. As a single scalar measure, volumes were used for
diagnosis group comparisons at baseline and follow-up [29] and displacement momentum vector
fields based on LDDMM were used for discrimination [57].
But the metric distances computed from LDDMM
has not hitherto been used in diagnosis group analysis. The metric
distance gives a single number reflecting the global morphometry
(i.e., the size and shape) while volume measurements only provide
information on size. So metric distances provide morphometric
information not conveyed by volume whereas momentum vector fields
also provide local information on shape changes. Further, the
morphometric information conveyed by the metric distance depends on
the choice of the template, while the morphometric information
conveyed by momentum vector fields is independent of the template
chosen. That is, although the vector fields change when the template
changes, the morphometric information they convey is the same.

Previously, it has been shown that hippocampal volume loss and shape
deformities observed in subjects with DAT distinguished them from
both elderly and younger control subjects [10].
The pattern of hippocampal deformities
in subjects with DAT was largely symmetric and suggested damage to
the CA1 hippocampal subfield [74].
Hippocampal shape changes were
also observed in healthy elderly subjects, which distinguished them
from healthy younger subjects. These shape changes occurred in a
pattern distinct from the pattern seen in DAT and were not
associated with substantial volume loss [75].

Furthermore, Wang et al. [45] analyzed
the baseline hippocampi of the same data set and showed that the
very mild DAT subjects showed significant inward variation in the
left and right lateral zones (LZ) the left and right intermediate
zone (IMZ), but not in the left and right superior zones (SZ) as
compared to CDR0 subjects. In their logistic regression analysis,
inward variation of the left and right LZ or IMZ by 0.1 mm relative
to the average of the nondemented subjects increased the odds of the
subject being a very mild DAT subject rather than being a
nondemented subject. The odds ratios for the left and right SZ were
not significant. These results represented a replication of their
previous findings [10] and suggest that
inward deformities of the hippocampal surface in proximity to the
CA1 subfield and subiculum can be used to distinguish subjects with
very mild DAT from CDR0 subjects [45].
However, although momentum vector fields obtained by the LDDMM
algorithm can be used to detect such local (i.e., location specific)
morphometric changes (as in the CA1 subfield), metric distance does
not provide such local information, hence fails to indicate any type
of laterality differences.

The main results in this paper are that although metric distances
did not detect any significant difference in morphometry at baseline
(see Table \ref{tab:comparison-lddmm-distance}), follow-up metric distances for the
right hippocampus in CDR0.5 (i.e., mildly demented) subjects are
found to be significantly larger than those in CDR0 (i.e.,
non-demented) subjects (see Table \ref{tab:comparison-lddmm-distance}).
Wang et al. also analyzed the velocity vector fields for the baseline hippocampi of
the same data set and found that the left hippocampus in the DAT
group shows significant shape abnormality and the right hippocampus
shows similar pattern of abnormality [57].
Again, the reason for the metric failing to detect such abnormality in the
baseline hippocampi is that metric distance is a compound and
oversummarizing measure of global morphometry. From baseline to
follow-up, metric distances for CDR0.5 subjects significantly
increase while those in CDR0 subjects do not (see
Table \ref{tab:comparison-lddmm-distance}). That is, the morphometry (shape and size) of
hippocampus in CDR0.5 subjects changes significantly over time, but
not in CDR0 subjects. Atrophy -- over two years -- might occur with
aging, and this is captured by metric distances (see
Table \ref{tab:comparison-lddmm-distance}). However the increase in the metric distances
in CDR0 subjects is not found to be statistically significant.

Such differences in morphometry between diagnosis groups or
morphometric changes over time can be detected by metric distances
computed via LDDMM and could potentially serve as a biomarker for
the disease. Previously, the volumes and velocity vector fields
associated with the same data set (i.e., baseline and follow-up of
both groups) were also analyzed and it was found that hippocampal
volume loss over time was significantly greater in the CDR0.5
subjects (left = 8.3\%, right = 10.2\%) than in the CDR0
subjects (left = 4.0\%, right = 5.5\%) (ANOVA, $F$ = 7.81, $p$
 = 0.0078). Using singular-value decomposition and logistic
regression models, [45] quantified
hippocampal shape change across time within individuals, and this
shape change in the CDR0.5 and CDR0 subjects was found to be
significantly different (Wilks' $\lambda $, p = 0.014). Further, at
baseline, CDR0.5 subjects, in comparison to CDR0 subjects, showed
inward deformation over 38\% of the hippocampal surface; after 2
years this difference grew to 47\%. Also, within the CDR0
subjects, shape change between baseline and follow-up was largely
confined to the head of the hippocampus and subiculum, while in the
CDR0.5 subjects, shape change involved the lateral body of the
hippocampus as well as the head region and subiculum. These results
suggest that different patterns of hippocampal shape change in time
as well as different rates of hippocampal volume loss distinguish
very mild DAT from healthy aging [29].

In regard to statistical analysis, as a compound but brief measure
of morphometry, metric distances can thus serve as a first step to
identify the morphometric differences, and can be used as a pointer
to which direction a clinician or data analyst could go.
The importance of the CDR0 versus CDR0.5 contrast analyzed here is that
it tests a necessary but not sufficient condition for the eventual
goal of discriminating CDR0s who subsequently convert to CDR0.5,
from CDR0s who subsequently stay CDR0. As subjects are followed
longitudinally and some convert, we have shown that cross-sectional
measures of the hippocampal structure can be used to predict those
who convert. Metric distances may also be used in this way. When
baseline and followup of converted and nonconverted nondemented
subjects were analyzed, it was found that the inward variation of
the lateral zone and left hippocampal volume significantly predicted
conversion to CDR0.5 in separate Cox proportional hazards models.
When hippocampal surface variation and volume were included in a
single model, inward variation of the lateral zone of the left
hippocampal surface was selected as the only significant predictor
of conversion. The pattern of hippocampal surface deformation
observed in nondemented subjects who later converted to CDR0.5 was
similar to the pattern of hippocampal surface deformation previously
observed to discriminate subjects with very mild DAT and nondemented
subjects. These results suggest that inward deformation of the left
hippocampal surface in a zone corresponding to the CA1 subfield is
an early predictor of the onset of DAT in nondemented elderly
subjects [74]. This appears to contradict our finding that the
morphometric changes in CDR0.5 right hippocampi from baseline to
follow-up is significantly larger than those of CDR0.5 left
hippocampi ($p = 0.044$). The morphometric changes in CDR0 right
hippocampi from baseline to follow-up are not significantly
different from those of CDR0 left hippocampi ($p = 0.382$). The
morphometric changes in CDR0.5 left hippocampi from baseline to
follow-up are not significantly different from those in CDR0 left
hippocampi ($p = 0.134$), while the morphometric changes in CDR0.5
right hippocampi from baseline to follow-up are significantly larger
than those of CDR0 right hippocampi ($p = 0.007$). Therefore, over
time, DAT may alter the (global) morphometry of the right
hippocampus. However, note that the finding in [74] are concerned
with changes in (local) subregions of hippocampi, while metric
distance is concerned with overall morphometric changes. That is,
DAT might implicate CA1 of the left hippocampi, yet the overall
change in morphometry of right hippocampi might be more substantial.
Moreover, in [74] the converted (from CDR0 to CDR0.5) subjects
were analyzed separately, which we do not consider such conversion
in our analysis. Also, metric distance results agree with the volume
comparisons of [29], hence volume
(i.e., scale) might be highly dominating the morphometric changes in
the hippocampi. In other words, the significant volume reduction in
left and right hippocampi might dominate the change in shape, when
morphometry is measured by metric distances. To remove the size
influence so as to measure the shapes only, one can perform scaling
on the hippocampi and then apply LDDMM to normalize the size
differences.

Differences and changes (over time) in morphometry can also be used
for diagnostic discrimination of subjects in non-demented or
demented groups. Many discrimination techniques such as Fisher's
linear discriminant functions, support vector machines, and logistic
discrimination can be applied to the metric distances, together with
other qualitative variables. In this study we applied logistic
discrimination based on metric distances, as logistic regression not
only provides a means for classification, but also yields a
probability estimate for having DAT. Furthermore, one can optimize
the threshold probability for a particular cost function for the
entire training data set, or by a cross-validation technique. The
correct classification rate of the hippocampi was about 70\% in
our logistic regression analysis.
In [57] PCA of the initial momentum of the same data set led to correct classification
of 12 out of 18 (i.e., 67\% of the) demented subjects and 22 out
of 26 (i.e., 85\% of the) control subjects.
Metric distances can be used to distinguish AD from normal aging quantitatively; however,
to be able to use it for diagnostic purposes, the method should be
improved to a greater extent.

We perform a principal component analysis on metric distances
and hippocampus, brain, and intracranial volumes.
Considering the variable loadings, we conclude that
volumes are mostly measures of size and partly related to shape,
while the metric distance is mostly a measure of shape and partly related to size.

We also compare the cross-sectional, longitudinal, and
discrimination results of LDDMM distances with those of volumes.
We observe that cross-sectional and longitudinal analysis give similar results,
although metric distances increase and volumes decrease by time.
The metric distance, being an extremely condensed summary
measure give very similar results as the hippocampal volume.
That is, neither volume nor metric distance discriminated left baseline
(LB), right baseline (RB), or left followup (LF) between CDR0 and CDR0.5;
volume reduction and metric distances differences are both
significant for CDR0.5 subjects,
but neither of them are significant for CDR0 subjects;
and ANOVA suggested a significant diagnosis
group-by-timepoint interaction for both measures.
On the other hand, we obtain better classification results with using volumes compared
to metric distances.
When volume and LDDMM distances are used together, the classification results improve compared to results
based on volume or distance only.
Furthermore, the differential volume and distance changes are measured by annual percentage rate
of change (APC) for the two year period in the study.
Similar to the results of [71],
we found that APC in volumes may be a good indicator for early stage of DAT.
However, APC in LDDMM distances do not provide a good performance in
classification of CDR0 versus CCDR0.5 hippocampi.
Comparing the discrimination results, we found that the classifier based on volume, distance,
and APC in volume has the best performance.
Hence these measures may constitute a reliable biomarker when used together.

In conclusion, we have presented detailed statistical analysis of
metric distances computed with LDDMM and show that this is
potentially a powerful tool in detecting morphometric changes
between diagnosis groups or changes in morphometry over time.
Furthermore, we avoid the single subject analysis, which might be of
greater interest clinically. Metric distances depend on the choice
of template anatomy used. However, in this article we do not address
the issue of template selection for optimal differentiation between
hippocampus morphometry.


\section*{References}
1. Davis, D.G., et al., {\it Alzheimer neuropathologic alterations
in aged cognitively normal subjects.} Journal of Neuropathology and
Experimental Neurology, 1999. {\bf 58}(4): p. 376-388.

2. Haroutunian, V., et al., {\it Regional distribution of neuritic
plaques in the nondemented elderly and subjects with very mild
Alzheimer Disease.} Archives of Neurology, 1998. {\bf 55}(9): p.
1185-1191.

3. Thompson, P.M., et al., {\it Cortical change in Alzheimer's
disease detected with a disease-specific population-based brain
atlas.} Cerebral Cortex, 2001. {\bf 11}(1): p. 1-16.

4. Troncoso, J.C., et al., {\it Neuropathology in controls and
demented subjects from the Baltimore longitudinal study of aging.}
Neurobiology of Aging, 1996. {\bf 17}(3): p. 365-371.

5. Braak, H. and E. Braak, {\it Staging of Alzheimer's
disease-related neurofibrillary changes.} Neurobiol Aging, 1995.
{\bf 16}(3): p. 271-8; discussion 278-84.

6. Braak, H. and E. Braak, {\it Staging of Alzheimer-related
cortical destruction.} Int Psychogeriatr, 1997. {\bf 9 Suppl 1}: p.
257-61; discussion 269-72.

7. Braak, H., E. Braak, and J. Bohl, {\it Staging of
Alzheimer-related cortical destruction.} Eur Neurol, 1993. {\bf
33}(6): p. 403-8.

8. Price, J.L., et al., {\it Neuron number in the entorhinal cortex
and CA1 in preclinical alzheimer disease.} Archives of Neurology,
2001. {\bf 58}(9): p. 1395-1402.

9. Convit, A., et al., {\it Hippocampal atrophy in early Alzheimer's
disease: anatomic specificity and validation.} Psychiatric
Quarterly, 1993. {\bf 64}(4): p. 371-387.

10. Csernansky, J.G., et al., {\it Early DAT is distinguished from
aging by high-dimensional mapping of the hippocampus.} Neurology,
2000. {\bf 55}(11): p. 1636-1643.

11. Krasuski, J.S., et al., {\it Relation of medial temporal lobe
volumes to age and memory function in nondemented adults with Down's
syndrome: Implications for the prodromal phase of Alzheimer's
disease.} American Journal of Psychiatry, 2002. {\bf 159}(1): p.
74-81.

12. Mega, M.S., et al., {\it Hippocampal atrophy in persons with
age-associated memory impairment: Volumetry within a common space.}
Psychosomatic Medicine, 2002. {\bf 64}(3): p. 487-492.

13. Mu, Q., et al., {\it A quantitative MR study of the hippocampal
formation, the amygdala, and the temporal horn of the lateral
ventricle in healthy subjects 40 to 90 years of age.} American
Journal of Neuroradiology, 1999. {\bf 20}(2): p. 207-211.

14. Scheltens, P. and F. Barkhof, {\it Structural neuroimaging
outcomes in clinical dementia trials, with special reference to
disease modifying designs.} Journal of Nutrition, Health and Aging,
2006. {\bf 10}(2): p. 123-128.

15. Gosche, K.M., et al., {\it Hippocampal volume as an index of
Alzheimer neuropathology: Findings from the Nun study.} Neurology,
2002. {\bf 58}(10): p. 1476-1482.

16. Fox, N.C., et al., {\it Imaging of onset and progression of
Alzheimer's disease with voxel-compression mapping of serial
magnetic resonance images.} The Lancet, 2001. {\bf 358}(9277): p.
201-205.

17. Chan, D., et al., {\it Rates of global and regional cerebral
atrophy in AD and frontotemporal dementia.} Neurology, 2001. {\bf
57}(10): p. 1756-1763.

18. Christensen, G.E., R.D. Rabbitt, and M.I. Miller, {\it
Deformable templates using large deformation kinematics.} IEEE
Transactions on Image Processing, 1996. {\bf 5}(10): p. 1435-1447.

19. Miller, M.I., et al., {\it Mathematical textbook of deformable
neuroanatomies.} Proceedings of the National Academy of Sciences of
the United States of America, 1993. {\bf 90}(24): p. 11944-11948.

20. Hogan, R.E., et al., {\it MRI-based high-dimensional hippocampal
mapping in mesial temporal lobe epilepsy.} Brain, 2004. {\bf
127}(8): p. 1731-1740.

21. Miller, M.I., {\it Computational anatomy: Shape, growth, and
atrophy comparison via diffeomorphisms.} NeuroImage, 2004. {\bf 23}:
p. S19-S33.

22. Thompson, P.M., et al., {\it Mapping cortical change in
Alzheimer's disease, brain development, and schizophrenia.}
NeuroImage, 2004. {\bf 23}: p. S2-S18.

23. Grenander, U. and M.I. Miller, {\it Computational anatomy: An
emerging discipline.} Quarterly of Applied Mathematics, 1998. {\bf
56}(4): p. 617-694.

24. Toga, A.W. and P.M. Thompson, {\it Brain atlases of normal and
diseased populations.} Int Rev Neurobiol., 2005. {\bf 66}: p. 1-54.

25. Toga, A.W., {\it Computational biology for visualization of
brain structure.} Anatomy and Embryology, 2005. {\bf 210}(5-6): p.
433-438.

26. Grenander, U., {\it General Pattern Theory}. 1993, Oxford:
Clarendon Press.

27. Grenander, U. and M.I. Miller, {\it Representations of knowledge
in complex systems.} J. R. Statist. Soc. B, 1994. {\bf 56}(3): p.
549-603.

28. Wang, L., et al., {\it Statistical analysis of hippocampal
asymmetry in schizophrenia.} NeuroImage, 2001. {\bf 14}(3): p.
531-545.

29. Wang, L., et al., {\it Changes in hippocampal volume and shape
across time distinguish dementia of the Alzheimer type from healthy
aging.} NeuroImage, 2003. {\bf 20}(2): p. 667-682.

30. Fox, N.C. and P.A. Freeborough, {\it Brain atrophy progression
measured from registered serial MRI: Validation and application to
Alzheimer's disease.} Journal of Magnetic Resonance Imaging, 1997.
{\bf 7}(6): p. 1069-1075.

31. Fox, N.C., P.A. Freeborough, and M.N. Rossor, {\it Visualisation
and quantification of rates of atrophy in Alzheimer's disease.}
Lancet, 1996. {\bf 348}(9020): p. 94-97.

32. Killiany, R.J., et al., {\it MRI measures of entorhinal cortex
vs hippocampus in preclinical AD.} Neurology, 2002. {\bf 58}(8): p.
1188-1196.

33. Wang, D., et al., {\it MR image-based measurement of rates of
change in volumes of brain structures. Part II: Application to a
study of Alzheimer's disease and normal aging.} Magnetic Resonance
Imaging, 2002. {\bf 20}(1): p. 41-48.

34. Yamaguchi, S., et al., {\it Five-year retrospective changes in
hippocampal atrophy and cognitive screening test performances in
very mild Alzheimer's disease: The Tajiri project.} Neuroradiology,
2002. {\bf 44}(1): p. 43-48.

35. Crum, W.R., R.I. Scahill, and N.C. Fox, {\it Automated
hippocampal segmentation by regional fluid registration of serial
MRI: Validation and application in Alzheimer's disease.} NeuroImage,
2001. {\bf 13}(5): p. 847-855.

36. Leow, A.D., et al., {\it Longitudinal stability of MRI for
mapping brain change using tensor-based morphometry.} NeuroImage,
2006. {\bf 31}(2): p. 627-640.

37. Apostolova, L.G., et al., {\it Conversion of mild cognitive
impairment to alzheimer disease predicted by hippocampal atrophy
maps.} Archives of Neurology, 2006. {\bf 63}(5): p. 693-699.

38. Mungas, D., et al., {\it Longitudinal volumetric MRI change and
rate of cognitive decline.} Neurology, 2005. {\bf 65}(4): p.
565-571.

39. Dickerson, B.C. and R.A. Sperling, {\it Neuroimaging biomarkers
for clinical trials of disease-modifying therapies in Alzheimer's
disease.} NeuroRx, 2005. {\bf 2}(2): p. 348-360.

40. Ewers, M., S.J. Teipel, and H. Hampel, {\it Update of structural
MRI-based methods for the early detection of Alzheimer's disease}
{\it [Aktuelle entwicklungen der strukturellen MRT zur
fruhdiagnostik der Alzheimer-demenz].} Nervenheilkunde, 2005. {\bf
24}(2): p. 113-119.

41. Barnes, J., et al., {\it Does Alzheimer's disease affect
hippocampal asymmetry? Evidence from a cross-sectional and
longitudinal volumetric MRI study.} Dementia and Geriatric Cognitive Disorders, 2005. {\bf 19}(5-6): p. 338-344.

42. Beg, M.F., et al., {\it Computing large deformation metric
mappings via geodesic flows of diffeomorphisms.} International
Journal of Computer Vision, 2005. {\bf 61}(2): p. 139-157.

43. Miller, M.I., A. Trouve, and L. Younes, {\it On the metrics and
Euler-Lagrange equations of computational anatomy.} Annual Review of
Biomedical Engineering, 2002. {\bf 4}: p. 375-405.

44. Miller, M.I., et al., {\it Collaborative Computational Anatomy:
The Perfect Storm for MRI Morphometry Study of the Human Brain via
Diffeomorphic Metric Mapping, Multidimensional Scaling and Linear
Discriminant Analysis.} 2006.

45. Wang, L., et al., {\it Abnormalities of hippocampal surface
structure in very mild dementia of the Alzheimer type.} NeuroImage,
2006. {\bf 30}(1): p. 52-60.

46. Younes, L., {\it Jacobi fields in groups of diffeomorphisms and
applications.} Quart. Appl. Math., 2007. {\bf 65}: p. 113-134.

47. Qiu, A., et al., {\it Parallel transport in diffeomorphisms
distinguishes the time-dependent pattern of hippocampal surface
deformation due to healthy aging and the dementia of the Alzheimer's
type.} NeuroImage, 2008. {\bf 40}(1): p. 68-76.

48. Morris, J.C., {\it The Clinical Dementia Rating (CDR): Current
version and scoring rules.} Neurology, 1993. {\bf 43}(11): p.
2412-2414.

49. Cohen, J., {\it A coefficient for agreement for nominal scales.}
Educational and Psychological Measurement, 1960. {\bf 20}: p. 37-46.

50. Burke, W.J., et al., {\it Reliability of the Washington
University Clinical Dementia Rating.} Archives of Neurology, 1988.
{\bf 45}(1): p. 31-32.

51. Morris, J.C., et al., {\it Clinical dementia rating training and
reliability in multicenter studies: The Alzheimer's Disease
Cooperative Study experience.} Neurology, 1997. {\bf 48}(6): p.
1508-1510.

52. Berg, L., et al., {\it Clinicopathologic studies in cognitively
healthy aging and Alzheimer disease: Relation of histologic markers
to dementia severity, age, sex, and apolipoprotein E genotype.}
Archives of Neurology, 1998. {\bf 55}(3): p. 326-335.

53. Morris, J.C., et al., {\it Cerebral amyloid deposition and
diffuse plaques in ``normal" aging: Evidence for presymptomatic and
very mild Alzheimer's disease.} Neurology, 1996. {\bf 46}(3): p.
707-719.

54. Price, J.L. and J.C. Morris, {\it Tangles and plaques in
nondemented aging and `preclinical' Alzheimer's disease.} Annals of
Neurology, 1999. {\bf 45}(3): p. 358-368.

55. Petersen, R.C., et al., {\it Current concepts in mild cognitive
impairment.} Archives of Neurology, 2001. {\bf 58}(12): p.
1985-1992.

56. Storandt, M., et al., {\it Longitudinal course and
neuropathologic outcomes in original vs revised MCI and in pre-MCI.}
Neurology, 2006. {\bf 67}(3): p. 467-73.

57. Wang, L., et al., {\it Large Deformation Diffeomorphism and
Momentum Based Hippocampal Shape Discrimination in Dementia of the
Alzheimer Type.} IEEE Trans. Medical Imaging, 2007. {\bf 26}: p.
462-470.

58. Haller, J.W., et al., {\it Three-dimensional hippocampal MR
morphometry with high-dimensional transformation of a neuroanatomic
atlas.} Radiology, 1997. {\bf 202}(2): p. 504-510.

59. Dupuis, P., U. Grenander, and M.I. Miller, {\it Variational
problems on flows of diffeomorphisms for image matching.} Quarterly
of Applied Mathematics, 1998. {\bf 56}(3): p. 587-600.

60. Trouve, A., {\it Diffeomorphisms groups and pattern matching in
image analysis.} International Journal of Computer Vision, 1998.
{\bf 28}(3): p. 213-221.

61. Box, G.E.P., G.M. Jenkins, and R. G.C., {\it Time Series
Analysis: Forecasting and Control}. 3rd ed. 1994: Holden-Day.

62. Littel, R.C., et al., {\it SAS Systems for Mixed Models}. 1996:
SAS Institute.

63. Venables, W.N. and B.D. Ripley, {\it Modern Applied Statistics
with S-PLUS}. 2nd ed. 1997: Springer-Verlag.

64. Burnham, K.P. and D. Anderson, {\it Model Selection and
Multi-Model Inference}. 2003, New York: Springer.

65. Thode Jr, H.C., {\it Testing for Normality}. 2002, New York:
Marcel Dekker. 368.

66. Seber, G.A.F. and A.J. Lee,{\it  Linear Regression Analysis}.
Wiley Series in Probability and Statistics. 2003, New York: Wiley
{\&} Sons.

67. Conover, W.J., {\it Practical Nonparametric Statistics}. 3rd ed.
1999, New York: John Wiley {\&} Sons.

68. Zar, J.H., {\it Biostatistical Analysis}. 1984, New Jersey:
Prentice Hall. 718.

69. Dalgaard, P., {\it Introductory Statistics with R}. 2002:
Springer-Verlag.

70. Johnson, D.E., {\it Applied Multivariate Methods for Data
Analysis}. 1998, California: Duxbury Press.

71. Du, A.T., et al., {\it Higher atrophy rate of entorhinal cortex
than hippocampus in AD.} Neurology, 2004. {\bf 62}(3): p. 422-427.

72. Baringhaus, L. and C. Franz, {\it On a new multivariate
two-sample test.} J. Multivariate Analysis, 2004. {\bf 88}: p.
190-206.

73. Du, A.T., et al., {\it Atrophy rates of entorhinal cortex in AD
and normal aging.} Neurology, 2003. {\bf 60}(3): p. 481-6.

74. Csernansky, J.G., et al., {\it Preclinical detection of
Alzheimer's disease: Hippocampal shape and volume predict dementia
onset in the elderly.} NeuroImage, 2005. {\bf 25}(3): p. 783-792.

75. Csernansky, J.G., et al., {\it Hippocampal morphometry in
schizophrenia by high dimensional brain mapping.} Proceedings of the
National Academy of Sciences of the United States of America, 1998.
{\bf 95}(19): p. 11406-11411.

76. Mardia, K. V., J. T. Kent and J. M. Bibby, {\it Multivariate Analysis.}
1979, London: Academic Press.

\section*{Tables and Figures}

\begin{table}[htbp]
\centering
\begin{tabular}{|c|c|c|c|c|}
\hline
\multicolumn{5}{|c|}{I- Summary Information of Subjects}  \\
\hline
 &Gender (M/F) & Age (years)    & Scan interval       & Education (years) \\
 &Gender (M/F) &(mean $\pm$ SD) & (years) ([Min-Max]) & (mean $\pm$ SD) \\
\hline
CDR0 & 12/14 & 75.2$\pm$7.0 & 2.2 [1.4-4.1] & 14.8$\pm$ 2.7 \\
\hline
CDR0.5 & 11/7 & 75.7$\pm$4.4 & 2.0 [1.0-2.6] & 13.7$\pm$ 2.8 \\
\hline
overall & 23/21 & 75.4$\pm$6.1 & 2.1 [1.0-4.1] & 14.3$\pm$ 2.8 \\
\hline
$p_L $ & NA & 0.4224 & NA & 0.0001 \\
\hline
$p_W $ & NA & 0.8202 & NA & 0.2101 \\
\hline
\end{tabular}
\\
\begin{tabular}{|c|c|c|c|c|c|c|}
\hline
\multicolumn{7}{|c|}{II- Summary Statistics for Metric Distances at Baseline and Follow-up}  \\
\hline & Mean $\pm$ SD & Min & $Q_1$ & Median & $Q_3 $ &
Max \\
\hline Left-$d_k^b $ (LDB) & 3.40 $\pm$ 0.68 & 1.97 & 3.00 & 3.30 &
3.65 &
5.08 \\
\hline Left-$d_k^f $ (LDF) & 3.57 $\pm$ 0.77 & 2.26 & 2.99 & 3.48 &
4.02 &
6.03 \\
\hline Right-$d_k^b $(RDB) & 3.65 $\pm$ 0.67 & 1.73 & 3.32 & 3.53 &
4.09 &
4.98 \\
\hline Right-$d_k^f $ (RDF) & 4.05 $\pm$ 0.67 & 2.96 & 3.72 & 3.95 &
4.34 &
5.71 \\
\hline
\end{tabular}
\\
\begin{tabular}{|c|c|c||c|c|}
\hline
\multicolumn{5}{|c|}{III- Mean $\pm$ SD Values of Brain and Intracranial Volumes}  \\
\hline
 & BV1 & BV3 & ICV1 & ICV3 \\
\hline
CDR0 & 1006892$\pm$ 104214.0 & 1003319.4$\pm$ 101129.0 & 1407972 $\pm$ 156067.1 & 1464494$\pm$ 177496.0 \\
\hline
CDR0.5 & 1003850$\pm$ 92293.4 & 993380.8 $\pm$ 95425.0 & 1408507$\pm$ 134912.6 & 1454966$\pm$ 138931.2 \\
\hline
overall & 1005647$\pm$ 98408.2 & 999253.6 $\pm$ 97828.6 & 1408191$\pm$ 146140.3 & 1460596$\pm$ 161152.7 \\
\hline
$p_L $ & 0.2302 & 0.0079 & 0.0503 & $_{0.1070}$ \\
\hline
$p_W $ & 0.5192 & 0.3277 & 0.7929 & 0.8299 \\
\hline
\end{tabular}
\\
\begin{tabular}{|c|c|c||c|c|}
\hline
\multicolumn{5}{|c|}{IV- Mean $\pm$ SD of Hippocampal Volumes}  \\
\hline
 & LB & LF & RB & RF \\
\hline
CDR0 & 2081.4 $\pm$ 354.8 & 2081.4 $\pm$ 354.8 & 2081.4 $\pm$ 354.8 & 2081.4 $\pm$ 354.8 \\
\hline
CDR0.5 & 1717.6 $\pm$ 224.8 & 1717.6 $\pm$ 224.8 & 1717.6 $\pm$ 224.8 & 1717.6 $\pm$ 224.8 \\
\hline
overall & 1932.6 $\pm$ 354.8 & 1932.6 $\pm$ 354.8 & 1932.6 $\pm$ 354.8 & 1932.6 $\pm$ 354.8 \\
\hline
$p_L $ & 0.3528 & 0.0268 & 0.2001 & 0.2359 \\
\hline
$p_W $ & 0.0003 & $<\quad 0.0001$ & 0.0149 & 0.0004 \\
\hline
\end{tabular}
\\
\begin{tabular}{|c|c|c||c|c|}
\hline
\multicolumn{5}{|c|}{V- Mean $\pm$ SD of Metric Distances}  \\
\hline
 & LB & LF & RB & RF \\
\hline
CDR0 & 3.34 $\pm$ 0.62 & 3.41 $\pm$ 0.54 & 3.63 $\pm$ 0.57 & 3.83 $\pm$ 0.47 \\
\hline
CDR0.5 & 3.48 $\pm$ 0.76 & 3.82 $\pm$ 0.98 & 3.68 $\pm$ 0.81 & 4.37 $\pm$ 0.78 \\
\hline
overall & 3.40 $\pm$ 0.68 & 3.57 $\pm$ 0.77 & 3.65 $\pm$ 0.67 & 4.05 $\pm$ 0.67 \\
\hline
$p_L $ & 0.0498 & 0.4718 & 0.2891 & $_{0.1084}$ \\
\hline
$p_W $ & 0.5994 & 0.1590 & 0.9145 & $_{0.02058}$ \\
\hline
\end{tabular}
\caption{
\label{tab:summary-stat-dist}
Summary information of subjects (I);
summary statistics for metric distances at baseline and follow-up, where SD stands for
standard deviation, $Q_1$ and $Q_3 $ stand for the first and third quartiles (II);
means and SDs of brain and intracranial volumes by diagnosis group (III);
means and SDs of hippocampal volumes by diagnosis group (IV);
and means and SDs of metric distances by diagnosis group (V).
$p_L $: p-value based on Lilliefor's test of normality,
$p_W $: p-value based on Wilcoxon rank sum test.
NA: not applicable;
BV1 (BV3): brain volume at baseline (followup);
ICV1(ICV3): intracranial volume at baseline (followup);
LB: left baseline;
LF: left followup;
RB: right baseline;
and RF: right followup.}
\end{table}

\begin{table}[htbp]
\centering
\begin{tabular}{|c|c|c|c|c|}
\hline
\multicolumn{5}{|c|}{Importance of Components}  \\
\hline
 & PC1 & PC2 & PC3 & PC4 \\
\hline
Prop. Var & .9877 & .0123 & $\sim 0.0$ & $\sim 0.0$ \\
\hline
Cum. Prop & .9877 & $\sim 1.0$ & $\sim 1.0$ & 1.0 \\
\hline
\multicolumn{5}{|c|}{Variable Loadings}  \\
\hline
 & PC1 & PC2 & PC3 & PC4 \\
\hline
HLV1 & $\sim 0.0$ & $\sim 0.0$ & 1.00 & $\sim 0.0$ \\
\hline
HLM1 & $\sim 0.0$ & $\sim 0.0$ & $\sim 0.0$ & 1.00 \\
\hline
BV1 & .55 & -.83 & $\sim 0.0$ & $\sim 0.0$ \\
\hline
ICV1 & .83 & .55 & $\sim 0.0$ & $\sim 0.0$ \\
\hline
\end{tabular}
\caption{
\label{tab:PCA-left-covariance}
The importance of principal components and variable loadings
from the principal component analysis of metric distances and volumes of
left hippocampi at baseline with eigenvalues based on the covariance matrix.
PCi stands for principal component $i$ for $i=1,2,3,4$;
Prop.Var: proportion of variance explained by the principal components;
Cum.Prop: cumulative proportion of the variance explained by the particular principal component;
HLV1: volume of left hippocampus at baseline;
HLM1: metric distance of left hippocampus at baseline;
BV1: brain volume at baseline;
ICV1: intracranial volume at baseline.
}
\end{table}

\begin{table}[htbp]
\centering
\begin{tabular}{|c|c|c|c|c|c|c|c|c|}
\hline
\multicolumn{5}{|c|}{Baseline} & \multicolumn{4}{|c|}{Followup}  \\
\hline
\multicolumn{5}{|c|}{Importance of Components} & \multicolumn{4}{|c|}{Importance of Components}  \\
\hline
 & PC1 & PC2 & PC3 & PC4 & PC1 & PC2 & PC3 & PC4 \\
\hline
Prop. Var & .57 & .27 & .15 & .01 & .54 & .32 & .12 & .02 \\
\hline
Cum. Prop. & .57 & .84 & .99 & 1.0 & .54 & .86 & .98 & 1.0 \\
\hline
\multicolumn{5}{|c|}{Variable Loadings} & \multicolumn{4}{|c|}{Variable Loadings}  \\
\hline
 & PC1 & PC2 & PC3 & PC4 & PC1 & PC2 & PC3 & PC4 \\
\hline
HLV & .37 & .59 & .71 & $\sim 0.0$ & .41 & .55 & .73 & $\sim 0.0$ \\
\hline
HLM & .22 & -.80 & .55 & $\sim 0.0$ & $\sim 0.0$ & -.80 & .60 & $\sim 0.0$ \\
\hline
BV & .64 & $\sim 0.0$ & -.27 & -.72 & .65 & -.16 & -.18 & -.72 \\
\hline
ICV & .63 & $\sim 0.0$ & -.33 & .70 & .64 & -.17 & -.29 & .69 \\
\hline
\end{tabular}
\caption{
\label{tab:PCA-left-correlation}
The importance of principal components and variable loadings
from the principal component analysis of metric distances and volumes of
left hippocampi at baseline and followup with eigenvalues based on the
correlation matrix.
HLV: volume of left hippocampus;
HLM: metric distance of left hippocampus;
BV: brain volume;
ICV: intracranial volume.
The other abbreviations are as in Table \ref{tab:PCA-left-covariance}.}
\end{table}

\begin{table}[htbp]
\centering
\begin{tabular}{|c|c|c|c|c|c|c|c|c|}
\hline
\multicolumn{5}{|c|}{Baseline} & \multicolumn{4}{|c|}{Followup}  \\
\hline
\multicolumn{5}{|c|}{Importance of Components} & \multicolumn{4}{|c|}{Importance of Components}  \\
\hline
 & PC1 & PC2 & PC3 & PC4 & PC1 & PC2 & PC3 & PC4 \\
\hline
Prop. Var & .54 & .33 & .12 & .01 & .57 & .35 & .07 & .02 \\
\hline
Cum. Prop. & .54 & .87 & .99 & 1.0 & .57 & .92 & .98 & 1.0 \\
\hline
\multicolumn{5}{|c|}{Variable Loadings} & \multicolumn{4}{|c|}{Variable Loadings}  \\
\hline
 & PC1 & PC2 & PC3 & PC4 & PC1 & PC2 & PC3 & PC4 \\
\hline
HRV & .35 & .61 & .71 & $\sim 0.0$ & .50 & .46 & .72 & .13 \\
\hline
HRM & $\sim 0.0$ & -.78 & .62 & $\sim 0.0$ & -.30 & -.70 & .64 & $\sim 0.0$ \\
\hline
BV & .66 & $\sim 0.0$ & -.20 & -.71 & .58 & -.37 & $\sim 0.0$ & -.72 \\
\hline
ICV & .66 & -.13 & -.26 & .70 & .57 & -.40 &-.25 & .67 \\
\hline
\end{tabular}
\caption{
\label{tab:PCA-right-correlation}
The importance of principal components and variable loadings
from the principal component analysis of metric distances and volumes of
right hippocampi at baseline and followup with eigenvalues based on the
correlation matrix.
HRV: volume of right hippocampus;
HRM: metric distance of right hippocampus;
BV: brain volume;
ICV: intracranial volume.
The other abbreviations are as in Table \ref{tab:PCA-left-covariance}.}
\end{table}

\begin{table}[htbp]
\centering
\begin{tabular}{|c|c|}
\hline
Compound Symmetry & Unstructured \\
\par $\left[ {{\begin{array}{*{20}c}
 {\sigma^2} \hfill & \hfill & \hfill & \hfill \\
 {\sigma_1} \hfill & {\sigma^2} \hfill & \hfill & \hfill \\
 {\sigma_1} \hfill & {\sigma_1} \hfill & {\sigma^2} \hfill & \hfill \\
 {\sigma_1} \hfill & {\sigma_1} \hfill & {\sigma_1} \hfill & {\sigma^2} \hfill \\
\end{array}}} \right]$
 &
\par $\left[ {{\begin{array}{*{20}c}
 {\sigma_1^2} \hfill & \hfill & \hfill & \hfill \\
 {\sigma_{21}} \hfill & {\sigma_2^2} \hfill & \hfill & \hfill \\
 {\sigma_{31}} \hfill & {\sigma_{32}} \hfill & {\sigma_3^2} \hfill & \hfill \\
 {\sigma_{41}} \hfill & {\sigma_{42}} \hfill & {\sigma_{43}} \hfill & {\sigma_4^2} \hfill \\
\end{array}}} \right]$ \\
\hline
Autoregressive & Autoregressive Hetero-\\
 & geneous Variances\\
\par $\left[ {{\begin{array}{*{20}c}
 {\sigma^2} \hfill & \hfill & \hfill & \hfill \\
 {\sigma \rho} \hfill & {\sigma^2} \hfill & \hfill & \hfill \\
 {\sigma \rho^2} \hfill & {\sigma \rho} \hfill & {\sigma^2} \hfill & \hfill \\
 {\sigma \rho ^3} \hfill & {\sigma \rho^2} \hfill & {\sigma \rho} \hfill & {\sigma^2} \hfill \\
\end{array}}} \right]$
 &
\par $\left[{{\begin{array}{*{20}c}
 {\sigma_1^2} \hfill & \hfill & \hfill & \hfill \\
 {\sigma \rho} \hfill & {\sigma_2^2} \hfill & \hfill & \hfill \\
 {\sigma \rho^2} \hfill & {\sigma \rho} \hfill & {\sigma_3^2} \hfill & \hfill \\
 {\sigma \rho ^3} \hfill & {\sigma \rho^2} \hfill & {\sigma \rho} \hfill & {\sigma_4^2} \hfill \\
\end{array}}} \right]$ \\
\hline
\end{tabular}
\caption{
\label{tab:var-cov-for-RepeatedMeasuresANOVA}
The Var-Cov structures for the repeated measures ANOVA
analysis on the metric distances; $\sigma^2$ is the common variance
term, $\sigma_i^2$ is the variance for repeated factor $i$,
$\sigma_{ij}$ is the covariance between repeated factors $i$ and $j$, and
$\rho $ is the correlation coefficient for first order in an
autoregressive model.
}
\end{table}

\begin{table}[htbp]
\centering
\begin{tabular}{|c|c|c|c|c|c|c|c|c|}
\hline Model & Var-Cov & $df$ & AIC & BIC & Log Likelihood & Test & L.Ratio & $p$-value \\
\hline 1 & CS & 10 & 362.8 & 394.1 & -171.4 & --- & --- & --- \\
\hline 2 & UN & 18 & 352.9 & 409.1 & -158.4 & 1 vs 2 & 25.9 & 0.0011 \\
\hline 3 & ARH & 13 & 350.9 & 391.5 & -162.5 & 1 vs 3 & 17.92 & $< 0.0001$ \\
\hline 4 & AR & 10 & 347.1 & 378.4 & -163.6 & --- & --- & --- \\
\hline
\end{tabular}
\caption{
\label{tab:model-selection-criteria}
Model selection criteria results for models with compound
symmetry (CS), unstructured (US), autoregressive (AR), and
autoregressive heterogeneous (ARH) Var-Cov structures.
$df$ = error degree of freedom, AIC = Akaike information criteria, BIC = Bayesian
information criteria, L.Ratio = likelihood ratio.
}
\end{table}

\begin{table}[htbp]
\centering
\begin{tabular}{|c|c|c|c|c|c|c|}
\hline
\multicolumn{7}{|c|}{Independent Group Comparisons of LDDMM Distances}  \\
\hline & \multicolumn{3}{|c|}{$p$-values for $t$-test} &
\multicolumn{3}{|c|}{$p$-values for Wilcoxon test}  \\
\hline Groups & 2-sided & $1^{st}<2^{nd}$ & $1^{st}>2^{nd}$ & 2-sided &
$1^{st}<2^{nd}$ &
$1^{st}>2^{nd}$ \\
\hline LB-CDR0.5,LB-CDR0 & .5362 & .7319 & .2681 & .6078 & .7044 &
.3039 \\
\hline LF-CDR0.5,LF-CDR0 & .1179 & .9410 & .0590 & .1625 & .9223 &
.0813 \\
\hline RB-CDR0.5,RB-CDR0 & .8176 & .5912 & .4088 & .9239 & .5475 &
.462 \\
\hline RF-CDR0.5,RF-CDR0 & .0148* & .9926 & .0074* & .0212* & .99 &
.0106* \\
\hline
\multicolumn{7}{|c|}{Dependent Group Comparisons of LDDMM Distances}  \\
\hline & \multicolumn{3}{|c|}{$p$-values for paired $t$-test}
 &
\multicolumn{3}{|c|}{$p$-values for paired Wilcoxon test}  \\
\hline Groups & 2-sided & $1^{st}<2^{nd}$ & $1^{st}>2^{nd}$ & 2-sided &
$1^{st}<2^{nd}$ &
$1^{st}>2^{nd}$ \\
\hline LB-CDR0.5,LF-CDR0.5 & .0259* & .0129* & .9871 & .0311* & .0155* &
.9861 \\
\hline RB-CDR0.5,RF-CDR0.5 & .0002* & .0001* & .9999 & .0005* & .0002* &
.9998 \\
\hline LB-CDR0,LF-CDR0 & .5958 & .2979 & .7021 & .7127 & .3563 &
.6531 \\
\hline RB-CDR0,RF-CDR0 & .1241 & .0621 & .9379 & .1244 & .0622 &
.9409 \\
\hline
\end{tabular}
\caption{
\label{tab:comparison-lddmm-distance}
The $p$-values based on independent sample t-test (top) and
Wilcoxon rank sum test (middle) for both left and right metric
distances and $p$-values based on paired t-tests for both left and
right metric distances (bottom). Notice that we use t-tests when the
assumptions (such as normality or homogeneity of variances hold),
otherwise we use Wilcoxon test. Significant $p$-values at 0.05 level
are marked with an asterisk (*).
}
\end{table}

\begin{table}[htbp]
\centering
\begin{tabular}{|c|c|c|c|}
\hline
\multicolumn{4}{|c|}{Correlation Coefficients of Baseline vs Follow-up Distances}\\
\hline
Groups & $r_P$ & $\rho_S$ & $\tau_K$ \\
\hline
LDB,LDF & .6642 ($<$.0001*) & .5686 ($<$.0001*) & .3843 (.0001*) \\
\hline
RDB,RDF & .5076 (.0002*) & .3835 (.0053*) & .2754 (.0042*) \\
\hline
LB-CDR0.5,LF-CDR0.5 & .7988 ($<$.0001*) & .8147 ($<$.0001*) & .6164 (.0002*) \\
\hline
RB-CDR0,RF-CDR0.5 & .6995 (.0006*) & .7028 (.0008*) & .5556 (.0004*) \\
\hline
LB-CDR0,LF-CDR0 & .4929 (.0053*) & .3455 (.0420*) & .2102 (.0661) \\
\hline
RB-CDR0,RF-CDR0 & .2812 (.0820) & .1888 (.1767) & .1418 (.1549) \\
\hline
\end{tabular}
\caption{
\label{tab:correl-baseline-followup}
The correlation coefficients and the associated $p$-values
for the one-sided (correlation greater than zero) alternatives.
$r_P$ = Pearson's correlation coefficient, $\rho_S$ = Spearman's rank
correlation coefficient, and $\tau_K$ = Kendall's rank correlation
coefficient. Significant $p$-values at 0.05 level are marked with an
asterisk (*).
}
\end{table}

\begin{table}[htbp]
\centering
\begin{tabular}{|c|c|c|c|}
\hline
\multicolumn{4}{|c|}{Correlation Coefficients of Distances of Left vs Right Hippocampi}  \\
\hline
Groups & $r_P$ & $\rho_S$ & $\tau_K$ \\
\hline
LDB,RDB & .4017 (.0034*) & .27 (.0382*) & .1749 (.0471*) \\
\hline
LDF,RDF & .3441 (.0111*) & .2009 (.0952) & .1241 (.1175) \\
\hline
LB-CDR0.5,RB-CDR0 & .3312 (.0492*) & .3813 (.0277*) & .2191 (.0582) \\
\hline
LF-CDR0.5,RF-CDR0.5 & .1033 (.3078) & .0400 (.4225) & .0340 (.4039) \\
\hline
LB-CDR0,RB-CDR0 & .3312 (.0492*) & .3813 (.0277*) & .2191 (.0582) \\
\hline
LF-CDR0,RF-CDR0 & .1033 (.3078) & .0400 (.4225) & .0340 (.4039) \\
\hline
\end{tabular}
\caption{
\label{tab:correl-left-right}
The correlation coefficients and the associated $p$-values
for the one-sided (correlation greater than zero) alternatives.
$r_P$, $\rho_S$, and $\tau_K$ stand for Pearson's, Spearman's, and
Kendall's correlation coefficients, respectively.
Significant $p$-values at 0.05 level are marked with an asterisk (*).
}
\end{table}

\begin{table}[htbp]
\centering
\begin{tabular}{|c|c|c|c||c|c|}
\hline &
\multicolumn{5}{|c|}{$p$-values for cdf comparisons of Distances}  \\
\hline
Groups & $p_{KS}$ (2-s) & $p_{KS}$ (l) & $p_{KS}$ (g) & $p_C$ & $p_{CvM}$ \\
\hline
LB-CDR0.5,LB-CDR0 & .6932 & .364 & .5706 & .4325 & .5112 \\
\hline
RB-CDR0,RB-CDR0 & .8997 & .5204 & .5875 & .6684 & .7098 \\
\hline
LF-CDR0.5,LF-CDR0 & .1208 & .0604 & .5706 & .0495* & .0665 \\
\hline
RF-CDR0.5,RF-CDR0 & .0517* & .0259* & .9365 & .0095* & .0235* \\
\hline
\end{tabular}
\caption{
\label{tab:cdf-compare-distance}
The $p$-values for the K-S, Cram\'{e}r's, and
Cram\'{e}r-von Mises tests. $p_{KS}$ (2-s), $p_{KS}$ (l),
and $p_{KS}$ (g) stand for the $p$-values based on K-S test for the
two sided alternative, first cdf less than the second, and first cdf
greater than the second alternatives, respectively, $p_C$, and
$p_{CvM}$ stands for the $p$-values for Cram\'{e}r's test and
Cram\'{e}r-von Mises test, respectively. Significant $p$-values at
0.05 level are marked with an asterisk (*).
}
\end{table}

\begin{table}[htbp]
\centering
\begin{tabular}{|c|c|}
\hline

\begin{tabular}{|c|c|c|c|c|}
\hline A &
\multicolumn{4}{|c|}{Truth}  \\
\hline
\raisebox{-4.50ex}[0cm][0cm]{Predict} & & CDR0 & CDR0.5 & Total \\
\cline{2-5}
 & CDR0 & 95 & 52 & 147 \\
\cline{2-5}
 & CDR0.5 & 9 & 20 & 29 \\
\cline{2-5}
 & Total & 104 & 72 & 176 \\
\hline
\end{tabular}
 &
\begin{tabular}{|c|c|c|c|c|}
\hline B &
\multicolumn{4}{|c|}{Truth}  \\
\hline \raisebox{-4.50ex}[0cm][0cm]{Predict} & & CDR0 & CDR0.5 &
Total \\
\cline{2-5}
 & CDR0 & 18 & 8 & 26 \\
\cline{2-5}
 & CDR0.5 & 8 & 10 & 18 \\
\cline{2-5}
 & Total & 26 & 18 & 44 \\
\hline
\end{tabular}
 \\
\hline
\begin{tabular}{|c|c|c|c|c|}
\hline C &
\multicolumn{4}{|c|}{Truth}  \\
\hline \raisebox{-4.50ex}[0cm][0cm]{Predict} & & CDR0 & CDR0.5 &
Total \\
\cline{2-5}
 & CDR0 & 22 & 8 & 30 \\
\cline{2-5}
 & CDR0.5 & 4 & 10 & 14 \\
\cline{2-5}
 & Total & 26 & 18 & 44 \\
\hline
\end{tabular}
 &
\begin{tabular}{|c|c|c|c|c|}
\hline D &
\multicolumn{4}{|c|}{Truth}  \\
\hline \raisebox{-4.50ex}[0cm][0cm]{Predict} & & CDR0 & CDR0.5 &
Total \\
\cline{2-5}
 & CDR0 & 22 & 10 & 32 \\
\cline{2-5}
 & CDR0.5 & 4 & 8 & 12 \\
\cline{2-5}
 & Total & 26 & 18 & 44 \\
\hline
\end{tabular}
 \\
\hline
\end{tabular}
\caption{
\label{tab:classification-matrices}
The classification matrices using metric distances in
logistic regression with threshold p = 0.5: A = classification matrix
of all hippocampi using logistic model $M_{II}(D)$ using
hippocampal LDDMM metric distances in Equation (9); B =
classification matrix of subjects using logistic model $M_{II}(D)$
in Equation (9) with one hippocampus MRI classified as CDR0.5 being
sufficient to label CDR0.5; C = classification matrix of subjects
using logistic model $M_{III}(D)$ in Equation (10) that only uses
follow-up hippocampus MRIs and one hippocampus sufficient to label
CDR0.5; D = classification matrix of subjects using logistic model
$M_{IV}(D)$ in Equation (11) that only uses follow-up right
hippocampus MRIs.
}
\end{table}

\begin{table}[htbp]
\centering
\begin{tabular}{|c|c|c|c|c|}
\hline
& A & B & C$^*$ & D \\
\hline
$P_{CCR}$ & 65\% & 64\% & 73\% & 68\% \\
\hline
$P_{sens}$ & 28\% & 56\% & 56\% & 44\% \\
\hline
$P_{spec}$ & 91\% & 69\% & 85\% & 85\% \\
\hline
\end{tabular}
\caption{
\label{tab:class-rates-for-matrices}
The correct classification rates ($P_{CCR}$),
sensitivity ($P_{sens}$), and specificity ($P_{spec}$) percentages
with $p_o = 0.50$ for the classification procedures A-D in
Table \ref{tab:classification-matrices}.
The model with the best classification performance is marked with an asterisk ($^*$).
}
\end{table}

\begin{table}[htbp]
\centering
\begin{tabular}{|c|c|c|c|c|c|c|c|c|}
\hline
& \multicolumn{4}{|c|}{$p_o = 1/2$} & \multicolumn{4}{|c|}{$p_o = 18/44$}  \\
\hline
& $M_I(D)$ & $M_{II}(D)$ & $M_{III}(D)^*$ & $M_{IV}(D)$ & $M_I(D)$ & $M_{II}(D)$ & $M_{III}(D)$ & $M_{IV}(D)^*$ \\
\hline
$P_{CCR}$ & 66\% & 64\% & 73\% & 68\% & 57\% & 47\% & 66\% & 68\% \\
\hline
$P_{sens}$ & 56\% & 56\% & 56\% & 44\% & 83\% & 67\% & 67\% & 61\% \\
\hline
$P_{spec}$ & 73\% & 69\% & 85\% & 85\% & 38\% & 35\% & 65\% & 73\% \\
\hline
\multicolumn{9}{|c|}{Using optimum $p_o$ based on cost function $C_1(p_o ,w_1 ,w_2$) with}  \\
\hline
& \multicolumn{4}{|c|}{$w_1 = w_2 = 1$} & \multicolumn{4}{|c|}{$w_1 = 1,\;w_2 = 3$}  \\
\hline
& $M_I(D)$ & $M_{II}(D)$ & $M_{III}(D)$ & $M_{IV}(D)^*$ & $M_I(D)$ & $M_{II}(D)$ & $M_{III}(D)$ & $M_{IV}(D)^*$ \\
\hline
$p_{opt}$ & .51 & .50 & .45 & .38 & .51 & .47 & .36,.37 & .38 \\
\hline
$P_{CCR}$ & 68\% & 64\% & 70\% & 70\% & 68\% & 57\% & 68\% & 70\% \\
\hline
$P_{sens}$ & 56\% & 56\% & 67\% & 72\% & 56\% & 61\% & 78\% & 72\% \\
\hline
$P_{spec}$ & 77\% & 69\% & 73\% & 69\% & 77\% & 54\% & 61\% & 69\% \\
\hline
\multicolumn{9}{|c|}{Using optimum $p_o$ based on cost function $C_2(p_o$) with}  \\
\hline
& \multicolumn{4}{|c|}{$\eta_1 = \eta_2 = 0.5$} & \multicolumn{4}{|c|}{$\eta_1 = .3,\;\eta_2 = 0.7$}  \\
\hline
& $M_I(D)$ & $M_{II}(D)$ & $M_{III}(D)$ & $M_{IV}(D)^*$ & $M_I(D)$ & $M_{II}(D)$ & $M_{III}(D)^*$ & $M_{IV}(D)$ \\
\hline
$p_{opt}$ & .81-.82 & .76-.78 & .50-.52 & .38 & .37 & .37 & .33-.34 & .22-.29 \\
\hline
$P_{CCR}$ & 75\% & 73\% & 73\% & 70\% & 59\% & 61\% & 66\% & 55\% \\
\hline
$P_{sens}$ & 39\% & 39\% & 56\% & 72\% & 95\% & 100\% & 89\% & 89\% \\
\hline
$P_{spec}$ & 100\% & 96\% & 85\% & 69\% & 35\% & 35\% & 50\% & 31\% \\
\hline
\end{tabular}
\caption{
\label{tab:class-rates-dist}
The correct classification rates ($P_{CCR}$),
sensitivity ($P_{sens}$), and specificity ($P_{spec}$) percentages
for the classification procedures based on models
$M_I(D)-M_{IV}(D)$ using hippocampal LDDMM metrics and volumes with threshold
probabilities $p_o = 1/2$ and $p_o = 18/44$ (top); with optimum
threshold values $p_o = p_{opt}$ based on the cost function
$C_1(p_o,w_1 ,w_2$) with $w_1 = w_2 = 1$ and $w_1 = 1,\;w_2 = 3$ (middle); and
with optimum threshold values $p_o = p_{opt}$ based on the cost
function $C_2(p_o ,\eta_1 ,\eta_2$) with $\eta_1 = \eta_2 = 0.5$
and $\eta_1 = .3,\;\eta_2 = 0.7$ (bottom).
The model with the best classification performance is marked with an asterisk ($^*$).
}
\end{table}


\begin{table}[htbp]
\centering
\begin{tabular}{|c|c|c|c|c|c|c|}
\hline
Volume comparisons & \multicolumn{3}{|c|}{$p$-values for $t$-test} & \multicolumn{3}{|c|}{$p$-values for Wilcoxon test}  \\
\hline
Groups & 2-sided & $1^{st}<2^{nd}$ & $1^{st}>2^{nd}$ & 2-sided & $1^{st}<2^{nd}$ & $1^{st}>2^{nd}$ \\
\hline
LB-CDR0.5,LB-CDR0 & $.0002\ast$ & $.0001\ast$ & $.9999$ & $.0002\ast$ & $.0001\ast$ & $.9999$ \\
\hline
LF-CDR0.5,LF-CDR0 & $<.0001\ast$ & $<.0001\ast$ & $\approx 1.000$ & $<.0001\ast$ & $<.0001\ast$ & $\approx 1.000$ \\
\hline
RB-CDR0.5,RB-CDR0 & $.0025\ast$ & $.0012\ast$ & $.9998$ & $.0143\ast$ & $.0071\ast$ & $.9933$ \\
\hline
RF-CDR0.5,RF-CDR0 & $.0001\ast$ & $<.0001\ast$ & $\approx 1.000$ & $.0002\ast$ & $.0001\ast$ & $.9999$ \\
\hline
& \multicolumn{3}{|c|}{$p$-values for paired $t$-test} & \multicolumn{3}{|c|}{paired Wilcoxon test}  \\
\hline
Groups & 2-sided & $1^{st}<2^{nd}$ & $1^{st}>2^{nd}$ & 2-sided & $1^{st}<2^{nd}$ & $1^{st}>2^{nd}$ \\
\hline
LB-CDR0.5,LF-CDR0.5 & $<.0001\ast$ & $\approx 1.000$ & $<.0001\ast$ & $.0001\ast$ & $\approx 1.000$ & $<.0001\ast $ \\
\hline
RB-CDR0.5,RF-CDR0.5 & $<.0001\ast$ & $\approx 1.000$ & $<.0001\ast$ & $<.0001\ast$ & $\approx 1.000$ & $<.0001\ast $ \\
\hline
LB-CDR0,LF-CDR0 & $<.0001\ast$ & $\approx 1.000$ & $<.0001\ast$ & $.0001\ast$ & $.9999$ & $<.0001\ast $ \\
\hline
RB-CDR0,RF-CDR0 & $<.0001\ast$ & $.9999$ & $.0001\ast$ & $.0002\ast$ & $.9999$ & $.0001\ast $ \\
\hline
LB-CDR0.5,RB-CDR0.5 & $<.0001\ast$ & $<.0001\ast$ & $\approx 1.000$ & $<.0001\ast$ & $<.0001\ast$ & $\approx 1.000$ \\
\hline
LF-CDR0.5,RF-CDR0.5 & $<.0001\ast$ & $<.0001\ast$ & $\approx 1.000$ & $.0001\ast$ & $.0001\ast$ & $\approx 1.000$ \\
\hline
LB-CDR0,RB-CDR0 & $<.0001\ast$ & $<.0001\ast$ & $\approx 1.000$ & $<.0001\ast$ & $<.0001\ast$ & $\approx 1.000$ \\
\hline
LF-CDR0,RF-CDR0 & $<.0001\ast$ & $<.0001\ast$ & $\approx 1.000$ & $<.0001\ast$ & $<.0001\ast$ & $\approx 1.000$ \\
\hline
\end{tabular}
\caption{
\label{tab:comparison-volume}
The $p$-values based on independent sample t-test (top) and
Wilcoxon rank sum test (middle) for both left and right hippocampus
volumes and $p$-values based on paired t-tests for both left and right
hippocampus volumes (bottom). Significant $p$-values at 0.05 level are
marked with an asterisk (*).
}
\end{table}

\begin{table}[htbp]
\centering
\begin{tabular}{|c|c|c|c|c|c|c|c|c|}
\hline
& \multicolumn{4}{|c|}{$p_o = 1/2$} & \multicolumn{4}{|c|}{$p_o = 18/44$}  \\
\hline
& $M_I(V)$ & $M_{II}(V)$ & $M_{III}(V)^*$ & $M_{IV}(V)$ & $M_I(V)$ & $M_{II}(V)$ & $M_{III}(V)$ & $M_{IV}(V)^*$ \\
\hline
$P_{CCR}$ & 68\% & 70\% & 73\% & 80\% & 70\% & 68\% & 64\% & 78\% \\
\hline
$P_{sens}$ & 83\% & 89\% & 83\% & 72\% & 89\% & 89\% & 100\% & 78\% \\
\hline
$P_{spec}$ & 58\% & 58\% & 65\% & 85\% & 58\% & 54\% & 38\% & 77\% \\
\hline
\multicolumn{9}{|c|}{Using optimum $p_o$ based on cost function $C_1(p_o ,w_1 ,w_2$) with}  \\
\hline
& \multicolumn{4}{|c|}{$w_1 = w_2 = 1$} & \multicolumn{4}{|c|}{$w_1 = 1,\;w_2 = 3$}  \\
\hline
& $M_I(V)$ & $M_{II}(V)$ & $M_{III}(V)$ & $M_{IV}(V)^*$ & $M_I(V)$ & $M_{II}(V)$ & $M_{III}(V)$ & $M_{IV}(V)^*$ \\
\hline
$p_{opt}$ & .64-.66 & .62-.63 & .55-.58 & .35-.42 & .61-.62 & .58-.60 & .55-.58 & .35-.36 \\
\hline
$P_{CCR}$ & 82\% & 75\% & 77\% & 80\% & 80\% & 73\% & 77\% & 80\% \\
\hline
$P_{sens}$ & 78\% & 78\% & 83\% & 89\% & 83\% & 83\% & 83\% & 89\% \\
\hline
$P_{spec}$ & 85\% & 73\% & 73\% & 73\% & 77\% & 65\% & 73\% & 73\% \\
\hline
\multicolumn{9}{|c|}{Using optimum $p_o$ based on cost function $C_2(p_o ,\eta_1 ,\eta_2$) with}  \\
\hline
& \multicolumn{4}{|c|}{$\eta_1 = \eta_2 = 0.5$} & \multicolumn{4}{|c|}{$\eta_1 = .3,\;\eta_2 = 0.7$}  \\
\hline
& $M_I(V)$ & $M_{II}(V)$ & $M_{III}(V)$ & $M_{IV}(V)^*$ & $M_I(V)$ & $M_{II}(V)$ & $M_{III}(V)$ & $M_{IV}(V)^*$ \\
\hline
$p_{opt}$ & .64-.66 & .70 & .69 & .35-.36 & .31 & .31 & .25-.32 & .26 \\
\hline
$P_{CCR}$ & 82\% & 80\% & 82\% & 80\% & 70\% & 66\% & 70\% & 77\% \\
\hline
$P_{sens}$ & 78\% & 61\% & 61\% & 89\% & 100\% & 100\% & 94\% & 94\% \\
\hline
$P_{spec}$ & 85\% & 92\% & 96\% & 73\% & 50\% & 42\% & 54\% & 65\% \\
\hline
\end{tabular}
\caption{
\label{tab:class-rates-vol}
The correct classification rates ($P_{CCR}$),
sensitivity ($P_{sens}$), and specificity ($P_{spec}$) percentages
for the classification procedures based on models
$M_I(V)-M_{IV}(V)$ using hippocampal LDDMM metrics and volumes with threshold
probabilities $p_o = 1/2$ and $p_o = 18/44$ (top); with optimum
threshold values $p_o = p_{opt}$ based on the cost function
$C_1(p_o,w_1 ,w_2$) with $w_1 = w_2 = 1$ and $w_1 = 1,\;w_2 = 3$ (middle); and
with optimum threshold values $p_o = p_{opt}$ based on the cost
function $C_2(p_o ,\eta_1 ,\eta_2$) with $\eta_1 = \eta_2 = 0.5$
and $\eta_1 = .3,\;\eta_2 = 0.7$ (bottom).
The model with the best classification performance is marked with an asterisk ($^*$).
}
\end{table}

\begin{table}[htbp]
\centering
\begin{tabular}{|c|c|c|c|c|c|c|c|c|}
\hline
& \multicolumn{4}{|c|}{$p_o = 1/2$} & \multicolumn{4}{|c|}{$p_o = 18/44$}  \\
\hline
& $M_I(V,D)$ & $M_{II}(V,D)$ & $M_{III}(V,D)$ & $M_{IV}(V,D)^*$ &
  $M_I(V,D)$ & $M_{II}(V,D)$ & $M_{III}(V,D)$ & $M_{IV}(V,D)^*$ \\
\hline
$P_{CCR}$ & 75\% & 66\% & 75\% & 82\% & 66\% & 66\% & 73\% & 77\% \\
\hline
$P_{sens}$ & 89\% & 89\% & 83\% & 78\% & 89\% & 89\% & 89\% & 83\% \\
\hline
$P_{spec}$ & 65\% & 50\% & 69\% & 85\% & 50\% & 50\% & 62\% & 73\% \\
\hline
\multicolumn{9}{|c|}{Using optimum $p_o$ based on cost function $C_1(p_o ,w_1 ,w_2$) with}  \\
\hline
& \multicolumn{4}{|c|}{$w_1 = w_2 = 1$} & \multicolumn{4}{|c|}{$w_1 = 1,\;w_2 = 3$}  \\
\hline
& $M_I(V,D)^*$ & $M_{II}(V,D)$ & $M_{III}(V,D)$ & $M_{IV}(V,D)$ &
  $M_I(V,D)^*$ & $M_{II}(V,D)$ & $M_{III}(V,D)$ & $M_{IV}(V,D)$ \\
\hline
$p_{opt}$ & .64-.65 & .66 & .48-.58 & .48 & .64-.65 & .66 & .48-.54 & .28-.33 \\
\hline
$P_{CCR}$ & 84\% & 84\% & 75\% & 84\% & 84\% & 84\% & 75\% & 80\% \\
\hline
$P_{sens}$ & 89\% & 83\% & 83\% & 83\% & 89\% & 83\% & 83\% & 89\% \\
\hline
$P_{spec}$ & 81\% & 85\% & 69\% & 85\% & 81\% & 85\% & 69\% & 73\% \\
\hline
\multicolumn{9}{|c|}{Using optimum $p_o$ based on cost function $C_2(p_o ,\eta_1 ,\eta_2$) with}  \\
\hline
& \multicolumn{4}{|c|}{$\eta_1 = \eta_2 = 0.5$} & \multicolumn{4}{|c|}{$\eta_1 = .3,\;\eta_2 = 0.7$}  \\
\hline
& $M_I(V,D)^*$ & $M_{II}(V,D)$ & $M_{III}(V,D)$ & $M_{IV}(V,D)$ & $M_I(V,D)^*$ & $M_{II}(V,D)$ & $M_{III}(V,D)$ & $M_{IV}(V,D)$ \\
\hline
$p_{opt}$ & .64-.65 & .66 & .68-.72 & .48 & .64-.65 & .66 & .23 & .20-.22 \\
\hline
$P_{CCR}$ & 84\% & 84\% & 80\% & 84\% & 84\% & 84\% & 70\% & 75\% \\
\hline
$P_{sens}$ & 89\% & 83\% & 61\% & 83\% & 89\% & 83\% & 100\% & 94\% \\
\hline
$P_{spec}$ & 81\% & 85\% & 92\% & 85\% & 81\% & 85\% & 50\% & 61\% \\
\hline
\end{tabular}
\caption{
\label{tab:class-rates-Vol-Dist}
The correct classification rates ($P_{CCR}$),
sensitivity ($P_{sens}$), and specificity ($P_{spec}$) percentages
for the classification procedures based on models
$M_I(V,D)-M_{IV}(V,D)$ using hippocampal LDDMM metrics and volumes with threshold
probabilities $p_o = 1/2$ and $p_o = 18/44$ (top); with optimum
threshold values $p_o = p_{opt}$ based on the cost function
$C_1(p_o,w_1 ,w_2$) with $w_1 = w_2 = 1$ and $w_1 = 1,\;w_2 = 3$ (middle); and
with optimum threshold values $p_o = p_{opt}$ based on the cost
function $C_2(p_o ,\eta_1 ,\eta_2$) with $\eta_1 = \eta_2 = 0.5$
and $\eta_1 = .3,\;\eta_2 = 0.7$ (bottom).
The model with the best classification performance is marked with an asterisk ($^*$).
}
\end{table}

\begin{table}[htbp]
\centering
\begin{tabular}{|c|c|c|c|c|c|c|}
\hline
& \multicolumn{3}{|c|}{$p_o = 1/2$} & \multicolumn{3}{|c|}{$p_o = 18/44$}  \\
\hline
& $M_I \left( {V^{APC}} \right)^*$ & $M_{II} \left( {V^{APC}}\right)$ & $M_{III} \left( {V^{APC}} \right)$ & $M_I \left( {V^{APC}}
\right)$ & $M_{II} \left( {V^{APC}} \right)$ & $M_{III} \left( {V^{APC}} \right)$ \\
\hline
$P_{CCR}$ & 75\% & 80\% & 75\% & 64\% & 73\% & 73\% \\
\hline
$P_{sens}$ & 72\% & 61\% & 61\% & 83\% & 61\% & 61\% \\
\hline
$P_{spec}$ & 77\% & 92\% & 85\% & 50\% & 81\% & 81\% \\
\hline
\multicolumn{7}{|c|}{Using optimum $p_o$ based on cost function $C_1(p_o ,w_1 ,w_2$) with}  \\
\hline
& \multicolumn{3}{|c|}{$w_1 = w_2 = 1$} & \multicolumn{3}{|c|}{$w_1 = 1,\;w_2 = 3$}  \\
\hline
& $M_I \left( {V^{APC}} \right)^*$ & $M_{II} \left( {V^{APC}}\right)$ & $M_{III} \left( {V^{APC}} \right)$ & $M_I \left( {V^{APC}}
\right)^*$ & $M_{II} \left( {V^{APC}} \right)$ & $M_{III} \left( {V^{APC}} \right)$ \\
\hline
$p_{opt}$ & .55-.55 & .38-.39 & .27-.28 & .54-.55 & .34 & .25 \\
\hline
$P_{CCR}$ & 80\% & 75\% & 75\% & 80\% & 73\% & 73\% \\
\hline
$P_{sens}$ & 72\% & 72\% & 78\% & 72\% & 83\% & 83\% \\
\hline
$P_{spec}$ & 85\% & 81\% & 73\% & 85\% & 65\% & 65\% \\
\hline
\multicolumn{7}{|c|}{Using optimum $p_o$ based on cost function $C_2(p_o ,\eta_1 ,\eta_2$) with}  \\
\hline
& \multicolumn{3}{|c|}{$\eta_1 = \eta_2 = 0.5$} & \multicolumn{3}{|c|}{$\eta_1 = .3,\;\eta_2 = 0.7$}  \\
\hline
& $M_I \left( {V^{APC}} \right)^*$ & $M_{II} \left( {V^{APC}} \right)$ & $M_{III} \left( {V^{APC}} \right)$ & $M_I \left( {V^{APC}}
\right)$ & $M_{II} \left( {V^{APC}} \right)$ & $M_{III} \left( {V^{APC}} \right)$ \\
\hline
$p_{opt}$ & .54-.55 & .82-.85 & .61-.69 & .39 & .34 & .18-.21 \\
\hline
$P_{CCR}$ & 80\% & 82\% & 82\% & 64\% & 73\% & 70\% \\
\hline
$P_{sens}$ & 72\% & 56\% & 56\% & 89\% & 83\% & 100\% \\
\hline
$P_{spec}$ & 85\% & 100\% & 100\% & 46\% & 65\% & 50\% \\
\hline
\end{tabular}
\caption{
\label{tab:class-rates-APCinVolumes}
The correct classification rates ($P_{CCR}$),
sensitivity ($P_{sens}$), and specificity ($P_{spec}$) percentages
for the classification procedures based on models
$M_I \left({V^{APC}} \right)-M_{III} \left( {V^{APC}} \right)$ using APC in
hippocampal volumes with threshold probabilities $p_o = 1/2$ and
$p_o = 18/44$ (top); with optimum threshold values $p_o = p_{opt}$ based on
the cost function $C_1(p_o ,w_1 ,w_2$) with $w_1 = w_2 = 1$ and
$w_1 = 1,\;w_2 = 3$ (middle); and with optimum threshold values
$p_o = p_{opt}$ based on the cost function $C_2(p_o ,\eta_1 ,\eta_2$)
with $\eta_1 = \eta_2 = 0.5$ and $\eta_1 = .3,\;\eta_2 = 0.7$
(bottom). The model with the best classification performance is marked with an asterisk ($^*$).
}
\end{table}

\newpage
\begin{table}[htbp]
\centering
\begin{tabular}{|c|c|c|c|c|c|c|}
\hline
& \multicolumn{3}{|c|}{$p_o = 1/2$} & \multicolumn{3}{|c|}{$p_o = 18/44$}  \\
\hline
& $M_I \left( {D^{APC}} \right)$ & $M_{II} \left( {D^{APC}} \right)$ & $M_{III} \left( {D^{APC}} \right)$ & $M_I \left( {D^{APC}}
\right)$ & $M_{II} \left( {D^{APC}} \right)^*$ & $M_{III} \left( {D^{APC}} \right)$ \\
\hline
$P_{CCR}$ & 59\% & 61\% & 61\% & 57\% & 77\% & 64\% \\
\hline
$P_{sens}$ & 27\% & 28\% & 28\% & 72\% & 72\% & 56\% \\
\hline
$P_{spec}$ & 81\% & 85\% & 85\% & 46\% & 42\% & 69\% \\
\hline
\multicolumn{7}{|c|}{Using optimum $p_o$ based on cost function $C_1(p_o ,w_1 ,w_2$) with}  \\
\hline
& \multicolumn{3}{|c|}{$w_1 = w_2 = 1$} & \multicolumn{3}{|c|}{$w_1 = 1,\;w_2 = 3$}  \\
\hline
& $M_I \left( {D^{APC}} \right)^*$ & $M_{II} \left( {D^{APC}} \right)^*$ & $M_{III} \left( {D^{APC}} \right)$ & $M_I \left( {D^{APC}}
\right)$ & $M_{II} \left( {D^{APC}} \right)$ & $M_{III} \left( {D^{APC}} \right)$ \\
\hline
$p_{opt}$ & .45 & .45-.46 & .41 & .42 & .45-.46 & .36 \\
\hline
$P_{CCR}$ & 66\% & 66\% & 66\% & 61\% & 66\% & 64\% \\
\hline
$P_{sens}$ & 61\% & 61\% & 56\% & 67\% & 61\% & 78\% \\
\hline
$P_{spec}$ & 69\% & 69\% & 73\% & 58\% & 69\% & 54\% \\
\hline
\multicolumn{7}{|c|}{Using optimum $p_o$ based on cost function $C_2(p_o ,\eta_1 ,\eta_2$) with}  \\
\hline
& \multicolumn{3}{|c|}{$\eta_1 = \eta_2 = 0.5$} & \multicolumn{3}{|c|}{$\eta_1 = .3,\;\eta_2 = 0.7$}  \\
\hline
& $M_I \left( {D^{APC}} \right)$ & $M_{II} \left( {D^{APC}} \right)$ & $M_{III} \left( {D^{APC}} \right)^*$ & $M_I \left( {D^{APC}}
\right)$ & $M_{II} \left( {D^{APC}} \right)$ & $M_{III} \left( {D^{APC}} \right)$ \\
\hline
$p_{opt}$ & .45 & .45-.46 & .32 & .37 & .35-.36 & .29 \\
\hline
$P_{CCR}$ & 66\% & 66\% & 64\% & 52\% & 55\% & 59\% \\
\hline
$P_{sens}$ & 61\% & 61\% & 89\% & 100\% & 100\% & 100\% \\
\hline
$P_{spec}$ & 69\% & 69\% & 46\% & 19\% & 23\% & 31\% \\
\hline
\end{tabular}
\caption{
\label{tab:class-rates-APCinDistances}
The correct classification rates ($P_{CCR}$),
sensitivity ($P_{sens}$), and specificity ($P_{spec}$) percentages
for the classification procedures based on models $M_I \left(
{D^{APC}} \right)-M_{III} \left( {D^{APC}} \right)$ using APC in
hippocampal LDDMM metric distances with threshold probabilities $p_o
 = 1/2$ and $p_o = 18/44$ (top); with optimum threshold values $p_o
 = p_{opt}$ based on the cost function $C_1(p_o ,w_1 ,w_2$) with $w_1
 = w_2 = 1$ and $w_1 = 1,\;w_2 = 3$ (middle); and with optimum threshold
values $p_o = p_{opt}$ based on the cost function $C_2(p_o ,\eta_1
,\eta_2$) with $\eta_1 = \eta_2 = 0.5$ and $\eta_1 = .3,\;\eta_2
 = 0.7$ (bottom). The model with the best classification performance is marked with an asterisk ($^*$).
}
\end{table}


\begin{sidewaystable}[htbp]
\centering
\begin{tabular}{|c|c|c|c|c|c|c|}
\hline
& \multicolumn{3}{|c|}{$p_o =1/2$} & \multicolumn{3}{|c|}{$p_o =18/44$}  \\
\hline
& $M_I \left( V^{APC},D \right)$ & $M_{II} \left( V^{APC},D \right)$ & $M_{III} \left( V^{APC},D \right)$&
$M_I \left( V^{APC},D \right)$ & $M_{II} \left( V^{APC},D \right)$ & $M_{III} \left( V^{APC},D \right)^*$ \\
\hline
$P_{CCR} $ & 73\% & 77\% & 77\% & 68\% & 75\% & 80\% \\
\hline
$P_{sens} $ & 83\% & 67\% & 56\% & 83\% & 72\% & 87\% \\
\hline
$P_{spec} $ & 65\% & 85\% & 92\% & 58\% & 77\% & 85\% \\
\hline
\multicolumn{7}{|c|}{Using optimum $p_o $ based on cost function $C_1 (p_o ,w_1 ,w_2 )$with}  \\
\hline
 & \multicolumn{3}{|c|}{$w_1 =w_2 =1$}  & \multicolumn{3}{|c|}{$w_1 =1,\;w_2 =3$}  \\
\hline
 & $M_I \left( V^{APC},D \right)$ & $M_{II} \left( V^{APC},D \right)$ & $M_{III} \left( V^{APC},D \right)^*$&
$M_I \left( V^{APC},D \right)^*$ & $M_{II} \left( V^{APC},D \right)$ & $M_{III} \left( V^{APC},D \right)$ \\
\hline
$p_{opt} $ & .61 & .53-.73 & .35-.40 & .56 & .35-.36 & .31-.32 \\
\hline
$P_{CCR} $ & 84\% & 82\% & 80\% & 80\% & 70\% & 75\% \\
\hline
$P_{sens} $ & 72\% & 67\% & 78\% & 78\% & 78\% & 83\% \\
\hline
$P_{spec} $ & 92\% & 92\% & 81\% & 81\% & 65\% & 69\% \\
\hline
\multicolumn{7}{|c|}{Using optimum $p_o $ based on cost function $C_2 (p_o ,\eta _1 ,\eta _2 )$ with}  \\
\hline
 & \multicolumn{3}{|c|}{$\eta _1 =\eta _2 =.5$}  & \multicolumn{3}{|c|}{$\eta _1 =.3,\;\eta _2 =.7$}  \\
\hline
 & $M_I \left( V^{APC},D \right)$ & $M_{II} \left( V^{APC},D \right)$ & $M_{III} \left( V^{APC},D \right)$&
$M_I \left( V^{APC},D \right)^*$ & $M_{II} \left( V^{APC},D \right)$ & $M_{III} \left( V^{APC},D \right)$ \\
\hline
$p_{opt} $ & .61 & .53-.73 & .49 & .56 & .23-.25 & .31-.32 \\
\hline
$P_{CCR} $ & 84\% & 82\% & 82\% & 80\% & 55\% & 75\% \\
\hline
$P_{sens} $ & 72\% & 67\% & 67\% & 78\% & 100\% & 83\% \\
\hline
$P_{spec} $ & 92\% & 92\% & 92\% & 81\% & 23\% & 69\% \\
\hline
\end{tabular}
\caption{
\label{tab:class-rates-APCinVolume-Distance}
The correct classification rates ($P_{CCR}$),
sensitivity ($P_{sens}$), and specificity ($P_{spec}$) percentages
for the classification procedures based on models
$M_I \left( V^{APC},D \right)-M_{III} \left( V^{APC},D \right)$ using metric distance,
and APC in hippocampal volumes with threshold probabilities
$p_o = 1/2$ and $p_o = 18/44$ (top); with optimum threshold values
$p_o = p_{opt}$ based on the cost function $C_1(p_o ,w_1 ,w_2$) with $w_1
 = w_2 = 1$ and $w_1 = 1,\;w_2 = 3$ (middle); and with optimum threshold
values $p_o = p_{opt}$ based on the cost function
$C_2(p_o ,\eta_1 ,\eta_2$) with $\eta_1 = \eta_2 = 0.5$ and $\eta_1 = .3,\;\eta_2  = 0.7$ (bottom).
The model with the best classification performance is marked with an asterisk ($^*$).
}
\end{sidewaystable}

\begin{sidewaystable}[htbp]
\centering
\begin{tabular}{|c|c|c|c|c|c|c|}
\hline
& \multicolumn{3}{|c|}{$p_o =1/2$} & \multicolumn{3}{|c|}{$p_o =18/44$}  \\
\hline
& $M_I \left( V, V^{APC},D \right)$ & $M_{II} \left( V, V^{APC},D \right)^*$ & $M_{III} \left( V, V^{APC},D \right)^*$&
$M_I \left( V, V^{APC},D \right)$ & $M_{II} \left( V, V^{APC},D \right)$ & $M_{III} \left( V, V^{APC},D \right)^*$ \\
\hline
$P_{CCR} $ & 86\% & 89\% & 89\% & 80\% & 80\% & 91\% \\
\hline
$P_{sens} $ & 94\% & 89\% & 89\% & 94\% & 94\% & 94\% \\
\hline
$P_{spec} $ & 81\% & 88\% & 88\% & 69\% & 69\% & 88\% \\
\hline
\multicolumn{7}{|c|}{Using optimum $p_o $ based on cost function $C_1 (p_o ,w_1 ,w_2 )$with}  \\
\hline
 & \multicolumn{3}{|c|}{$w_1 =w_2 =1$}  & \multicolumn{3}{|c|}{$w_1 =1,\;w_2 =3$}  \\
\hline
 & $M_I \left( V, V^{APC},D \right)$ & $M_{II} \left( V, V^{APC},D \right)$ & $M_{III} \left( V, V^{APC},D \right)^*$&
$M_I \left( V, V^{APC},D \right)$ & $M_{II} \left( V, V^{APC},D \right)$ & $M_{III} \left( V, V^{APC},D \right)^*$ \\
\hline
$p_{opt} $ & .56-.57 & .48-.51 & .39-.42 & .56-.57 & .48-.51 & .39-.42 \\
\hline
$P_{CCR} $ & 89\% & 89\% & 91\% & 89\% & 89\% & 91\% \\
\hline
$P_{sens} $ & 94\% & 89\% & 94\% & 94\% & 89\% & 94\% \\
\hline
$P_{spec} $ & 85\% & 88\% & 88\% & 85\% & 88\% & 88\% \\
\hline
\multicolumn{7}{|c|}{Using optimum $p_o $ based on cost function $C_2 (p_o ,\eta _1 ,\eta _2 )$ with}  \\
\hline
 & \multicolumn{3}{|c|}{$\eta _1 =\eta _2 =.5$}  & \multicolumn{3}{|c|}{$\eta _1 =.3,\;\eta _2 =.7$}  \\
\hline
 & $M_I \left( V, V^{APC},D \right)$ & $M_{II} \left( V, V^{APC},D \right)$ & $M_{III} \left( V, V^{APC},D \right)^*$&
$M_I \left( V, V^{APC},D \right)$ & $M_{II} \left( V, V^{APC},D \right)$ & $M_{III} \left( V, V^{APC},D \right)^*$ \\
\hline
$p_{opt} $ & .56-.57 & .48-.51 & .39-.42 & .56-.57 & .48-.51 & .39-.42 \\
\hline
$P_{CCR} $ & 89\% & 89\% & 91\% & 89\% & 89\% & 91\% \\
\hline
$P_{sens} $ & 94\% & 89\% & 94\% & 94\% & 89\% & 94\% \\
\hline
$P_{spec} $ & 85\% & 88\% & 88\% & 85\% & 88\% & 88\% \\
\hline
\end{tabular}
\caption{
\label{tab:class-rates-Volume-Distance-APCinVolume}
The correct classification rates ($P_{CCR}$),
sensitivity ($P_{sens}$), and specificity ($P_{spec}$) percentages
for the classification procedures based on models
$M_I \left( V, V^{APC},D \right)-M_{III} \left( V, V^{APC},D \right)$ using metric distance,
and APC in hippocampal volumes with threshold probabilities
$p_o = 1/2$ and $p_o = 18/44$ (top); with optimum threshold values
$p_o = p_{opt}$ based on the cost function $C_1(p_o ,w_1 ,w_2$) with $w_1
 = w_2 = 1$ and $w_1 = 1,\;w_2 = 3$ (middle); and with optimum threshold
values $p_o = p_{opt}$ based on the cost function
$C_2(p_o ,\eta_1 ,\eta_2$) with $\eta_1 = \eta_2 = 0.5$ and $\eta_1 = .3,\;\eta_2  = 0.7$ (bottom).
The model with the best classification performance is marked with an asterisk ($^*$).
}
\end{sidewaystable}


\begin{figure}[htbp]
\centering
\rotatebox{-90}{ \resizebox{2.5 in}{!}{\includegraphics{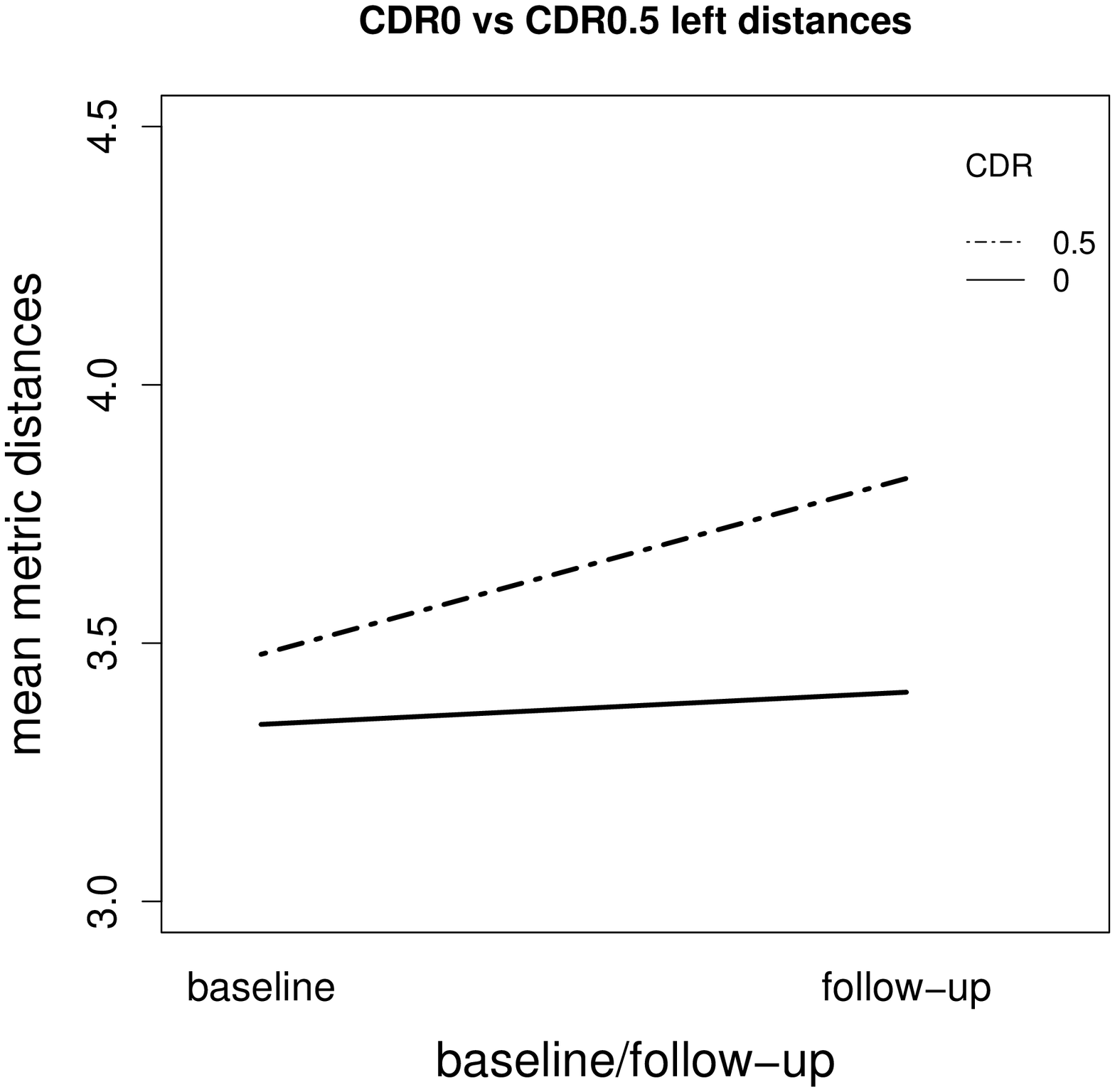}}}
\rotatebox{-90}{ \resizebox{2.5 in}{!}{\includegraphics{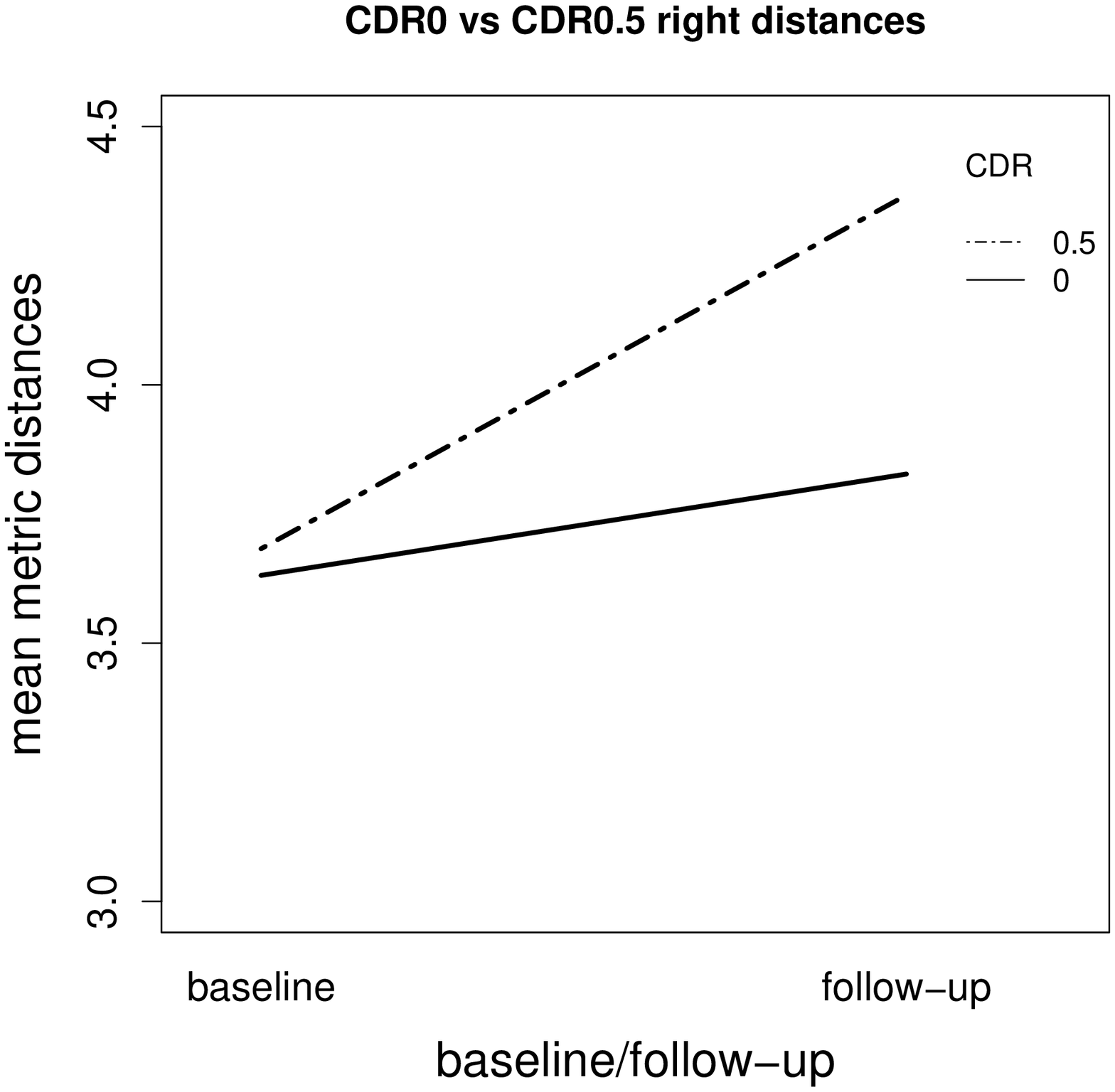}}}
\caption{
\label{fig:interaction-plot-CDR0vsCDR0.5inLeftandRight}
Interaction plots for diagnosis levels over the timepoint
levels for left and right metric distances. Although the slopes are
different for both left and right hippocampi, the difference in the
right seems to be much larger.
}
\end{figure}

\end{document}